\documentclass[pre,graphicx,twocolumn,reprint]{revtex4}

\usepackage{amsmath}
\usepackage{enumerate}
\usepackage{graphicx}
\usepackage{mathbbol}
\usepackage{amsfonts}
\usepackage{natbib}

\newcommand{\nn}{\nonumber}

\newcommand{\half}{\frac{1}{2}}

\newcommand{\rmd}{{\mathrm{d}}}
\newcommand{\gras}[1]{\boldsymbol{#1}}
\newcommand{\grasrm}[1]{\boldsymbol{\mathrm{#1}}}
\newcommand{\boldlbd}{\boldsymbol\lambda}
\newcommand{\boldmu}{\boldsymbol\mu}

\graphicspath{{/Users/joachim/These/Watermelons/notes/figures_physrev/}
}

\begin{document}


\title{Distribution of the time at which $N$ vicious walkers reach their maximal height} 



\author{Joachim Rambeau}
\email[]{joachim.rambeau@th.u-psud.fr}
\affiliation{Laboratoire de Physique Th\'eorique d'Orsay, Universit\'e Paris Sud 11 and CNRS}

\author{Gr\'egory Schehr}
\email[]{gregory.schehr@th.u-psud.fr}
\affiliation{Laboratoire de Physique Th\'eorique d'Orsay, Universit\'e Paris Sud 11 and CNRS}


\date{\today}

\begin{abstract}
We study the extreme statistics of $N$ non-intersecting Brownian motions (vicious walkers) over a unit time interval in one dimension. 
Using path-integral techniques we compute exactly the joint distribution of the maximum $M$ and of the time $\tau_M$ at which this maximum
is reached. We focus in particular on non-intersecting Brownian bridges ("watermelons without wall") and non-intersecting Brownian excursions
("watermelons with a wall"). We discuss in detail the relationships between such vicious walkers models in watermelons configurations and 
 stochastic growth models in curved geometry on the one hand and the directed polymer in a disordered medium (DPRM) with one free end-point on the other hand. We also check our results
 using numerical simulations of Dyson's Brownian motion and confront them with numerical simulations of the Polynuclear Growth Model (PNG) and of a model of DPRM on a discrete lattice. Some of the results presented here were announced in a recent letter [J. Rambeau and G. Schehr, Europhys. Lett. {\bf 91}, 60006 (2010)].

\end{abstract}

\pacs{}

\maketitle 

\section{Introduction and Motivations}

Extreme value statistics (EVS) is at the heart of optimization problems and, as such, it plays a crucial role in the theory of 
complex and disordered systems \cite{Bouchaud97, LeDoussal03}. For instance, to characterize the thermodynamical properties of a disordered or glassy system at low temperature, one is often interested in computing its ground state, {\it i.e.} the configuration with {\it lowest} energy. Similarly, the low temperature dynamics of such systems is determined, at large times, by the largest energy barriers of the underlying free energy landscape. Thus the distribution of relaxation times is directly related to the statistics of the {\it largest} barriers in the systems. Hence, EVS has been widely studied in various models of
statistical physics during the last few years~\cite{Dean01, Carpentier01, Gyorgyi03, Bertin06, Fyodorov08, Fyodorov09}. 
This renewal of interest for EVS is certainly not restricted to physics but extends far beyond to biology~\cite{Omalley09}, finance~\cite{Embrecht97, Majumdar_Bouchaud08} or environmental sciences~\cite{Katz02} where extreme events may have drastic consequences.   
 
EVS of a collection of ${\cal N}$ random variables $X_1, \cdots, X_{\cal N}$, which are identical and independent, or weakly correlated, is very well understood, thanks to the identification of three different universality classes in the large ${\cal N}$ (thermodynamical) limit \cite{Gumbel58}. In that case, the distribution of the maximum $X_{\max}$ (or the minimum $X_{\min}$), properly shifted and scaled, converges to either (i) the Gumbel, or (ii) the Fr\'echet or (iii) the Weibull distribution, depending on the large argument behavior of the parent distribution of the $X_i$'s. By contrast, much less is known about the EVS of {\it strongly} correlated variables. In this context, recent progress has been done in the study of EVS of one-dimensional stochastic processes
which provide instances of sets of strongly correlated variables whose extreme statistics can be studied analytically. These include Brownian motion (BM) and its variants -- which can be often mapped onto elastic lines in $1+1$ dimensions -- \cite{Ray01, Majumdar04, Majumdar05, Schehr06, Gyorgyi07, Majumdar08, Rambeau09, Randon09, Majumdar10_convex, Krapivsky10}, continuous time random walks and Bessel processes \cite{Schehr10}, or the random acceleration process (RAP) \cite{Burkhardt07, Majumdar10}. Incidentally, it was shown that extreme value questions for such one-dimensional processes have nice applications in the study of random convex geometry in two dimensions. For instance, the extreme statistics of one-dimensional Brownian motion enters into the study of the convex hull of planar Brownian motion \cite{Randon09, Majumdar10_convex} while extreme statistics of the RAP appears in the Sylvester's problem where one studies the probability $p_n$ that $n$ points randomly ch
 osen in the unit disc are the vertices of a convex $n$-sided polygon \cite{Hilhorst08}. These stochastic processes involve a single degree of freedom, or several non-interacting degrees of freedom as in Ref. \cite{Randon09, Majumdar10_convex, Krapivsky10}, where $N$ independent Brownian motions are considered. A natural way to include the effects of interactions in a "minimal", albeit non trivial, model is to constraint these $N$ Brownian motions not to cross: this yields a model of $N$ non-intersecting Brownian motions, which we will focus on in this paper.

\begin{figure}[h]
\includegraphics[width=\linewidth]{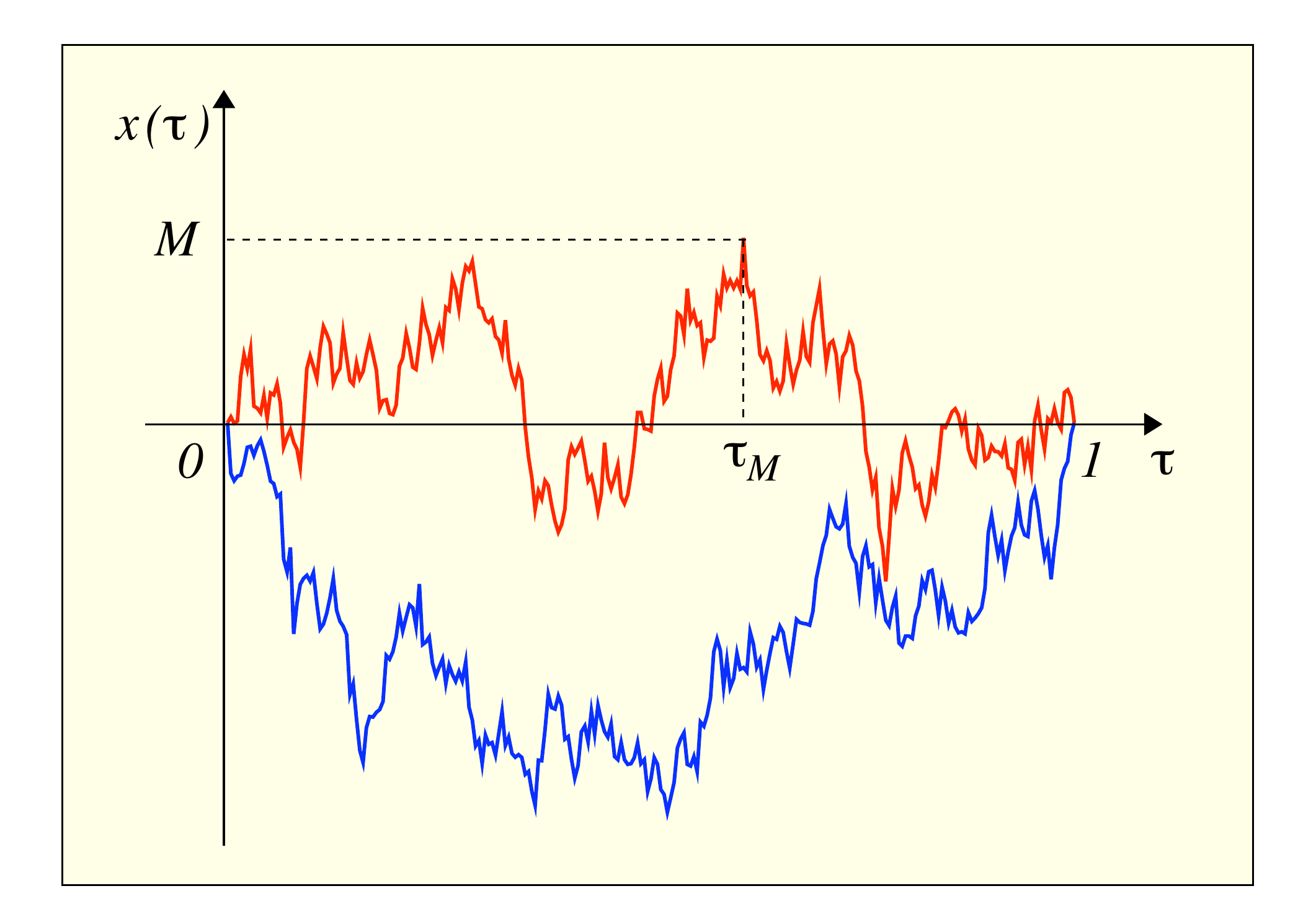}
\caption{One realization of $N=2$ non-intersecting bridges, {\it i.e.} a watermelon configuration, 
with the maximum $M$ and the time $\tau_M$ at which this maximum is reached.}
\label{real_bridge2}
\end{figure}

Here we thus consider $N$ non-colliding Brownian motions 
\begin{eqnarray}
x_1(\tau)<\cdots < x_N(\tau) \;, \; \forall \; \tau \in [0,1] \;,
\end{eqnarray}
on the unit time interval, $\tau \in [0,1]$. We will consider different types of configurations of such vicious walkers, which are conveniently represented in the $(x,t)$ plane. In the first case, called "watermelons", all the walkers start, at time $\tau =0$ and end, at time  $\tau =1$, at the origin (see Fig.~\ref{real_bridge2}). They thus correspond to $N$ non-intersecting Brownian {\it bridges}. In the second case, we will consider such watermelons with the additional constraint that the positions of the walkers have to stay positive (see Fig. \ref{pdf_star2}): we will call these configurations ''watermelons with a wall'' and they thus correspond to non-intersecting Brownian excursions. Finally, we will also consider "stars" configuration, where the endpoints of the Brownian motions in $\tau = 1$ are free. They correspond to non-intersecting free Brownian motions, a realization of which is shown in Fig.~\ref{real_star2} for $N=2$. 

Motivated by extreme value questions, we will compute here the joint distribution $P_N(M, \tau_M)$ of the maximal height and the time at which this maximum is reached (see Fig.~\ref{real_bridge2}): 
\begin{eqnarray}
M = \max_{0 \leq \tau \leq 1} x_N(\tau) \;, \; x_N(\tau_M) = M \;.
\end{eqnarray}
The (marginal) distribution of $M$ has recently been studied by several authors \cite{Feierl07, Katori08, Schehr08, Kobayashi08, Feierl09, Izumi10, Forrester10}, while we have announced exact results for this joint distribution in the two first cases (for bridges and excursions) in a recent Letter~\cite{Rambeau10}. The goal of this paper is to give a detailed account of the method and the computations leading to these results. On this route we will also provide several new results, including for instance results for the star configuration.

In the physics literature, such non-intersecting Brownian motions were first introduced by de Gennes in the context of fibrous polymers~\cite{deGennes68}. They were then widely studied after the seminal work of Fisher~\cite{Fisher84}, who named them "vicious walkers", as the process is killed if two paths cross each other. These models have indeed found many applications in statistical physics, ranging from 
wetting and melting transitions~\cite{Fisher84}, commensurate-incommensurate transitions~\cite{Huse84}, networks of polymers~\cite{Essam95} to persistence properties in nonequilibrium systems~\cite{Bray04}. 

Vicious walkers have also very interesting connections with random matrix theory (RMT), in particular through Dyson's Brownian motion~\cite{Dyson62}. For example, for watermelons configurations, it can be shown that the positions of the $N$ random walkers at a fixed time $\tau$, correctly scaled by a $\tau$ dependent factor, are distributed like the $N$ eigenvalues of the random $N \times N$ Hermitian matrices of the Gaussian Unitary Ensemble of RMT corresponding to $\beta = 2$~\cite{Mehta91}. If one denotes by $P_{\rm joint}({\mathbf x}, \tau) \equiv P_{\rm joint}(x_1, \cdots, x_N, \tau)$ the joint distribution of the positions of the walkers at a given time $\tau$, in the watermelons configuration, one has indeed (see for instance Ref. \cite{Schehr08} for a rather straightforward derivation of this result):
\begin{eqnarray}
P_{\rm joint}({\mathbf x}, \tau) = Z_N^{-1} \sigma(\tau)^{-N^2} \prod_{1\leq i<j \leq N} (x_i - x_j)^2 e^{- \frac{{\mathbf x}^2}{2 \sigma^2(\tau)}} \;, 
\end{eqnarray}
where $Z_N$ is a normalization constant and $\sigma(\tau) = \sqrt{\tau(1-\tau)}$ and where we use the notation ${\mathbf x}^2 = \sum_{i=1}^N x_i^2$. This means in particular for the top path that $x_N(\tau)/\sqrt{\tau(1-\tau)}$ is, at fixed $\tau$, distributed like the largest eigenvalue of GUE random matrices, which means that in the large $N$ limit one has \cite{Tracy94-96}
\begin{eqnarray}\label{top_path}
\frac{x_N(\tau)}{\sqrt{2} \, \sigma(\tau)} = \sqrt{ 2 N} + \frac{1}{\sqrt{2}} N^{-1/6} \chi_2 \;,
\end{eqnarray}
where $\chi_2$ is distributed according to the Tracy-Widom (TW) distribution for $\beta=2$, ${\cal F}_2$, namely $\Pr[\chi_2 \leq x] = {\cal F}_2(x)$. In the case of watermelons with a wall, the positions of the random walkers are instead related to the eigenvalues of Wishart matrices \cite{Schehr08, Nadal09} and the fluctuations of the top path $x_N(\tau)$ are again described by ${\cal F}_2(x)$.       

\begin{figure}
\begin{center}
\includegraphics[width = \linewidth]{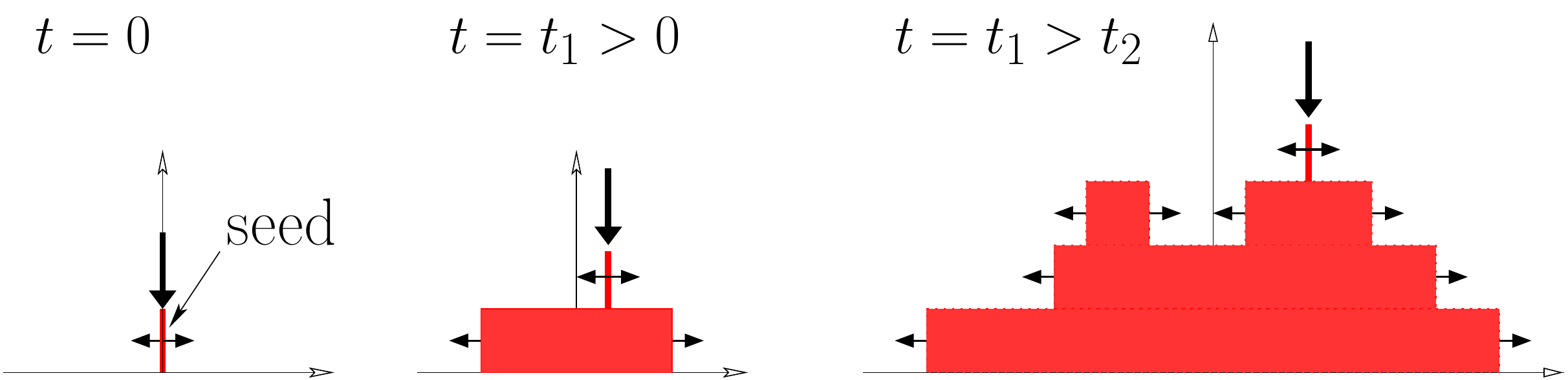}
\caption{Dynamical rules  of the PNG model (in continuous time) as described in the text.}\label{cartoon_png}
\end{center}
\end{figure}

Yet another reason why there is currently a rekindled interest in vicious walkers problems is because of their connection with
stochastic growth processes in the Kardar-Parisi-Zhang (KPZ) universality class \cite{Kardar86, Krug91} in $1+1$ dimensions. This connection
is believed to hold for any systems belonging to the KPZ universality class \cite{Sasamoto10}, and it can be rigorously shown on one particular model of stochastic growth, the so-called polynuclear growth model (PNG) \cite{Franck74, Krug89}. It is defined as follows (see Fig.~\ref{cartoon_png}). At time $t=0$ a single island starts spreading on a flat substrate at the origin $x=0$ with unit velocity. Seeds of negligible size then nucleate randomly at a constant rate $\rho=2$ per unit length and unit time and then grow laterally also at unit velocity. When two islands on the same layer meet they coalesce. Meanwhile, nucleations continuously generate additional layers. In the {\it flat geometry} nucleations can occur at any point $x$ while  
in {\it the droplet geometry}, which we will focus on, nucleations only occur above previously formed layers. 
Therefore, for the droplet geometry, 
denoting by $h_{\rm drop}(x,t)$ the height of the interface at point $x$ and time $t$, one has $h_{\rm drop}(x,t) = 0$ for $|x| > t$.  
On the other hand, in the long time limit, the profile for $|x| \leq t$ becomes droplet-like $\langle h_{\rm drop}(x,t) \rangle \sim 2t \sqrt{1-(x/t)^2}$ \cite{praeho_thesis}, but there remain height fluctuations around this mean value (see Fig.~\ref{fig_intro} left). The standard way to characterize these fluctuations is to look at the width of the interface  $W_L (t)= \langle (h_{\rm drop}(x,t) - \langle h_{\rm drop}(x,t) \rangle)^2 \rangle^{\frac{1}{2}}  \sim L^{\zeta}{\cal W}(t/L^z)$ with universal exponents 
$\zeta = \frac{1}{2}$ and $z = \frac{3}{2}$ \cite{Dhar87, Gwa92}, in agreement with the fact that this model belongs to the KPZ universality class. More recently, it was shown for several models belonging to the KPZ universality class~\cite{Praehofer00, Praehofer02, Johansson00, Gravner01, Majumdar_Nechaev04, Majumdar06} that, in the growth regime $t_{0} \ll t \ll L^z$ (where $t_{0}$ is a microscopic time scale) universality extends far beyond
the exponents $\zeta$ and $z$ but also applies to full distribution functions of physical observables. In particular, the scaled cumulative distribution of the height field at a given point coincides with the TW distribution ${\cal F}_{\beta}$ with $\beta = 2$ (respectively $\beta = 1$) for the curved geometry (respectively for the flat one), which describes the edge of the spectrum of random matrices in the Gaussian Unitary Ensemble (respectively of the Gaussian Orthogonal Ensemble) \cite{Tracy94-96}. Height fluctuations were measured in experiments, both in planar \cite{Miettinen05} and more recently in curved geometry in the electroconvection of nematic liquid crystals~\cite{Takeuchi10} and a good quantitative agreement with TW distributions was found.

 Although the relation with random matrices were initially achieved \cite{Praehofer00} through the longest increasing subsequence of random permutations \cite{Aldous99, Majumdar06}, using in particular the results of the seminal paper by Baik, Deift and Johansson \cite{Baik99}, it was then realized that the PNG model in the droplet geometry is actually directly related to the vicious walkers problem in the watermelon configuration~\cite{Praehofer02}. This was shown through an extension of the PNG model to the so called multi-layer PNG model, where non-intersecting paths naturally appear. Indeed, one can show that the fluctuations of the height field of the PNG model in the droplet geometry $h_{\rm drop}(x,t)$ are related, in the large time limit to the fluctuations of the top path $x_N(\tau)$ for the vicious walkers problem in the watermelon geometry in the large $N$ limit. This mapping, for $N, t \gg 1$ reads~\cite{Praehofer02, Ferrari08} : 
\begin{align}\label{def_airy}
 \frac{h_{\rm drop}(u t ^{\frac{2}{3}},t) - 2 t}{t^{\frac{1}{3}}} &\equiv 
 \frac{2 \left( x_N(\frac{1}{2}+ \frac{u}{2N^{\frac{1}{3}}}) - N^{\frac{1}{2}} \right)}{N^{-\frac{1}{6}}} \nn \\
 &\equiv {\cal A}_2(u) - u^2,
\end{align}
where ${\cal A}_2(u)$ is the Airy$_2$ process \cite{Praehofer02} which is a stationary, and non-Markovian, process. In particular, 
$\Pr[{\cal A}_2(0) \leq x] = {\cal F}_2(x)$, which is consistent with Eq.~(\ref{top_path}). Hence, from Eq.~(\ref{def_airy}) $x_N$ and $\tau$ map onto $h$
and $x$ in the growth model while $N$ plays essentially the role of $t$ (to make the correspondence between $N$ and $t$ exact one has to consider a watermelon configuration in the interval $\tau \in [0,N]$). To characterize the fluctuations of the height profile in the droplet geometry beyond the standard roughness $W_L(t)$ it is natural to consider the maximal height $M$ and its position $X_M$~\cite{Rambeau10} (see Fig.~\ref{fig_intro}). According to KPZ scaling, one expects $M-2t \sim t^{1/3}$ while $X_M \sim t^{2/3}$. On the other hand from Eq. (\ref{def_airy}), the joint distribution $P_t(M, X_M)$ of $M$ and $X_M$ can be written as
\begin{eqnarray}\label{rel_png_airy}
P_t(M, X_M) \sim t^{-1} {\cal P}_{\rm Airy}\left((M-2t)t^{-\frac{1}{3}}, X_M t^{-\frac{2}{3}} \right) \;,
\end{eqnarray}  
where ${\cal P}_{\rm Airy}(y,x)$ is the joint distribution of the maximum $y$ and its position $x$ for the process ${\cal A}_2(u) - u^2$. Finally, the relation (\ref{def_airy}) also gives us some interesting information for the vicious walkers problem. In the large $N$ limit, one has indeed from Eq. (\ref{def_airy}) that the joint distribution $P_N(M, \tau_M)$ for the watermelons configuration is also given by ${\cal P}_{\rm Airy}(y,x)$. One has indeed, for $N \gg 1$:
\begin{multline}\label{rel_wt_airy}
P_N(M, \tau_M) \\
\sim 4N^{\frac{1}{2}} {\cal P}_{\rm Airy}\left(2(M-\sqrt{N})N^{\frac{1}{6}}, 2(\tau_M - \frac{1}{2})N^{\frac{1}{3}} \right) \;,
\end{multline}
and therefore Eq. (\ref{rel_png_airy}) and Eq. (\ref{rel_wt_airy}) show that one can obtain the distribution of $P_t(M, X_M)$ for the growth model from the large $N$ limit of the joint distribution $P_N(M, \tau_M)$, which we compute here for any finite $N$. 
\begin{widetext}
\begin{center}
\begin{figure}[h]
\begin{center}
\begin{minipage}{0.45\linewidth}
\begin{center}
\includegraphics[width= \linewidth]{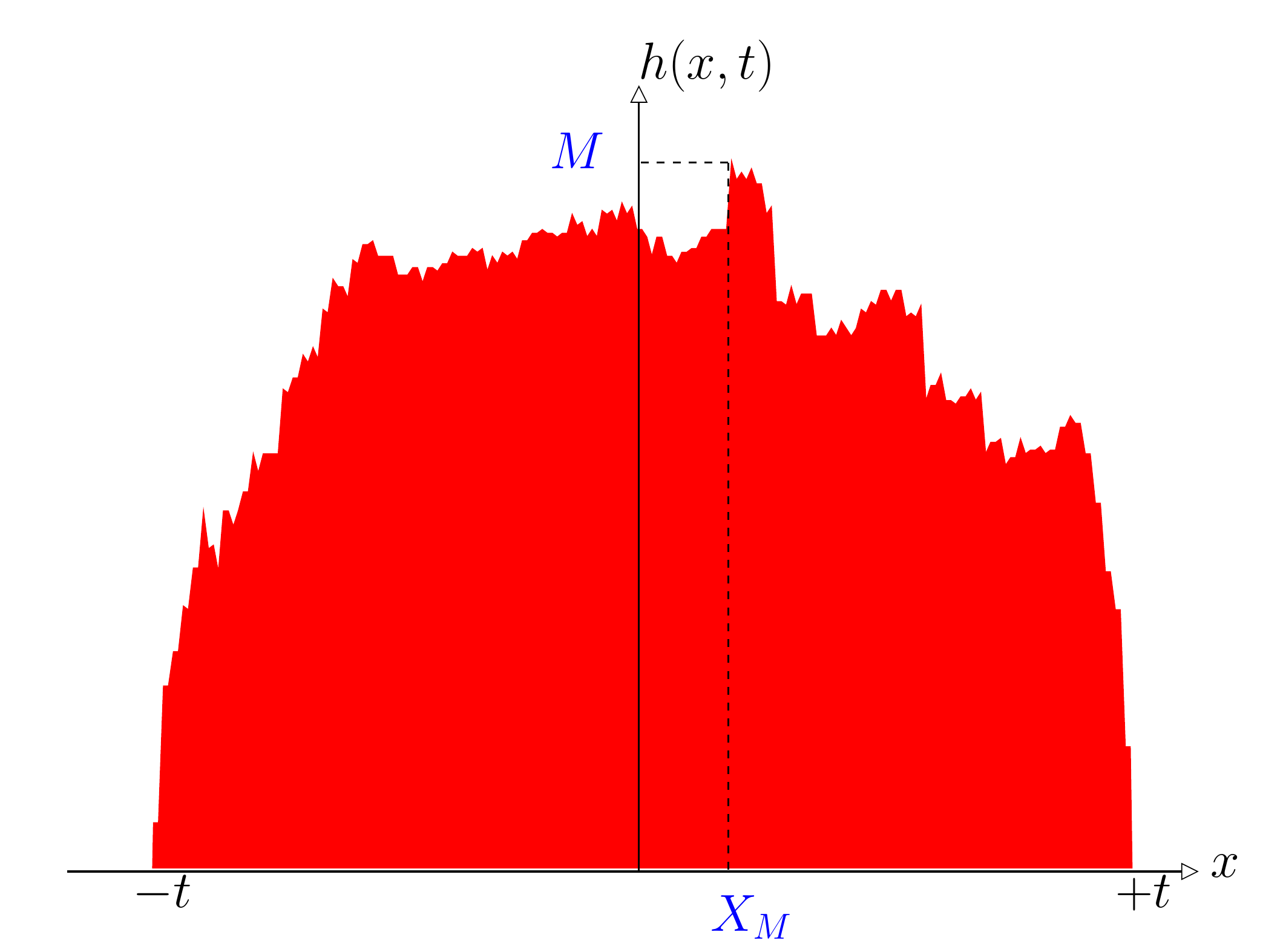}
\end{center}
\end{minipage}\hfill
\begin{minipage}{0.45 \linewidth}
\begin{center}
\vspace*{1.2cm}
\includegraphics[width= \linewidth]{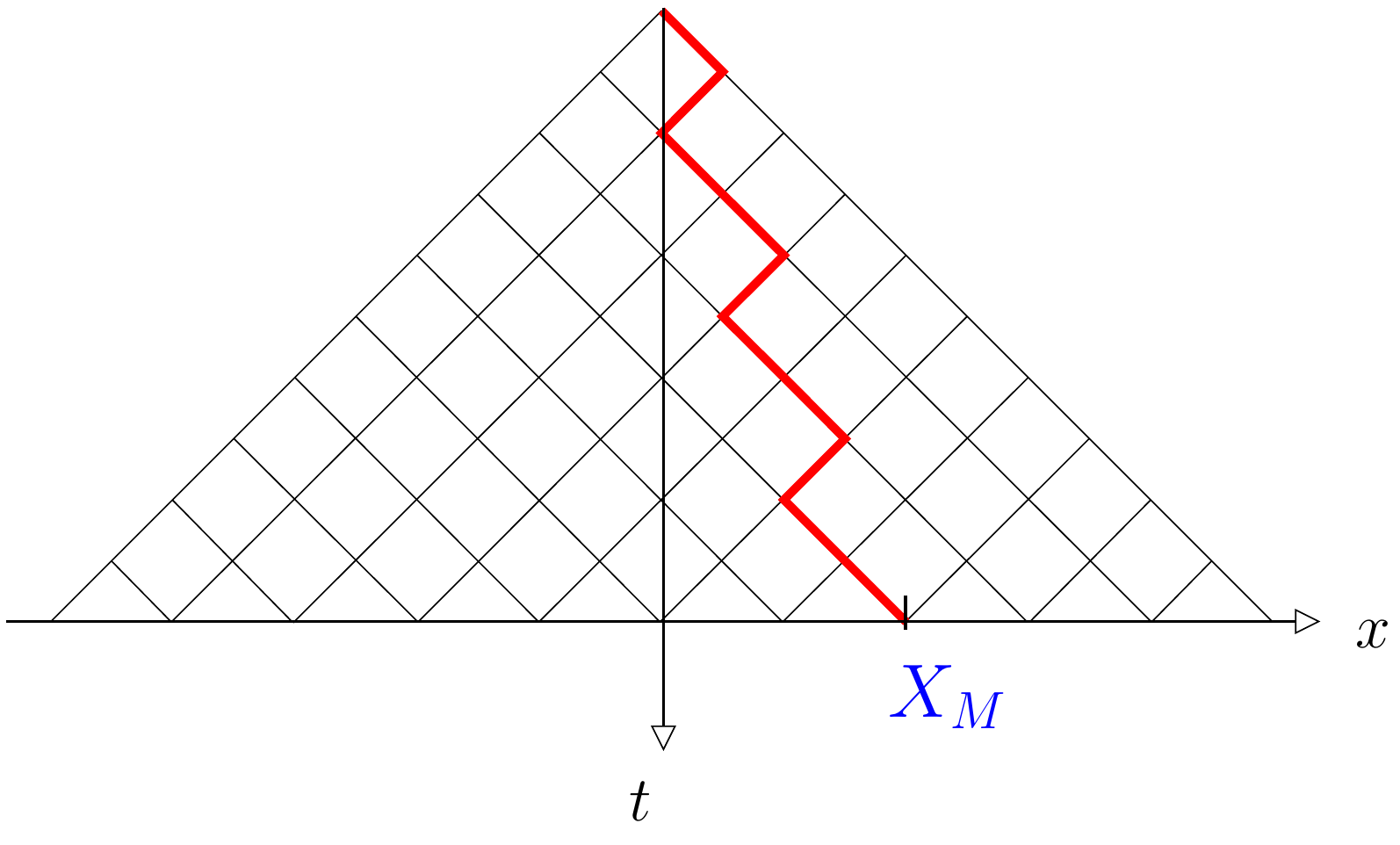}
\end{center}
\end{minipage}
\caption{{\bf Left :} Height profile $h_{\rm drop}(x,t)$ at fixed time $t$, as a function of $x$ for the PNG model in the droplet geometry. $X_M$ is the position at which the maximal height $M$ is reached. {\bf Right : } Directed polymer with one free end. Here $M$ corresponds to the energy of the optimal polymer, and $M - 2{t} \sim {\cal O}(t^{1/3})$ while $X_M \sim {\cal O}(t^{2/3})$ corresponds to the transverse coordinate of the end point of this optimal polymer.}\label{fig_intro}
\end{center}
\end{figure}
\end{center}
\end{widetext}

It is also well known that stochastic growth models in the KPZ universality class can be mapped onto the model of the directed polymer in a disordered medium (DPRM) \cite{Krug91, Halpin95}. This is also the case of the PNG model and to make this connection as clear as possible, we consider a discrete version of the PNG model introduced by Johansson \cite{Johansson03}. It is a growth model with discrete space {\it and} discrete time. The height function $h(x,t)$ is now an integer value $h(x,t) \in {\mathbb{N}}$ while $x \in {\mathbb Z}$ and $t \in {\mathbb{N}}$. The dynamics is defined as follows
\begin{eqnarray}\label{discrete}
h(x,t+1) &=& \max[h(x-1,t), h(x,t), h(x+1,t)] \nonumber \\
&+& \omega(x,t+1) \;,
\end{eqnarray}   
where the first term reproduces the lateral expansion of the islands (and also their coalescence when two of them meet) in the continuous PNG model while $\omega(x,t) \in {\mathbb N}$ is a random variable, distributed independently from site to site, which corresponds to the random nucleations (see Fig. \ref{cartoon_png}).

\begin{widetext}
\begin{center}
\begin{figure}
\begin{center}
\includegraphics[width = 0.8\linewidth]{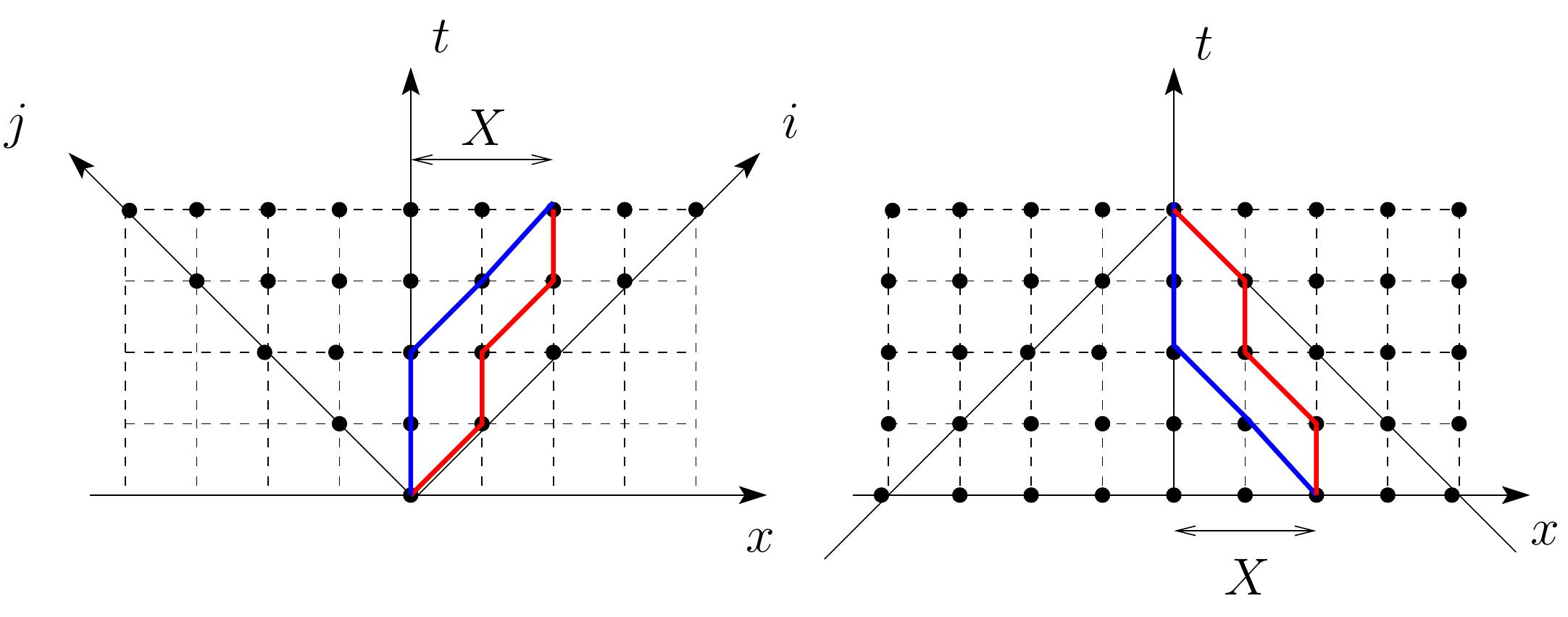}
\caption{{\bf Left} :  Model of a directed polymer in a random medium corresponding to the PNG in the droplet geometry~(\ref{drop_dprm}). On each site $\bullet$ there is a random energy variable $w(i,j)$ which corresponds to a nucleation in the growth model. In red and blue we have drawn two configurations of the polymer which belong to the ensemble ${\mathfrak P}_t(0,X)$. {\bf Right} : Model of a directed polymer in a random medium corresponding to the PNG in the flat geometry~(\ref{flat_dprm}). In red and blue we have drawn two configurations of the polymer which belong to the ensemble ${\mathfrak P}_t(X,0)$ which is equivalent to 
${\mathfrak P}_t(0,X)$  shown on the left panel.}\label{fig_mapping_dprm}
\end{center}
\end{figure}
\end{center}
\end{widetext}
In the droplet geometry, one has $\omega(x,t) = 0$ if $|x| > t$ while this constraint is removed in the flat geometry. To understand better the connection between this model (\ref{discrete}) and a model of directed polymer, we perform a rotation of the axis $(x,t)$ (Fig. \ref{fig_mapping_dprm}) and define
\begin{eqnarray}
\omega(i-j,i+j-1) \equiv w(i,j) \;.
\end{eqnarray}
One can then simply check (for instance by induction on $t$), that the height field $h_{\rm drop}(x,t)$ evolving with Eq.~(\ref{discrete}) in the droplet geometry, is given by
\begin{eqnarray}\label{drop_dprm}
h_{\rm drop}(X,t) = \max_{{\cal C} \in {\frak P}_t(0,X)} \left [\sum_{(i,j) \in {\cal C}} w(i,j) \right ] \;,
\end{eqnarray}
where ${\mathfrak P}_t(0,X)$ is the ensemble of directed paths of length $t$, starting in $x=0$ and terminating in $X$ (Fig. \ref{fig_mapping_dprm}). This formula (\ref{drop_dprm}) establishes a direct link between the PNG model and a model of directed polymer. Indeed, the height field $h_{\rm drop}(X,t)$ corresponds the energy of the optimal polymer (which is here the polymer with the highest energy) with both fixed ends in $0$ and $X$ and where the random energies, on each site, are given by $w(i,j)$. The aforementioned result for the distribution of $h_{\rm drop}(X,t)$ shows that the fluctuations of the energy of the optimal polymer are described in this case by ${\cal F}_2$, the Tracy-Widom distribution for $\beta = 2$. The study of this distribution, for the  DPRM in continuum space, has recently been the subject of several works, both in physics \cite{dprm_doussal} and in mathematics~\cite{dprm_math}.     

Similarly, one can also write the height field $h_{\rm flat}(0,t)$ in the flat geometry (here of course the fluctuations of the height field are invariant by a translation along the $x$ axis). One has indeed : 
\begin{equation}\label{flat_dprm}
h_{\rm flat}(0,t) = \max_{-t \leq X\leq t} \left[   \max_{{\cal C} \in {\frak P}_t(X,0)} \left [\sum_{(i,j) \in {\cal C}} w(i,j) \right ]      \right] \;,
\end{equation}  
where ${\mathfrak P}_t(X,0)$ is the ensemble of directed paths of length $t$, starting in $X$ and ending in $0$ (see Fig.~\ref{fig_mapping_dprm}). Given that these two ensembles ${\mathfrak P}_t(X,0)$ and ${\mathfrak P}_t(0,X)$ are obviously similar~(Fig. \ref{fig_mapping_dprm}), these two equations 
(\ref{drop_dprm}) and (\ref{flat_dprm}) allow to write~\cite{krug_dprm}
\begin{eqnarray}\label{rel_drop_flat}
h_{\rm flat}(0,t) = \max_{-t \leq X \leq t}  h_{\rm drop}(X,t)  \;.
\end{eqnarray}
For the directed polymer, $h_{\rm flat}(0,t)$ corresponds to the energy of the optimal polymer, of length $t$ and with one free end. From that Eq.~(\ref{rel_drop_flat}), one sees that the joint distribution of $M$ and $\tau_M$ in the vicious walkers problem corresponds to the joint distribution of the energy $M$ and the transverse coordinate $X_M$ of the free end of the optimal polymer. From the aforementioned results for the distribution of $h_{\rm flat}(0,t)$, we also conclude that the cumulative distribution of $M$ is given by ${\cal F}_1$, the Tracy-Widom distribution for $\beta = 1$. Although this result was obtained rather indirectly using this relation (\ref{rel_drop_flat}), it was recently shown directly by computing the cumulative distribution of the maximal height of $N$ non-intersecting excursions in the limit $N \to \infty$, in Ref.~\cite{forrester_npb}. We conclude this paragraph about the DPRM by noticing that this model, with one free end, has been widely studi
 ed in the context of disordered elastic systems~\cite{toy_model}, in particular using the approximation of the so-called "toy-model" where the Airy$_2$ process ${\cal A}_2(u)$ in (\ref{def_airy}) is replaced by a Brownian motion~\cite{LeDoussal03, toy_model}. A recent discussion of the toy model and its applications to the DPRM can be found in Ref.~\cite{vivien}. We also mention that physical observables related to $X_M$ have been studied for the DPRM in different geometry, like the winding number of the optimal polymer on a cylinder~\cite{brunet_cylindre}.

The paper is organized as follows. In section II we present in detail the method to compute the joint distribution of $M$ and $\tau_M$ for $N=2$ non-intersecting Brownian motions. We will show results for bridges, excursions and stars. In section III, we will extend this approach for any number $N$ of vicious walkers both for bridges and excursions. We will then confront, in section IV, our analytical results to numerical simulations, which were obtained
by simulating Dyson's Brownian motion, before we conclude in section V. Some technical details have been left in the appendices.

\section{Distribution of the position of the maximum for two vicious Brownian paths}
\label{TWO}
\subsection{Introduction}

Our purpose is to study $N$ simultaneous Brownian motions $x_1(\tau),x_2(\tau) \dots, x_N(\tau)$,
subjected to the condition that they do not cross each other. 
In the literature, such system has been studied first in the continuous case by de Gennes~\cite{deGennes68},
and in the discrete case by Fisher~\cite{Fisher84}. 
He named this kind of system vicious walkers, as the process is killed if two paths cross each others.
The name `vicious Brownian paths' will refer in the following to Brownian paths under the condition that they do not cross each other.
\par
\begin{figure}[h]
\includegraphics[width=\linewidth]{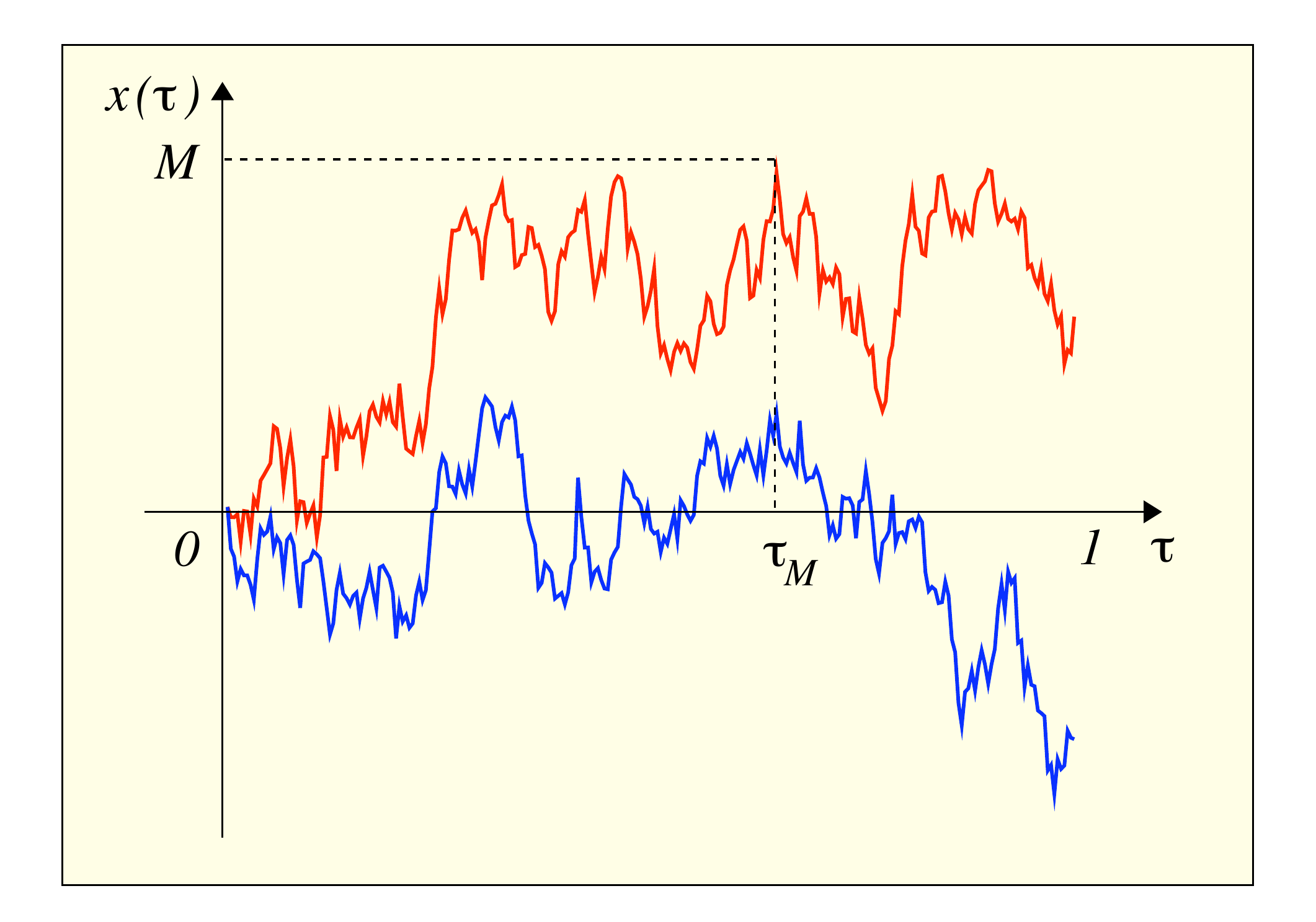}
\caption{One realization of $N=2$ non-intersecting free Brownian motions, {\it i.e.} in "star" configuration, 
with the maximum $M$ and the time $\tau_M$ at which this maximum is reached.}
\label{real_star2}
\end{figure}
\begin{figure}[t]
\includegraphics[width=\linewidth]{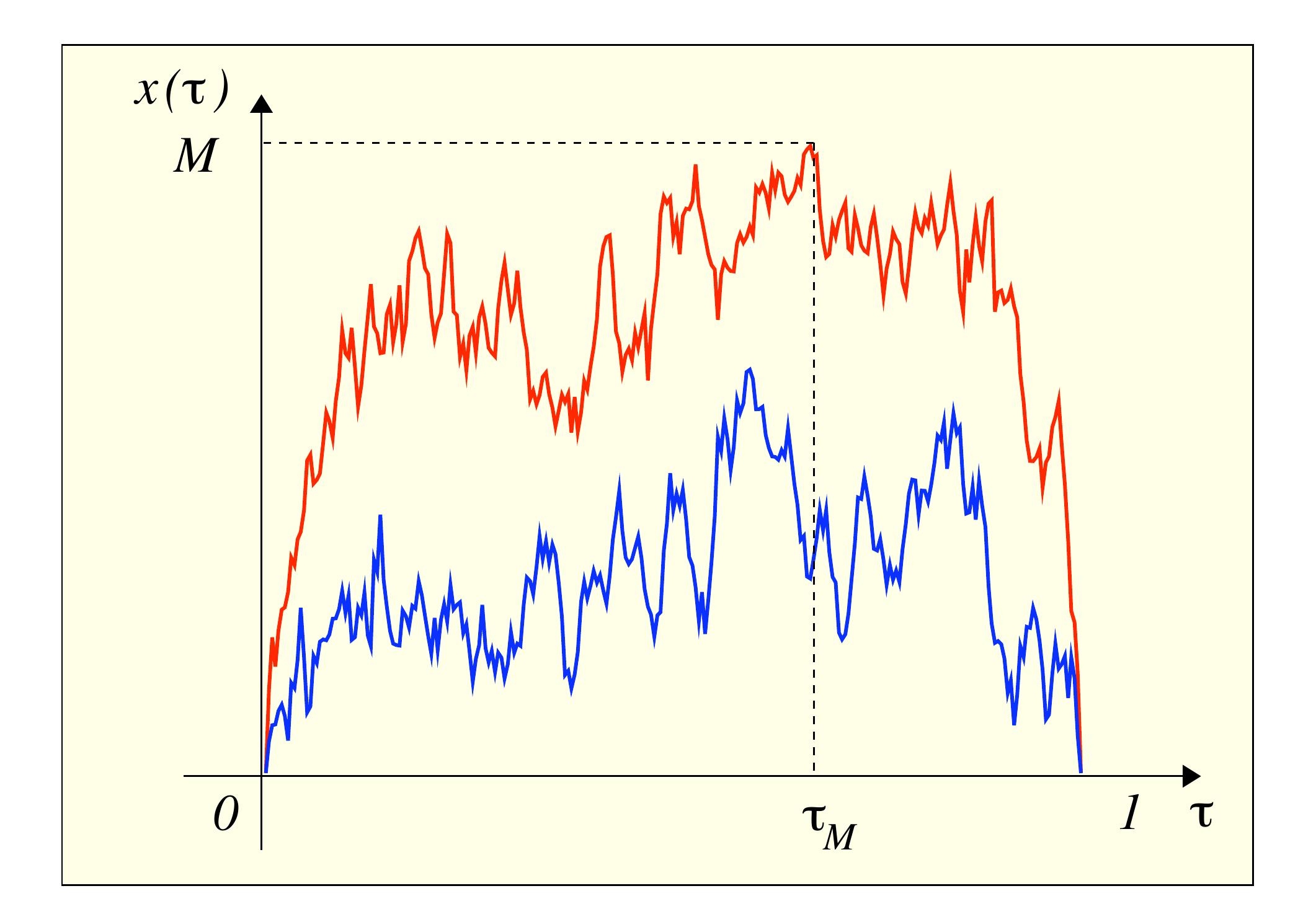}
\caption{One realization of the N=2 excursions configuration, with the maximum $M$ and the time to reach it $\tau_M$.}
\label{real_excur2}
\end{figure}

In this Section, we focus on the case $N=2$.
The two vicious Brownian paths start at $x_1(0)=x_2(0)=0$ and obey the non-crossing condition $x_1(\tau)<x_2(\tau)$
for $0<\tau<1$. We examine three different geometries:
\begin{enumerate}[A)]
\item [(i)] periodic boundary conditions where we consider non-intersecting bridges, also called ''watermelons'' configurations (see Fig.~\ref{real_bridge2}): this will be treated in subsection~E,
\item [(ii)] free boundary conditions where we consider non-intersecting free Brownian motions, also called ''stars'' configurations (see Fig.~\ref{real_star2}): this will be treated in subsection F, 
\item [(iii)] periodic boundary conditions, plus a positivity constraint where we consider non-intersecting excursions, also called ''watermelons with a wall'' configurations (see Fig.~\ref{real_excur2}): this will be treated in subsection G.
\end{enumerate}
\par
We compute, for each case, the joint distribution of the couple $(M,\tau_M)$ which are respectively the maximum $M$ 
of the uppermost vicious Brownian path, and its position $\tau_M$: $M=\max_{0\leq \tau \leq 1} x_2(\tau) =x_2(\tau_M)$.
Integrating over $M$, we find the probability distribution function (pdf) of the time to reach the maximum $\tau_M$.
\par
This computation is based on a path integral approach, using the link between the problem of vicious Brownian paths
and the quantum mechanics of fermions. This is described in subsection B. 
Then we present, in subsection C, the method to compute the joint probability distribution function of the maximum and the time to reach it
for vicious Brownian paths, with general boundary conditions.
An emphasis is put on the regularization scheme, essential to circumvent the basic problem of Brownian motion which is continuous both in space and time:
you cannot force it to be in a point without re-visiting it infinitely many times immediately after.
At this stage, general formulae will be obtained, and will be used directly in the paragraphs E, F and G.
The extension to a general $N$ number of paths will be treated straightforwardly in Section III.
\subsection{Path integral approach: treating the vicious Brownian paths as fermions}

Let us consider two Brownian motions, obeying the Langevin equations
\begin{equation}
\frac{\rmd x_j(\tau)}{\rmd \tau} = \xi_j(\tau), \quad (j=1,2),
\end{equation}
where $\xi_j(\tau)$'s are independent and identical Gaussian white noises of zero mean, {\it i.e.} $\langle \xi_j(\tau) \rangle=0$ and 
$\langle \xi_i(\tau) \xi_j(\tau') \rangle = \delta_{i,j} \delta(\tau-\tau')$.
Taking into account the non-crossing condition, the probability that the two paths end at $(x_1(t_b)=b_1,x_2(t_b)=b_2)$ at time $t_b$,
given that they started at $(x_1(t_a)=a_1,x_2(t_a)=a_2)$  at time $t_a$ is
\begin{multline}
\label{path_integral_start}
p_2(b_1,b_2,t_b|a_1,a_2,t_a)= \frac{1}{z} \int\limits_{x_1(t_a)=a_1}^{x_1(t_b)=b_1} \! \! \! \! \! \! \mathcal{D}x_1(\tau)
\int\limits_{x_2(t_a)=a_2}^{x_2(t_b)=b_2} \! \! \! \! \! \! \mathcal{D}x_2(\tau) \\
\times \Big\{ e^{-\int_{t_a}^{t_b} \frac{1}{2} \left(\frac{\rmd x_1(\tau)}{\rmd \tau}\right)^2 \rmd \tau}
e^{-\int_{t_a}^{t_b} \frac{1}{2} \left(\frac{\rmd x_2(\tau)}{\rmd \tau}\right)^2 \rmd \tau} \\
\times \prod_{t_a<\tau<t_b} \theta\left[x_2(\tau)-x_1(\tau)\right] \Big\} ,
\end{multline}
where the subscript '$2$' in the notation $p_2$ means that we treat the case of $N=2$ Brownian paths.
The exponential terms are the weight of each Brownian path, and the Heaviside theta function ensures that the paths do not cross.
In this Section, bold letters will denote vectors with two components, 
such as for example the positions at a given time of the two Brownian paths $\gras{x}(\tau)=(x_1(\tau),x_2(\tau))$ or
the starting points $\gras{a}=(a_1,a_2)$. In the previous formula~(\ref{path_integral_start}), $z$ is a normalization constant, such that
\begin{equation}
\int_{-\infty}^{+\infty} \rmd b_2 \int_{-\infty}^{b_2} \rmd b_1\  p_2(\gras{b},t_b|\gras{a},t_a)= 1 .
\end{equation}
Notice that the integration is ordered, and we will write it shortly 
$\int_{\text{ord}} \rmd \gras{b} \equiv \int_{-\infty}^{+\infty} \rmd b_2 \int_{-\infty}^{b_2} \rmd b_1$.
\par
We recognize in formula~(\ref{path_integral_start}) 
the path integral representation of the quantum propagator of two identical free particles,
subjected not to cross each other. In one dimension, this can be implemented
by requiring that the two particles are fermions~\cite{deGennes68, Schehr08}. Therefore the probability $p_2(\gras{b},t_b|\gras{a},t_a)$ reduces to the quantum propagator of two identical fermions in one dimension:
\begin{equation}
\label{intro_propagator}
p_2(\gras{b},t_b|\gras{a},t_a)= 
\langle \gras{b}| e^{- (t_b-t_a) H_{\text{free}}} | \gras{a} \rangle,
\end{equation}
with $H_{\text{free}}$ the total Hamiltonian of the two-particle system 
$H_{\text{free}}=H_{\text{free}}^{(1)} \otimes \mathbb{I}^{(2)} + \mathbb{I}^{(1)} \otimes H_{\text{free}}^{(2)}$,
$\mathbb{I}^{j}$ being the identity operator acting in the Hilbert space of the $j^{\text{th}}$ particle.
The one-particle Hamiltonians are
$H_{\text{free}}^{(j)}=-\frac{1}{2}\frac{\rmd^2}{\rmd {x_j}^2} $ (for $j=1,2$) in a position representation. 
The states $|a_1,a_2\rangle$ (and $|b_1,b_2\rangle$) are the tensorial product of the one-particle eigenstates of position
$|a_1,a_2 \rangle = |a_1\rangle \otimes |a_2 \rangle$, where each state $|a_j\rangle$ obeys the
eigenvalue equation $X^{(j)} |a_j\rangle = a_j | a_j \rangle$, with $X^{(j)}$ the position operator acting in the
Hilbert space of the $j^{\text{th}}$ particle, and $a_j$ the position of this particle.
\par
In our calculations we deal with more general Hamiltonians, but they will be written in the same way as the sum of individual
one-particle Hamiltonians $H=H^{(1)} \otimes \mathbb{I}^{(2)}+ \mathbb{I}^{(1)} \otimes H^{(2)}$ 
(no interaction between the two particles), and with $H^{(1)}(x,p)=H^{(2)}(x,p)$ (identical particles). 
Thus we will treat systems of two identical indistinguishable particles, so that one can write down a two-particle eigenstate
of the total Hamiltonian $H$ as an antisymmetric product of the one-particle eigenstates of $H^{(j)}$.
Explicitely, if $|E_1\rangle$ and $|E_2\rangle$ are two eigenstates of the one-particle Hamiltonian $H^{(j)}$,
with energies $E_1$ and $E_2$, then the eigenstate of the two-particle problem with energy $E=E_1+E_2$ is
\begin{equation}
|E\rangle =\frac{1}{\sqrt{2!}} \left( |E_1\rangle \otimes |E_2 \rangle - | E_2 \rangle \otimes |E_1 \rangle \right) \;. 
\end{equation}
Projecting onto the position basis $| \gras{x} \rangle = |x_1\rangle \otimes |x_2 \rangle$,
and using the general notation $\phi_{E_i}(x_j)= \langle x_i | E_j \rangle$ (lower-case letter $\phi$ for one particle wave function),
and $\Phi_{E}(\gras{x})=\langle \gras{x} | E \rangle$ (capital letter $\Phi$ for the two particle wave function), this is the usual Slater determinant
\begin{align}
\Phi_E(x_1,x_2)&= \frac{1}{\sqrt{2!}} \left( \phi_{E_1}(x_1) \phi_{E_2}(x_2) - \phi_{E_2}(x_1) \phi_{E_1}(x_2) \right) \nn \\
&= \frac{1}{\sqrt{2!}} \det_{1\leq i,j \leq 2} \phi_{E_i}(x_j).
\end{align}
Hence inserting the closure relation $\mathbb{I}=\sum_{ E} | E \rangle \langle E |$ into~(\ref{intro_propagator}), one
obtains the spectral decomposition of the propagator (with $E=E_1+E_2$)
\begin{equation}
\label{propagator}
p_2(\gras{b},t_b|\gras{a},t_a)=  \sum_{E_1,E_2} \Phi_E(\gras{b}) \Phi^*_E(\gras{a}) e^{- (E_1+E_2) (t_b-t_a)}.
\end{equation}
%
%
%
%

This result can also be found \textit{via} the Karlin-McGregor formula~\cite{karlin_mcgregor}.
It states that the probability of propagating without crossings is nothing but the determinant formed 
from one-particle propagators (named $p_1$):
\begin{equation}\label{k_mcg}
p_2(\gras{b},t_b|\gras{a},t_a) = \det_{1 \leq i,j \leq 2} p_1(b_i,t_b|a_j,t_a).
\end{equation}
The discrete version of this formula (\ref{k_mcg}) is given by the Lindstr\"om-Gessel-Viennot (LGV) theorem \cite{LGV}. The basic idea behind this formula is nothing but the method of images.
To recover our quantum mechanical expression obtained by considering fermions (\ref{propagator}), one has to express the one-particle
propagator as
\begin{equation}
p_1(b_i,t_b|a_j,t_a) = \sum_{E} \phi_{E}(b_i) \phi^*_{E}(a_j) e^{-E(t_b-t_a)},
\end{equation}
and use the Cauchy-Binet identity (\ref{nice_formula_det}).
\par
\subsection{Joint probability distribution of position and time}

The method to compute the distribution of the maximum and the time at which it is reached can be described as follows.
Let us consider the process $(x_1(\tau),x_2(\tau))$ of such Brownian paths with the non-intersecting condition
with $t_a=0$ and $t_b=T$. Keeping the same notations, the starting points are $x_1(0)=a_1$ and $x_2(0)=a_2$,
or $\gras{x}(0)=\gras{a}$, and the end-points are $x_1(T)=b_1$ and $x_2(T)=b_2$, or
$\gras{x}(T)=\gras{b}$. First we consider $\gras{a}$ and $\gras{b}$ as fixed.
Given a set of points $\gras{y}=(y_1,y_2)$, and an intermediate time $\tau_{\text{cut}}$, such as $0\leq \tau_{\text{cut}} \leq T$,
we ask what is the probability that $x_1(\tau_{\text{cut}}) \in [y_1, y_1 + \rmd y_1]$  
and $x_2(\tau_{\text{cut}}) \in [y_2, y_2 + \rmd y_2]$ ?
In other terms, we seek $Q_2(\gras{y},\tau_{\text{cut}}|\gras{b},\gras{a},T)$ the joint probability density of $(\gras{y},\tau_{\text{cut}})$, 
given that the paths start in $\gras{a}$ at $\tau=0$ and end in $\gras{b}$ at $\tau=T$.
The answer to this question is well known from quantum mechanics: cut the process in two time-intervals, one from $\tau=0$
to $\tau=\tau_{\text{cut}}$, and the second piece from $\tau=\tau_{\text{cut}}$ to $\tau=T$. Because of
the Markov property, these two parts are statistically independent, and one should take the product of propagators
to obtain the joint pdf of $(\gras{y},\tau_{\text{cut}})=(y_1,y_2,\tau_{\text{cut}})$, given that the initial and final conditions
are respectively $(\gras{a},0)$, and $(\gras{b},T)$:
\begin{equation}
\label{joint_positions_time}
Q_2(\gras{y},\tau_{\text{cut}}|\gras{b},\gras{a},T)=
\frac{p_2(\gras{b},T|\gras{y},\tau_{\text{cut}}) \times p_2(\gras{y},\tau_{\text{cut}}|\gras{a},0)}{Z_2(\gras{b},\gras{a},T)} \;,
\end{equation}
where $Z_2(\gras{b},\gras{a},T)$ is the normalization constant, obtained by integrating over all possible cuts $\tau_{\text{cut}}$, 
and all possible intermediate points
\begin{equation}
\label{normalization_general}
Z_2=\int\limits_0^T \rmd \tau_{\text{cut}} \int_{\text{ord}} \rmd \gras{y} \
p_2(\gras{b},T|\gras{y},\tau_{\text{cut}}) \times p_2(\gras{y},\tau_{\text{cut}}|\gras{a},0) \;,
\end{equation}
where we remind that the integration over $\gras{y}$ is ordered:
$\int_{\text{ord}} \rmd \gras{y}=\int_{-\infty}^{\infty} \rmd y_2 \int_{-\infty}^{y_2} \rmd y_1$.
Using the closure relation of the states $|\gras{y}\rangle$, one finds that
\begin{equation}
\label{normalization_general_propagator}
Z_2(\gras{b},\gras{a},T)=T\ p_2(\gras{b},T|\gras{a},0) \;.
\end{equation}
Because we will not care about the lowest path $x_1(\tau)$ and its intermediate position $y_1$ in the following, 
we compute the marginal by integrating over all possible values of $y_1=x_1(\tau_{\text{cut}})$, 
obtaining $P_2(y_2,\tau_{\text{cut}}|\gras{b},\gras{a},T)$ the joint probability density that the upper path is in $y_2$ at time $\tau_{\text{cut}}$ given that the paths start in ${\bf a}$ at time $\tau=0$ and end in ${\bf b}$ at time $\tau = T$:
\begin{multline}
\label{joint_position_time}
P_2(y_2,\tau_{\text{cut}}|\gras{b},\gras{a},T)=\int_{-\infty}^{y_2} \rmd y_1\ Q_2(\gras{y},\tau_{\text{cut}}|\gras{b},\gras{a},T) \\
= \frac{\int_{-\infty}^{y_2} \rmd y_1\  p_2(\gras{b},T|\gras{y},\tau_{\text{cut}}) \times p_2(\gras{y},\tau_{\text{cut}}|\gras{a},0)}
{T\ p_2(\gras{b},T|\gras{a},0)} \;.
\end{multline}
From now on we will set, for simplicity, $T=1$. This can be done without loss of generality because the Brownian scaling implies $M \propto T^{1/2}$
and $\tau_M \propto T$. A simple dimensional analysis allows to resinsert the time $T$ in our calculations.

\subsection{Maximum, and the regularization procedure}

The maximum $M$ and the time to reach it $\tau_M$ are defined by
\begin{equation}
\label{def_max}
M= \max_{0\leq \tau \leq 1} x_2(\tau) = x_2(\tau_M) \;.
\end{equation}
Hence, to compute the joint probability of the couple $(M,\tau_M)$,
we want to impose $y_2=M$ and $\tau_{\text{cut}}=\tau_M$, knowing that the upper path does not cross the line $x=M$.
This condition is implemented by inserting the product of Heaviside step functions
\begin{equation}
\prod_{0\leq \tau \leq 1} \theta[M-x_2(\tau)] \;,
\end{equation}
in the path integral formula~(\ref{path_integral_start}). The propagator
of two vicious Brownian paths, with the constraint that they do not cross $x=M$, reads
\begin{multline}
\label{path_integral_M}
p_{<M,2}(\gras{b},t_b|\gras{a},t_a)=\frac{1}{z_{<M}} \int\limits_{x_1(t_a)=a_1}^{x_1(t_b)=b_1} \! \! \! \! \! \! \mathcal{D}x_1(\tau)
\int\limits_{x_2(t_a)=a_2}^{x_2(t_b)=b_2} \! \! \! \! \! \! \mathcal{D}x_2(\tau) \\
\times \Big\{ 
\prod_{t_a <\tau<t_b} \theta(M-x_2(\tau)) \ \theta(M-x_1(\tau))\
\theta(x_2(\tau)-x_1(\tau)) \\
\times e^{-\int_{t_a}^{t_b} \frac{1}{2} \left(\frac{\rmd x_1(\tau)}{\rmd \tau}\right)^2 \rmd \tau} \
e^{-\int_{t_a}^{t_b} \frac{1}{2} \left(\frac{\rmd x_2(\tau)}{\rmd \tau}\right)^2 \rmd \tau}
 \Big\} ,
\end{multline}
for all $0\leq t_a<t_b\leq 1$, $a_1<a_2<M$ and $b_1<b_2<M$. The second Heaviside step function
does not change the value of the path integral (it does not select a smaller class of paths)
because for all $t_a<\tau<t_b$, if $x_1(\tau)<x_2(\tau)$ and $x_2(\tau)<M$, then $x_1(\tau)<M$ automatically.
Accordingly to our first analysis, this propagator~(\ref{path_integral_M}) reduces to the propagator of two identical fermions
without interaction between them, but with a hard wall in $x=M$. Hence the associated Hamiltonian is
\begin{subequations}
\label{hard_wall_in_M}
\begin{equation}
H_{<M}=H_{<M}^{(1)} \otimes \mathbb{I}^{(2)}+ \mathbb{I}^{(1)} \otimes H_{<M}^{(2)} \;,
\end{equation}
with the one particle Hamiltonian
\begin{equation}
\label{2bridges_Hamiltonian}
H_{<M}^{(j)}=-\half\frac{\rmd^2}{\rmd x_j^2} + V_{<M}(x_j),
\end{equation}
for $j=1,2$, where $V_{<M}(x)$ is the hard wall potential in $x=M$:
\begin{equation}
V_{<M}(x)=\begin{cases}
0 & \text{if $-\infty<x<M$ \;,}\\
+\infty & \text{if $x>M$ \;.}
\end{cases}
\end{equation}
This one particle Hamiltonian has eigenfunctions
\begin{equation}
\label{2bridges_eigenfunctions}
\phi_{k_j}(x_j)= \sqrt{\frac{2}{\pi}} \sin(k_j (M-x_j)) \;,
\end{equation}
with the associated energies
\begin{equation} 
\label{2bridges_energies}
E_{k_j}=\frac{{k_j}^2}{2}, \ k_j>0 \;.
\end{equation}
Hence, with these informations, one is able to compute the propagator $p_{<M}(\gras{b},t_b|\gras{a},t_a)$
using the spectral decomposition, as in Eq.~(\ref{propagator}):
\begin{multline}
\label{2bridges_propagator}
p_{< M,2}(\gras{b},t_b|\gras{a},t_a) = \langle \gras{b} | e^{- (t_b-t_a) H_{<M}} | \gras{a} \rangle \\
=\int\limits_0^\infty \rmd k_1 \int\limits_0^\infty \rmd k_2 \Phi_{\gras{k}}(\gras{b})  \Phi^*_{\gras{k}}(\gras{a}) e^{-(t_b-t_a) \frac{\gras{k}^2}{2}} \;,
\end{multline}
\end{subequations}
with $\Phi_{\gras{k}}(\gras{x})$ the Slater determinant built with the one particle eigenfunctions~(\ref{2bridges_eigenfunctions}).
\par
To obtain
the joint pdf of $M$ and $\tau_M$, one would insert this propagator in formulas~(\ref{joint_positions_time}-\ref{joint_position_time}), 
with $y_2=M$ and $\tau_{\text{cut}}=\tau_M$.
With this procedure one naively finds zero, because we impose the intermediate point $y_2$ to be
on the edge of the hard wall potential. This problem originates in the continuous nature of
the Brownian motion: once in a point, the Brownian motion explore its vicinity immediatly before and after~\cite{feller}.
This implies that we cannot prescribe the upper path to be in $x_2(\tau_M)=M$ without being
in $x_2(\tau)>M$ for $\tau$ close to $\tau_M$.  Another problem is that the normalization constant can not be written as
a single propagator. Indeed one can not insert the closure relation as we did from Eq.~(\ref{normalization_general})
to Eq.~(\ref{normalization_general_propagator}), 
because the potential depends actually on the position $y_2 \equiv M$.
\par
Hence one should prescribe a regularization scheme to avoid this phenomenon.
A common way to deal with it is to take $y_2=x_2(\tau_M)=M-\eta$, with $\eta>0$ a small regularization parameter,
compute all quantities as functions of $\eta$, and at the end take the limit $\eta \to 0$.
This regularization scheme is similar to that introduced in Ref.~\cite{Majumdar08}.
Following the same ideas, one obtains the joint pdf as
\begin{subequations}
\label{method_two}
\begin{equation}
\label{jointpdf_general}
P_2(M,\tau_M|\gras{b},\gras{a})=\lim_{\eta \to 0} \frac{W_2(M-\eta,\tau_M|\gras{b},\gras{a})}{Z_2(\eta|\gras{b},\gras{a})} \;,
\end{equation}
where $W_2(M-\eta,\tau_M| \gras{b},\gras{a})$ is the probability weight of all paths starting in $\gras{a}$ and ending in $\gras{b}$
in the unit time interval and such that $x_2(\tau_M)=M-\eta$, and $Z_2(\eta|\gras{b},\gras{a})$ is the normalization constant. Explicitly, one has 
\begin{multline}
\label{weight_2_general}
W_2(M-\eta,\tau_M| \gras{b},\gras{a}) =\int\limits_{-\infty}^{\infty} \rmd y_2 \int\limits_{-\infty}^{y_2} \rmd y_1 \ \Big\{\delta(y_2-(M-\eta)) \\
\times p_{<M,2}(\gras{b},1|\gras{y},\tau_M) 
\ p_{<M,2}(\gras{y},\tau_M|\gras{a},0) \Big\},
\end{multline}
for all $M>\max \{a_2,b_2\}$, and where we write a dummy integration over $y_2$ with the delta function
forcing $y_2$ to be $M-\eta$. The normalization depends on the regularization parameter $\eta$
\begin{equation}
\label{normalization_regularized}
Z_2(\eta|\gras{b},\gras{a})=
\int_{0}^{1} \rmd \tau_M \int\limits_{\max \{a_2,b_2\}}^{\infty} \! \! \! \! \! \! \rmd M \ W_2(M-\eta,\tau_M| \gras{b},\gras{a}) \;.
\end{equation}
\end{subequations}
\par
In the following paragraphs, all three configurations have their paths beginning at the origin. The problem of the continuous
Brownian paths forbids to take directly $\gras{a}=(0,0)$. Instead of that, we will separate
artificially the paths by an amount $\epsilon>0$. Details of this regularization are left in the concerned paragraphs.


\subsection{Periodic boundary condition: $N=2$ vicious Brownian bridges}
\label{2bridges}

In this case, the two paths start from the origin at $\tau=0$ 
and arrive also at the origin at final time $\tau=1$: $x_1(0)=x_2(0)=x_1(1)=x_2(1)=0$.
Without extra condition, this defines two Brownian bridges. Here we compute the
joint probability density function for two Brownian bridges, under the condition that they do not cross each other.
\par
As discussed before, there is again a regularization scheme to adopt in both starting and ending points.
We separate by an amount $\epsilon$  the two paths at the starting and ending points, 
taking $\gras{a}=\gras{b}=(0,\epsilon)=\gras{\epsilon}$.
At the end of the computation, we shall take the limit $\epsilon \to 0$.
Then the joint probability density function for two vicious Brownian bridges is
\begin{subequations}
\begin{equation}
P_{2,B}(M,\tau_M)= \lim_{\epsilon \to 0} P_2(M,\tau_M|\gras{\epsilon},\gras{\epsilon}) \;,
\end{equation}
where $P_2(M,\tau_M|\gras{b},\gras{b})$ is defined in Eq.~(\ref{jointpdf_general}). The numerator of Eq.~(\ref{jointpdf_general}), the probability
weight, reads 
\begin{equation}
W_{2,B}(M-\eta,\tau_M,\epsilon)=W_2(M-\eta,\tau_M|\gras{\epsilon},\gras{\epsilon}) \;,
\end{equation}
and the denominator of Eq.~(\ref{jointpdf_general}), {\it i.e.} the normalization, is given by
\begin{equation}
Z_{2,B}(\eta,\epsilon)=Z_2(\eta|\gras{\epsilon},\gras{\epsilon}), 
\end{equation}
where the subscript $'B'$ refers to `Bridges', and with the relation deduced from Eq.~(\ref{jointpdf_general})
\begin{equation}
\label{2bridges_limit}
P_{2,B}(M,\tau_M)=\lim_{\eta \to 0} \lim_{\epsilon \to 0} \frac{W_{2,B}(M-\eta,\tau_M,\epsilon)}{Z_{2,B}(\eta,\epsilon)} \;.
\end{equation}
\end{subequations}
We obtain this limit by expanding the numerator and the denominator in power series of $\eta$ and $\epsilon$,
keeping only the dominant term in each case.
\par
To compute the probability weight $W_{2,B}(M-\eta,\tau_M,\epsilon)$,
 we start with Eq.~(\ref{weight_2_general}), in which we put $\gras{a}=\gras{b}=\gras{\epsilon}$:
 \begin{multline}
 W_{2,B}(M-\eta,\tau_M,\epsilon)= \int_{\text{ord}} \rmd \gras{y} \ \Big\{\delta(y_2-(M-\eta)) \phantom{----}\\
\times p_{<M,2}(\gras{\epsilon},1|\gras{y},\tau_M) 
\ p_{<M,2}(\gras{y},\tau_M|\gras{\epsilon},0) \Big\} \;.
\end{multline}
Using the spectral decomposition~(\ref{2bridges_propagator}), one has
\begin{multline}
\label{2bridges_start}
W_{2,B}(M-\eta,\tau_M,\epsilon) = \int_{\text{ord}} \rmd \gras{y} \ \Big\{\delta(y_2-(M-\eta)) \phantom{----}\\
\times  \int_0^\infty \rmd \gras{k'}\ 
\Phi_{\gras{k'}}(\gras{\epsilon}) \Phi^*_{\gras{k'}}(\gras{y}) e^{-(1-\tau_M) \frac{\gras{k'}^2}{2}} \\
\times \int_0^\infty \rmd \gras{k} \
\Phi_{\gras{k}}(\gras{y}) \Phi^*_{\gras{k}}(\gras{\epsilon}) e^{-\tau_M \frac{\gras{k}^2}{2}} \Big\} \;.
\end{multline}
The integral over $y_2$ is easy due to the delta function.
By antisymmetry of Slater determinants in the exchange of $k_1$ and $k_2$,
the product $\Phi_{\gras{k}}(\gras{y}) \Phi^*_{\gras{k}}(\gras{\epsilon}) e^{-\tau_M \frac{\gras{k}^2}{2}} $ is symmetric in
the exchange of $k_1$ and $k_2$. Then one can replace $\Phi_{\gras{k}}(\gras{y})$
by twice the product of its diagonal terms which is
\begin{equation}
2 \times \frac{\phi_{k_1}(y_1) \phi_{k_2}(M-\eta)}{\sqrt{2!}} =\eta
\frac{2k_2}{\sqrt{\pi}} \phi_{k_1}(y_1) + \mathcal{O}(\eta^3) \;.
\end{equation}
The same operations applies to $\Phi_{\gras{k'}}(\gras{y})$.
\begin{multline}
W_{2,B}(M-\eta,\tau_M,\epsilon)=\eta^2 \frac{4}{\pi} \int_{-\infty}^{M-\eta} \rmd y_1 \\
\times \int_{0}^{\infty} \rmd \gras{k'} \ k'_2 \ \phi^*_{k'_1}(y_1) \ \Phi_{\gras{k'}}(\gras{\epsilon}) e^{-(1-\tau_M) \frac{\gras{k'}^2}{2}} \\
\times \int_{0}^{\infty} \rmd \gras{k} \ k_2 \ \phi_{k_1}(y_1) \ \Phi^*_{\gras{k}}(\gras{\epsilon}) e^{-\tau_M \frac{\gras{k}^2}{2}} \;. 
\end{multline}
Permuting the order of integrations, one can compute the integration with respect to $y_1$ as
\begin{equation}
\int_{-\infty}^{M-\eta} \rmd y_1\ \phi^*_{k'_1}(y_1) \phi_{k_1}(y_1) = \delta(k_1-k'_1) +\mathcal{O}(\eta) \;, 
\end{equation}
due to the orthonormalization of the eigenfunctions $\phi_k(x)$.
Performing the integration over $k'_1$ leads to, at lowest order in $\eta$,
\begin{multline}
\label{2bridges_middle}
W_{2,B}(M-\eta,\tau_M,\epsilon) =\frac{4}{\pi} \eta^2 \int\limits_0^\infty \rmd k_1 e^{-\frac{k_1^2}{2}} 
\int\limits_0^\infty \rmd k_2  \int\limits_0^\infty \rmd k'_2 \\
\times \left\{Êk'_2 k_2 \Phi_{k_1,k'_2}(\gras{\epsilon}) \Phi_{k_1,k_2}(\gras{\epsilon}) 
e^{-(1-\tau_M)\frac{{k'}_2^2}{2}} e^{- \tau_M \frac{k_2^2}{2}} \right \} \\
+ \mathcal{O}(\eta^3) \;.
\end{multline}
The next step is to expand in powers of $\epsilon$ inside the Slater determinants: 
the first column does not depend on $\epsilon$, and in the second column, 
we expand at order $\epsilon$, each of the two elements, 
sweeping the constant term by linear combination with the first column. This yields
\begin{align}
\Phi_{k_1,k_2}(\gras{\epsilon})=&\sqrt{\frac{1}{2!}}
\left|
\begin{array}{cc}
\phi_{k_1}(0) & \phi_{k_1}(\epsilon) \\
\phi_{k_2}(0) & \phi_{k_2}(\epsilon)
\end{array}
\right| \nn \\
=&\frac{\epsilon}{\sqrt{2}}
\left|
\begin{array}{cc}
\phi_{k_1}(0) & \phi'_{k_1}(0) \\
\phi_{k_2}(0) & \phi'_{k_2}(0)
\end{array}
\right| + \mathcal{O}(\epsilon^2)
\nn \\
\label{expansion_epsilon}
=& \epsilon \frac{\sqrt{2}}{\pi} \frac{1}{M} \Theta_2(q_1,q_2) +\mathcal{O}(\epsilon^2)
\end{align}
where we use the scaled variables $q_i=k_i M$ and the determinant
\begin{equation}\label{def_theta2}
\Theta_2(q_1,q_2)=\det_{1\leq i,j \leq 2} \left( q_i^{j-1} \cos\left(q_i-j\frac{\pi}{2} \right) \right) .
\end{equation}
Performing the same expansion in powers of $\epsilon$ in $\Phi_{k_1,k'_2}(\gras{\epsilon})$, with $q'_2=k'_2 M$,
one can identify the dominant term as a factor of the product of powers $\eta^2 \epsilon^2$, so that
\begin{equation}
W_{2,B}(M-\eta,\tau_M,\epsilon) = \eta^2 \epsilon^2 {\rm W}_{2,B}(M,\tau_M) +\mathcal{O}(\eta^3 \epsilon^2, \eta^2 \epsilon^3) \;,
\end{equation}
Inserting the expansion~(\ref{expansion_epsilon}) in (\ref{2bridges_middle}), and identifying the
the leading term, one obtains
\begin{multline}
{\rm W}_{2,B}(M,\tau_M) = \frac{8}{\pi^3 M^{7}}
\int_0^\infty \rmd q_1 e^{-\frac{q_1^2}{2 M^2}}  \\
\times \Bigg\{ \int_0^\infty \rmd q'_2\  q'_2 \Theta_{2}(q_1,q'_2) e^{-(1-\tau_M)\frac{{q'}_2^2}{2M^2}}   \\
\times \int_0^\infty \rmd q_2\ q_2 \Theta_{2}(q_1,q_2)  e^{- \tau_M \frac{q_2^2}{2M^2}} \Bigg\} \;.
\end{multline}
It can be written in the more compact form
\begin{multline}
\label{2bridges_factorized_upsilon}
W_{2,B}(M,\tau_M) = \frac{8}{\pi^3 M^7} \int_0^\infty \rmd q_1 e^{-\frac{q_1^2}{2M^2}} \\
\Upsilon_2(q_1|M,1-\tau_M) \Upsilon_2(q_1|M,\tau_M) ,
\end{multline}
with the help of the function $\Upsilon_2$
\begin{equation}\label{def_upsilon}
\Upsilon_2 \left(q_1|M,\tau_M\right) =
\int_0^\infty \rmd q_2\  q_2 e^{- \tau_M \frac{q_2^2}{2M^2}} \Theta_2(q_1,q_2) \;,
\end{equation}
which can also be written as a determinant:
\begin{multline}
\label{upsilon_det}
\Upsilon_2 \left(q_1|M,\tau_M\right) =
\sqrt{\pi}\left( \frac{M}{\sqrt{2\tau_M}} \right)^{2} e^{-\frac{M^2}{2 \tau_M}}  \\
\times
\left| \begin{array}{cc}
\cos\left(q_1-\frac{\pi}{2}\right) & q_1 \cos\left(q_1-\pi\right) \\
H_1\left(\frac{M}{\sqrt{2\tau_M}}\right) &
\left(\frac{M}{\sqrt{2\tau_M}}\right) H_2\left(\frac{M}{\sqrt{2\tau_M}}\right)
\end{array} \right| .
\end{multline}
$H_1(x)$ and $H_2(x)$ are the first and second Hermite polynomials. 
This expression is obtained by factorizing all the $q_2$-dependence in the last line of the determinant, 
and performing the integration directly in this last line. The Hermite polynomials then appear naturally, see Appendix~\ref{Hermite}.
Then one proceeds to the expansion of the determinants $\Upsilon_2$ with respect to their last lines, 
which permits the integration over $q_1$ (four terms, because there are two determinants $2\times2$),
and one obtains a simple, though long, expression (apart from the exponentials, only algebraic terms appear).
\par
The limit in Eq.~(\ref{2bridges_limit}) exists provided the normalization constant admits the following expansion
\begin{equation}
Z_{2,B}(\eta,\epsilon)=\eta^2 \epsilon^2 K_{2,B} + o(\eta^2 \epsilon^2) \;,
\end{equation}
where $K_{2,B}$ is a number (independent of $M$ and $\tau_M$). It can be computed using the normalization of the joint pdf
\begin{align}
1&=\int_0^\infty \rmd M \int_0^1 \rmd \tau_M \ P_{2,B}(M,\tau_M) \nn \\
&= \frac{1}{K_{2,B}} \int_0^\infty \rmd M \int_0^1 \rmd \tau_M \ W_{2,B}(M,\tau_M) \;.
\end{align}
One finds $K_{2,B}=1/\pi$. The joint pdf reads
\begin{multline}
P_{2,B}(M,\tau_M)=\sqrt{\frac{2}{\pi}} \frac{1}{[\tau_M(1-\tau_M)]^{3/2}} e^{-\frac{M^2}{2\tau_M(1-\tau_M)}}  \\
\times \Bigg\{ \frac{1}{4}H_2\left(\frac{M}{\sqrt{2\tau_M}}\right) H_2\left(\frac{M}{\sqrt{2(1-\tau_M)}}\right) (1-e^{-2M^2}) \\
+M^2e^{-2M^2}\left( H_2\left(\frac{M}{\sqrt{2\tau_M}}\right) + H_2\left(\frac{M}{\sqrt{2(1-\tau_M)}}\right) \right)\\
+M^2\left(1-\frac{1}{2}e^{-2M^2}H_2\left(\sqrt{2}M\right) \right)\Bigg\} ,
\end{multline}
where for compactness we use the Hermite polynomial $H_2(x)=4x^2-2$.
\begin{figure}
\includegraphics[width=\linewidth]{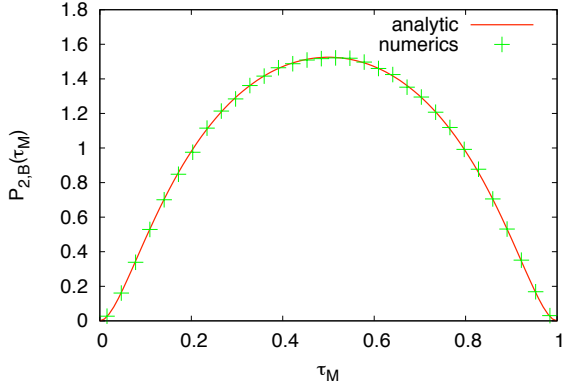}
\caption{Distribution of the time at which the maximum is reached for $N=2$ non-intersecting bridges. The solid line corresponds
to our analytical formula in Eq. (\ref{marginal_bridge_2}) while the symbols correspond to the results of our numerical simulations.}
\label{pdf_bridge2}
\end{figure}
A direct integration over $M$ gives the probability distribution function of $\tau_M$, the time at which the maximum is reached,
regardless to the value of the maximum:
\begin{equation}\label{marginal_bridge_2}
P_{2,B}(\tau_M)=4\left(1-\frac{1+10 \tau_M(1-\tau_M)}{\left[1+4\tau_M(1-\tau_M)\right]^{5/2}}\right) .
\end{equation}
The fact that $\tau_M$ enters in this expression only through the product $\tau_M(1-\tau_M)$ reflects the symmetry
of  the distribution around $\tau_M=1/2$, which of course is expected for periodic boundary conditions.
To center the distribution, over an unit length interval, one can use $\tau_M=1/2+u_M/2$,
and the distribution of $u_M$ is then
\begin{equation}
P_{2,B}^{\rm centered}(u_M)=2-\frac{5}{(2-u_M^2)^{3/2}} + \frac{3}{(2-u_M^2)^{5/2}} \;.
\end{equation}
Two asymptotic analysis can be made:
\begin{itemize}
\item for $\tau_M \simeq 0$ (or equivalently $\tau_M \simeq 1$), one has
\begin{equation}
P_{2,B}(\tau_M) \sim 120 \tau_M^2 + \mathcal{O}(\tau_M^3), \nn
\end{equation}
\item and for $\tau_M$ in the vicinity of $1/2$, better written in terms of the behaviour in $u_M=0$
of the centered distribution
\begin{equation}
P_{2,B}^{\rm centered}(u_M) \sim \left( 4 - \frac{7}{2\sqrt{2}} \right) - \frac{15 u_M^2}{8\sqrt{2}} + \mathcal{O}(u_M^3) \;. 
\end{equation}
\end{itemize}
\par
Furthermore, an integration of $P_{2,B}(M,\tau_M)$ with respect to $\tau_M$ gives the pdf of the maximum $F'_{2,B}(M)$
(the derivative of the cumulative distribution). Using the change of variables $\tau_M=(1+\sin(\varphi))/2$, the integral
gives
\begin{align}
F'_{2,B}(M)&=\int_0^1\rmd \tau_M\ P_{2,B}(M,\tau_M) \nn \\
&= 8M e^{-4M^2} - 8M e^{-2M^2} +16M^3 e^{-2M^2} \;, \nn
\end{align}
which coincides with the result computed in Ref.~\cite{Schehr08}, $F_{2,B}(x)=1-4x^2e^{-2x^2}-e^{-4x^2}$.



\subsection{Free boundary conditions: $N=2$ stars configuration}
\label{2stars}

In this subsection, we consider two vicious Brownian paths starting from the origin
with free end-points: this is the star configuration
(a subscript '$S$' will be used). The starting points are the same as before $\gras{a}=\gras{\epsilon}=(0,\epsilon)$,
but the endpoints are free, so that
for $M>0$ fixed, we sum over all endpoints $\gras{b}$ such that $b_1<b_2<M$.
Hence one has
\begin{equation}
P_{2,S}(M,\tau_M)=\lim_{\epsilon \to 0} \lim_{\eta \to 0} \frac{W_{2,S}(M-\eta,\tau_M,\epsilon)}{Z_{2,S}(\epsilon,\eta)} \;,
\end{equation}
where the probability weight is given by
\begin{align}
W_{2,S}&(M-\eta,\tau_M,\epsilon)=\int_{\text{ord}} \rmd \gras{b}\ W_2(M-\eta,\tau_M|\gras{b},\gras{\epsilon}) \nn \\
&= \int_{\text{ord}} \rmd \gras{b} \int_{\text{ord}} \rmd \gras{y}\ \Big\{ \delta(y_2-(M-\eta)) \\
& \phantom{ = \int_{\text{ord}} \rmd \gras{b} }
\times p_{<M,2}(\gras{b},1|\gras{y},\tau_M) \ p_{<M,2}(\gras{y},\tau_M|\gras{\epsilon},0) \Big\} \;, \nn
\end{align}
where the integrations satisfy $-\infty<b_1<b_2<M$ and $-\infty<y_1<y_2\leq M-\eta$ respectively.
The normalization is such that
\begin{equation}
Z_{2,S}(\eta,\epsilon)=\int_0^1 \rmd \tau_M \int_0^{\infty} \rmd M \, W_{2,S}(M-\eta,\tau_M,\epsilon) \;;.
\end{equation}
As in the previous subsection, we will use only the leading order term of the expansion of $W_{2,S}$ and $Z_{2,S}$ in powers of $\eta$ and $\epsilon$.
The computation of the probability weight can be done along the same lines as before~(\ref{2bridges_start}-\ref{2bridges_middle}),
keeping in mind that only the final points change. At lowest order in $\eta$, one has
\begin{multline}
W_{2,S}(M-\eta,\tau_M,\epsilon)=\int\limits_{-\infty}^M \rmd b_2 \int\limits_{-\infty}^{b_2}Ê\rmd b_1 \ 
\frac{4}{\pi} \eta^2 \int\limits_0^\infty \rmd \gras{k} \int\limits_0^\infty \rmd k'_2 \\
e^{-\frac{k_1^2}{2}} 
k'_2 k_2 \Phi_{k_1,k'_2}(\gras{b}) \Phi_{k_1,k_2}(\gras{\epsilon}) e^{-(1-\tau_M)\frac{{k'}_2^2}{2}} e^{- \tau_M \frac{k_2^2}{2}}\;, 
\end{multline}
where we have integrated over $y_1$ first, and then over $k'_1$, 
with the delta function coming from the closure relation of eigenfunctions.
For the eigenfunctions evaluated in the final points one obtains
\begin{multline}
\Phi_{k_1,k'_2}(\gras{b})= \sqrt{\frac{1}{2}}
\left|
\begin{array}{cc}
\phi_{k_1}(b_1) & \phi_{k_1}(b_2) \\
\phi_{k'_2}(b_1) & \phi_{k'_2}(b_2)
\end{array}
\right| \nn \\
= \frac{\sqrt{2}}{\pi} (\sin(q_1 \zeta_1) \sin(q'_2 \zeta_2) - \sin (q'_2 \zeta_1) \sin(q_1 \zeta_2) ) \;, 
\end{multline}
in terms of the previous $q_j=k_j M$, and with the variables $\zeta_j=\frac{M-b_j}{M}$ (for $j=1,2$).
Using the expansion in $\epsilon$ written in Eq.~(\ref{expansion_epsilon}), one finds the leading order to be ${\cal O}(\eta^2 \epsilon$)
so that the expansion is
\begin{equation}
W_{2,S}(M-\eta,\tau_M,\epsilon)=\eta^2 \epsilon \ {\rm W}_{2,S}(M,\tau_M) + \mathcal{O}(\eta^3 \epsilon,\eta^2 \epsilon^2) \;,
\end{equation}
with
\begin{multline}
{\rm W}_{2,S}(M,\tau_M)=\frac{8}{\pi^3 M^4} \int_0^\infty \rmd \zeta_2 \int_{\zeta_2}^\infty \rmd \zeta_1 \\
\int_0^\infty \rmd \gras{q} \int _0^\infty \rmd q'_2 \ q_2 q'_2 \Theta_2(q_1,q_2) 
e^{-\frac{q_1^2 +\tau_M q_2^2 + (1-\tau_M) {q'_2}^2}{2M^2}} \\
\times \left( \sin(q_1 \zeta_1) \sin(q'_2 \zeta_2) - \sin (q'_2 \zeta_1) \sin(q_1 \zeta_2) \right) \;,
\end{multline}
where $\Theta_2(q_1,q_2)$ is given in Eq. (\ref{def_theta2}). As before, the integration over the moments $q_2$ and $q'_2$, corresponding to the top walker, can be factorized in the determinants. From the part $[0,\tau_M]$, one recognizes $\Upsilon_2(q_1|M,\tau_M)$,
and for the part $[\tau_M,1]$, we introduce
\begin{multline}
\tilde{\Upsilon}_2(q_1,\gras{\zeta}|M,1-\tau_M)=\int_0^\infty \rmd q'_2\ q'_2 e^{-(1-\tau_M)\frac{{q'_2}^2}{2M^2}} \\
\times \left( \sin(q_1 \zeta_1) \sin(q'_2 \zeta_2) - \sin (q'_2 \zeta_1) \sin(q_1 \zeta_2) \right) .
\end{multline}
With the help of the two functions $\Upsilon_2$ and $\tilde{\Upsilon}_2$, one obtains
\begin{multline}
W_{2,S}(M,\tau_M)= \frac{8}{\pi^3 M^4}  \int_0^\infty \rmd \zeta_2 \int_{\zeta_2}^\infty \rmd \zeta_1\\
\int_0^\infty \rmd q_1\ \Upsilon_2(q_1|M,\tau_M) \tilde{\Upsilon}_2(q_1,\gras{\zeta}|M,1-\tau_M) .
\end{multline}
$\Upsilon_2$ and $\tilde{\Upsilon}_2$ can be written as determinants, as in Eq.~(\ref{upsilon_det}), and
\begin{multline}
\tilde{\Upsilon}_2(q_1,\gras{\zeta}|M,t)=\sqrt{\pi} \left( \frac{M}{\sqrt{2t}} \right)^2 \\
\times
\left|
\begin{array}{cc}
\sin(q_1 \zeta_1) & \sin(q_1 \zeta_2) \\
e^{-\frac{M^2}{2t}\zeta_1^2} H_1 \left( \frac{M}{\sqrt{2t}} \zeta_1\right) &
e^{-\frac{M^2}{2t}\zeta_2^2} H_1 \left( \frac{M}{\sqrt{2t}} \zeta_2\right)
\end{array}
\right| ,
\end{multline}
where we used the relation in Eq.~(\ref{hermite_zeta}), and where $t$ stands for $1-\tau_M$ (for compactness).
The fact that we do not have the periodic boundary conditions introduces an asymmetry, thus the two
determinants $\Upsilon_2$ and $\tilde{\Upsilon}_2$ do not have the same form.
\par
\begin{figure}
\centering
{\includegraphics[width=\linewidth]{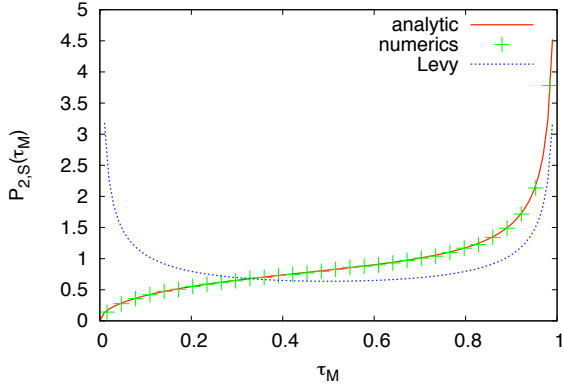}}
\caption{Distribution of the time at which the maximum is reached for $N=2$ non-intersecting free Brownian bridges: the solid line is our analytic result given in Eq. (\ref{margin_star}) while the symbols are the results of our numerical simulation. The dotted line describes the pdf of the time $\tau_M$ for one Brownian path (whose cumulative distribution is given by L\'evy's arcsine law).}
\label{pdf_star2}
\end{figure}

As before we also expand the normalization $Z_{2,S}(\eta,\epsilon)$ in powers of $\eta$ and $\epsilon$, the leading term being
\begin{equation}
Z_{2,S}(\eta,\epsilon)=\eta^2 \epsilon \ K_{2,S} + o(\eta^2 \epsilon) \;, \nn
\end{equation}
with $K_{2,S}$ a number, independent of $M$ and $\tau_M$ which can be computed by the normalization condition: 
\begin{align}
1&=\int_0^\infty \rmd M \int_0^1 \rmd \tau_M \ P_{2,S}(M,\tau_M) \nn \\
&= \frac{1}{K_{2,S}} \int_0^\infty \rmd M \int_0^1 \rmd \tau_M \ W_{2,S}(M,\tau_M)  \;.
\end{align}
One finds $K_{2,S}=2/\sqrt{\pi}$, and the joint probability distribution function, for two stars configuration is
\begin{multline}
P_{2,S}(M,\tau_M)=\sqrt{\frac{2}{\pi}} \frac{M^2}{[\tau_M(1-\tau_M)]^{3/2}} e^{-\frac{M^2}{2\tau_M}}\\
\times \Bigg\{
\frac{1-\tau_M}{2\sqrt{2}M^2} H_2\left(\frac{M}{\sqrt{2\tau_M}}\right) \textrm{Erf}\left(\frac{M}{\sqrt{2}}\right) \\
+\frac{1}{\sqrt{\pi}} \frac{1}{M} \frac{\tau_M(1-\tau_M)}{2-\tau_M} e^{-\frac{M^2}{2}}  \\
+ \sqrt{2} \left(\frac{1-\tau_M}{2-\tau_M}\right)^{3/2} \textrm{Erf}\left( \sqrt{\frac{1-\tau_M}{2-\tau_M}} \frac{M}{\sqrt{2}}\right) 
e^{-\frac{M^2}{2(2-\tau_M)}}\\
\times \left[1-\frac{2-\tau_M}{M^2} \frac{1}{2} H_2\left( \frac{M}{\sqrt{2\tau_M}}\right) \right ] \;.
 \Bigg\} 
\end{multline}
Integrating over $M$ one obtains the marginal, 
\textit{i.e.} the pdf of the time to reach the maximum
\begin{align}\label{margin_star}
&P_{2,S}(\tau_M) =\int_0^\infty \rmd M P_{2,S}(M,\tau_M) \nn \\
&= \frac{2}{\pi} \left( \arctan \left(\sqrt{\frac{\tau_M(1-\tau_M)}{2}}\right) +  \frac{1}{1+\tau_M} \sqrt{\frac{2 \tau_M}{1-\tau_M} } \right) \;,
\end{align}
which is plotted with a comparison with numerical simulation in Fig.~\ref{pdf_star2}.%
One easily obtains the asymptotic behavior of the distribution of the time $\tau_M$:
\begin{itemize}
\item for $\tau_M \to 0$, one has
\begin{equation}
P_{2,S}(\tau_M) \sim \frac{3 \sqrt{2}}{\pi} \tau_M^{1/2} + \mathcal{O}(\tau_M^{3/2}). \nn
\end{equation}
\item for $\tau_M \to 1$ the probability distribution diverges as
\begin{equation}
P_{2,S}(\tau_M) \sim \frac{2\sqrt{2}}{\pi} (1-\tau_M)^{-1/2} , \nn
\end{equation}
which can be compared to the divergence of the distribution of the time $\tau_M$ for one Brownian motion 
(the L\'evy arcsine law, plotted in Fig.~\ref{pdf_star2}) $P_{1,S}(\tau_M)=\frac{1}{\pi} \left[ \tau_M(1-\tau_M) \right]^{-1/2}$. 
The fact that we have another
uncrossing Brownian path below preserves the exponent of the divergence, but changes slightly the prefactor.
\end{itemize}
Notice finally the mean value is given by $\langle \tau_M \rangle_{2,S} = 7/2 - 2 \sqrt{2} = 0.671573\dots$ and that the sign of the derivative of $P_{2,S}(\tau_M)$ changes in $\tau_M \simeq 0.451175$.


\subsection{Periodic boundary condition and positivity constraint: $N=2$ excursions configuration}
\label{2excursions}

Now we come to the computation of the joint probability distribution of the maximum $M$ and its position $\tau_M$ for
the configuration of $2$ vicious Brownian paths subjected to stay in the half line $x>0$, in addition to the periodic condition:
the two paths start in $x=0$ and end in $x=0$ at $\tau=1$. This corresponds to two Brownian excursions,
with the non-crossing condition (we will use the subscript '$E$').
\par
The positivity constraint is easily implemented in the path integral formulation
by adding another hard wall in $x=0$. The potential associated to each of the two fermions then reads
\begin{subequations}
\begin{equation}
V_{\text{Box}}(x)=\begin{cases}
0 & \text{if $0<x<M$}\\
+\infty & \text{elsewhere} \;.
\end{cases}
\end{equation}
The eigenfunctions of the one-particle Hamiltonian
\begin{equation}
\label{2excursions_hamiltonian}
H^{(j)}_{\text{Box}}(x_j)=-\half \frac{\rmd^2}{\rmd x_j^2} +V_{\text{Box}}(x_j) 
\end{equation}
are given by
\begin{equation}\label{psi_n}
\psi_n(x)=\sqrt{\frac{2}{M}} \sin \left( \frac{n \pi}{M} x \right), \quad n=1,2,\dots
\end{equation}
with the associated eigenvalues
\begin{equation}
E_n=\half \frac{n^2 \pi^2}{M^2} \;.
\end{equation}
This allows to express the corresponding propagator, 
for any point $\gras{a}$, $\gras{b}$ and time $0\leq t_a<t_b\leq 1$, as: 
\begin{multline}
\label{2excursions_propagator}
p_{\text{Box}}(\gras{b},t_b|\gras{a},t_a) = \langle \gras{b} | e^{-(t_b-t_a)H_{\text{Box}} }|\gras{a} \rangle  \\
=\sum_{n_1,n_2>0} \Psi_{\gras{n}}(\gras{b}) \Psi^*_{\gras{n}}(\gras{a}) e^{-(t_b-t_a)\frac{\pi^2}{2M^2}\gras{n}^2} \;,
\end{multline}
\end{subequations}
where $\gras{n}=(n_1,n_2)$ is a couple of two positive integers, 
and in which we use the Slater determinant $\Psi_{\gras{n}}(\gras{a})=\det_{1 \leq i,j \leq 2} [ \psi_{n_i}(a_j) ] / \sqrt{2!}$.
\par
Here the regularization is necesary for the boundary values of both fermions, because the lowest path cannot start and end exactly in $x=0$.
Therefore we put $\gras{x}(0)=\gras{a}=(\epsilon,2 \epsilon)=\gras{\varepsilon}$ and $\gras{x}(1)=\gras{b}=\gras{\varepsilon}$, 
where the notation $\gras{\varepsilon}=(\epsilon,2 \epsilon)$ differs slightly from the previous one 
$\gras{\epsilon}=(0,\epsilon)$.
For this geometry, one uses Eqs.~(\ref{jointpdf_general}-\ref{normalization_regularized}) to obtain the probability weight
\begin{subequations}
\begin{align}
W_{2,E}&(M-\eta,\tau_M,\epsilon)\equiv W_2(M-\eta,\tau_M|\gras{\varepsilon},\gras{\varepsilon}) \nn \\
&= \int_{\text{ord}} \rmd \gras{y} \ \Big\{ \delta(y_2-(M-\eta)) \nn \\
\label{2excursions_weight_start}
&\times p_{\text{Box}}(\gras{\epsilon},1|\gras{y},\tau_M)
p_{\text{Box}}(\gras{y},\tau_M|\gras{\epsilon},0) \Big\} \;,
\end{align}
where the ordered integral covers the domain $0<y_1<y_2<M$.
The normalization function is
\begin{equation}
Z_{2,E}(\eta,\epsilon) = Z_2(\eta |\gras{\varepsilon},\gras{\varepsilon}) \;,
\end{equation}
so that
\begin{equation}
\label{2excursions_jpdf_limit}
P_{2,E}(M,\tau_M)=\lim_{\epsilon \to 0} \lim_{\eta \to 0} \frac{W_{2,E}(M-\eta,\tau_M,\epsilon)}{Z_{2,E}(\eta,\epsilon)} \;.
\end{equation}
\end{subequations}
Inserting the spectral decomposition of the propagator~(\ref{2excursions_propagator}) in Eq.~(\ref{2excursions_weight_start}), 
one has
\begin{multline}
\label{2excursions_weightstep1}
W_{2,E}(M-\eta,\tau_M,\epsilon)= \int_{\text{ord}} \rmd \gras{y}\  \delta(y_2-(M-\eta)) \\
\times \sum_{\gras{n}, \gras{n'}>0} \left[ \Psi_{\gras{n'}}(\gras{\varepsilon}) \Psi^*_{\gras{n'}}(\gras{y}) e^{-\frac{\pi^2(1-\tau_M) \gras{n'}^2}{2M^2} } \right. \\
\left.  \times \Psi_{\gras{n}}(\gras{y}) \Psi^*_{\gras{n}}(\gras{\varepsilon}) e^{-\frac{\pi^2 \tau_M  \gras{n}^2}{2M^2}} \right] \; .
\end{multline}
We follow the same procedure as before in Eqs.~(\ref{2bridges_start}-\ref{2bridges_middle}):
using the symmetry in the exchange of indices ($n_1 \leftrightarrow n_2$ and $n_1' \leftrightarrow n_2'$)
we express the two Slater determinants as
the product of their diagonal components
\begin{multline}
W_{2,E}(M-\eta,\tau_M,\epsilon)=\\
\sum_{\gras{n},\gras{n'}} \Big\{ \Psi_{\gras{n'}}(\gras{\varepsilon})\Psi^*_{\gras{n}}(\gras{\varepsilon}) 
e^{-\frac{\pi^2 \left[ (1-\tau_M) \gras{n'}^2 + \tau_M \gras{n}^2 \right]}{2M^2} } \\
\times \psi_{n_2}(M-\eta) \psi_{n'_2}(M-\eta) \int\limits_0^{M-\eta} \! \! \! \! \rmd y_1\ \psi^*_{n'_1}(y_1) \psi_{n_1}(y_1) \Big\}.
\end{multline}
An expansion at lowest order in $\eta$ of this expression, making use of
\begin{equation}
\label{box_wavefunction_expansion_eta}
\psi_{n}(M-\eta) =(-1)^{n+1} \sqrt{\frac{2}{M}} \frac{n \pi}{M} \eta + \mathcal{O}(\eta^3),
\end{equation}
together with the closure relation of the eigenfunctions
\begin{equation}
\int_0^M \rmd y_1\ \psi_{n_1}(y_1) \psi^*_{n'_1}(y_1) = \delta_{n_1,n'_1} , 
\end{equation}
yields
\begin{multline}
W_{2,E}(M-\eta,\tau_M,\epsilon)= \\
\eta^2 \frac{2 \pi^2}{M^3} 
\sum_{\gras{n},\gras{n'}} \Big\{ 
\Psi_{\gras{n'}}(\gras{\varepsilon}) \Psi^*_{\gras{n}}(\gras{\varepsilon})
e^{-\frac{\pi^2\left[(1-\tau_M)\gras{n'}^2+\tau_M \gras{n}^2\right]}{2 M^2}} \\
\times (-1)^{n_2+n'_2}\ n_2\ n'_2\ \delta_{n_1,n'_1} \Big\} \;. 
\end{multline}
The expansion to the lowest order in $\epsilon$ of the eigenfunctions reads
\begin{equation}
\label{box_wavefunction_expansion_epsilon}
\Psi_{\gras{n}}(\gras{\varepsilon})= (-1) \frac{4 \sqrt{2} \pi^4}{3 M^5}\ n_1 n_2\ \Delta_2(n_1^2,n_2^2)\  \epsilon^4 
+\mathcal{O}(\epsilon^6) \;,
\end{equation}
where $\Delta_2(\lambda_1,\lambda_2)$ is the $2\times2$ Van der Monde determinant
\begin{equation}
\Delta_2(\lambda_1,\lambda_2)=\left|\begin{array}{cc}
1&\lambda_1 \\
1 & \lambda_2
\end{array}
\right| =(\lambda_2- \lambda_1) . \nn
\end{equation}
Hence to lowest order, one has
\begin{eqnarray}\label{2excursions_num}
&& W_{2,E}(M-\eta,\tau_M,\epsilon)=\eta^2 \epsilon^8 \ {\rm W}_{2,E}(M,\tau_M) + {o}(\eta^2\epsilon^8) \nn \\
&& {\rm W}_{2,E}(M,\tau_M)= \frac{64 \pi^{10}}{9 M^{13}} \sum_{n_1>0} \Big\{ n_1^2 \ e^{-\frac{\pi^2}{2M^2} n_1^2} \\
&&\times \sum_{n_2>0} (-1)^{n_2}  n_2^2  \Delta_2(n_1^2,n_2^2) e^{-\frac{\pi^2}{2M^2}\tau_M n_2^2} \nn \\
&&\times \sum_{n_3>0} (-1)^{n_3}  n_3^2 \Delta_2(n_1^2,n_3^2)  e^{-\frac{\pi^2}{2M^2}(1-\tau_M) n_3^2} \Big\}\;. \nn
\end{eqnarray}
To ensure the joint pdf to exist, the normalization must have the same dominant term
\begin{equation}\label{2excursions_den}
Z_{2,E}(\eta,\epsilon)=\eta^2 \epsilon^8 \ K_{2,E} + o(\eta^2 \epsilon^8) \nn
\end{equation}
Combining these expansions in Eq.~(\ref{2excursions_num}) 
and in Eq.~(\ref{2excursions_den}) one obtains from Eq.~(\ref{2excursions_jpdf_limit}):
\begin{multline}
\label{2excursions_joint}
P_{2,E}(M,\tau_M)=\frac{\pi^{11}}{3M^{13}} \sum_{n_1,n_2,n_3>0} (-1)^{n_2+n_3} \ n_1^2 n_2^2 n_3^2 \\
\times (n_2^2-n_1^2) (n_3^2-n_1^2) e^{-\frac{\pi^2}{2 M^2}\left[n_1^2 + \tau_M n_2^2 + (1-\tau_M) n_3^2 \right]} .
\end{multline}
where the computation of normalization has been left in Appendix~\ref{normalization}. 

\begin{figure}
\centering
{\includegraphics[width=\linewidth]{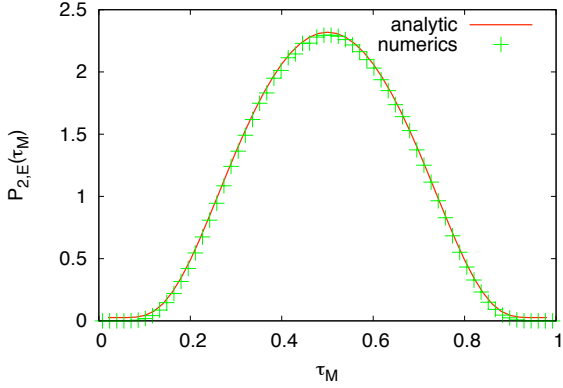}}
\caption{Distribution of the time at which the maximum is reached for $N=2$ non-intersecting excursions (watermelon with a wall): the solid line corresponds to our analytic result while the symbols are the results of our numerical simulations.}
\label{pdf_excur2}
\end{figure}

This formula is well suited for an integration over $M$ to deduce the marginal law of $\tau_M$:
\begin{multline}
P_{2,E}(\tau_M) = \int_0^\infty \rmd M \ P_{2,E}(M,\tau_M) \\
= \frac{1280}{\pi} \sum_{n_i>0} (-1)^{n_2+n_3} 
\frac{\left(n_1 n_2 n_3 \right)^2  (n_1^2-n_2^2)(n_1^2-n_3^2)}{\left[n_1^2+\tau_M n_2^2 +(1-\tau_M)n_3^2\right]^6} ,
\end{multline}
which is plotted in Fig.~\ref{pdf_excur2}.

Following the methods introduced in~\cite{Majumdar08}, 
one is able to compute the asymptotic behavior of $P_{2,E}(\tau_M)$ when $\tau_M \to 0$.
Starting from the expression of the joint distribution of $(M,\tau_M)$ in Eq.~(\ref{2excursions_joint}), one writes the integral over $M$ to get the marginal distribution $P_{2,E}(\tau_M)$ as
\begin{equation}
P_{2,E}(\tau_M)=\int_0^\infty \rmd M\ P_{2,E}(M,\tau_M), \nn
\end{equation}
and perform a change of variable
\begin{equation}
x=\sqrt{\tau_M} \frac{\pi^2}{2M^2}.
\end{equation}
One obtains
\begin{multline}
P_{2,E}(\tau_M)=\frac{2^5}{3 \pi} \frac{1}{\tau_M^3} \int_0^\infty \rmd x \ x^5 \\
\times \sum_{n_1,n_2,n_3} \Big\{ (-1)^{n_2+n_3} n_1^2 n_2^2 n_3^2 (n_2^2-n_1^2)(n_3^2-n_1^2) \\
\times e^{-\frac{x}{\sqrt{\tau_M}} \left(n_1^2+(1-\tau_M) n_3^2\right) } e^{-x\sqrt{\tau_M} n_2^2} \Big\} \;.
\end{multline}
When $\tau_M \to 0$, the argument of the first exponential becomes $-x(n_1^2+n_3^2)/\sqrt{\tau_M}$ at leading order,
and the dominant term in the sums over $n_1$ and $n_3$ is given by $(n_1=1,n_3=2)$ and $(n_1=2,n_3=1)$
because of the factor $(n_3^2-n_1^2)$, which enforces $n_1 \neq n_3$. Hence to leading order, one has
\begin{multline}
P_{2,E}(\tau_M) \simeq \frac{2^7}{\pi} \frac{1}{\tau_M^3} \int_0^\infty \rmd x \ x^5 \\
\times \sum_{n_2=1}^{\infty} (-1)^{n_2} n_2^2 \left(2 n_2^2-5 \right)e^{-\frac{5x}{\sqrt{\tau_M}}} e^{-x\sqrt{\tau_M}n_2^2} .
\end{multline}
By differenciating $N$ times the Jacobi identity 
\begin{multline}
1 + 2 \sum_{n=1}^\infty (-1)^n e^{-n^2 z} = 2 \left(\frac{\pi}{z} \right)^{1/2} \sum_{n=0}^\infty e^{- \pi^2 (n+\frac{1}{2})^2/z}
\end{multline}
with respect to $z$, one has at leading order when $z\to 0$
\begin{equation}
\sum_{n=1}^{\infty} (-1)^n n^{2N} e^{-n^2 z} \simeq \left(\frac{\pi}{z}\right)^{1/2} \left(\frac{\pi}{2}\right)^2N \frac{1}{z^{2N}} e^{-\frac{\pi^2}{4z}}.
\end{equation}
With $z=x\sqrt{\tau_M}$, one obtains at leading order
\begin{equation}
P_{2,E}(\tau_M) \simeq 2^4 \pi^{7/2} \tau_M^{-(5+\frac{1}{4})} 
\int_0^\infty \rmd x \ x^{1/2} e^{-\left(5x+\frac{\pi^2}{4x}\right) \frac{1}{\sqrt{\tau_M}}},
\end{equation}
which can be approximated by a saddle point method. At leading order when $\tau_M \to 0$, one obtains
\begin{equation}
\label{2excursions_asymptotic}
P_{2,E}(\tau_M) \simeq \frac{8 \pi^5}{5} \frac{1}{\tau_M^5} e^{-\frac{\pi \sqrt{5}}{\sqrt{\tau_M}}} .
\end{equation}
By symmetry one obtains the same behavior when $\tau_M \to 1$ by replacing $\tau_M$ by $1-\tau_M$ in the above formula.

As in the previous paragraphs, it is possible to have a compact formula for the joint pdf,
anticipating the general computation for $N$ vicious Brownian paths.
Let us introduce $\Omega_2$, the analogue of $\Upsilon_2$ (\ref{def_upsilon}) in this discrete case:
\begin{equation}  
\Omega_2(n_1|M,t)=\sum_{k>0}(-1)^k k^2 \Delta_2(n_1^2,k^2)e^{-\frac{\pi^2}{2M^2}t k^2},
\end{equation}
with $k$ a generic index standing for $n_2$ ($t=\tau_M$) or $n_3$ ($t=1-\tau_M$).
Factorizing the sum over $k$ in the last line of the Vandermonde determinant, and using sums defined by
\begin{equation}
\omega_i(M,t)=\sum_{k>0}(-1)^k k^{2 i}e^{-\frac{\pi^2}{2M^2}t k^2} ,
\end{equation}
one is able to write
\begin{equation}
\Omega_2(n_1|M,t)=\\ \left| \begin{array}{cc}
1 & n_1^2 \\
\omega_1(M,t) & \omega_2(M,t)
\end{array}
\right| .
\end{equation}
Hence the joint pdf can be expressed in a factorized way, reminiscent of the formula for the
watermelons case~(\ref{2bridges_factorized_upsilon})
\begin{multline}
P_{2,E}(M,\tau_M)=\frac{\pi^{11}}{3M^{13}} \sum_{n_1} n_1^2 \ e^{-\frac{\pi^2}{2M^2} n_1^2} \\
\times \Omega_2(n_1|M,\tau_M) Ê\Omega_2(n_1|M,1-\tau_M) .
\end{multline}
In the following we will extend these formula, for bridges and excursions, to $N$ particles, for generic $N$.

\section{Distribution of the position of the maximum for $N$ vicious Brownian paths}
 
In this section, we extend the previous method to any number $N$ of vicious Brownian paths,
ordered as $x_1(\tau)<x_2(\tau)<\dots<x_N(\tau)$ for $0<\tau<1$ (again we take a unit time interval, without loss of generality).
We want to compute the joint pdf of $(M,\tau_M)$, respectively the maximum and the time to reach it for different configurations, with
\begin{equation}
M=\max_{0\leq \tau \leq 1} x_N(\tau) = x_N(\tau_M) .
\end{equation}
In the same way as in the previous section, the non-colliding condition translates into treating indistinguishable fermions.
They do not interact between each other but are all subjected to the same potential 
(which is one or two hard wall(s) potential in our computations).
The Hamiltonian of the system is
\begin{equation}
\label{Hamiltonian_N}
H=\sum_{i=1}^{N} \mathbb{I}^{(1)} \otimes \dots \otimes H^{(i)} \otimes \dots \mathbb{I}^{(N)} \;,
\end{equation}
with identical one particle Hamiltonians $H^{(i)}$, for $1 \leq i \leq N$.
If $\psi_{E_i}(x_i)$ is the eigenfunction of $H^{(i)}$ associated to the energy $E_i$, 
one can write the eigenfunction of the $N$-particles problem as a Slater determinant
\begin{equation}
\Psi_{E}(\gras{x})=\frac{1}{\sqrt{N!}}\det_{1\leq i,j \leq N} \left( \psi_{E_i}(x_j) \right) \;,
\end{equation}
with $E=\sum_i E_i$ and where we used the bold type for $N$ vectors: $\gras{x}=(x_1,x_2,\dots,x_N)$.
Hence the propagator reads as in formula~(\ref{propagator}),
but with determinants of $N\times N$ matrices.
The method to compute the joint probability density function is exaclty the same as before, 
replacing $2$-vectors by $N$-vectors in formulas~(\ref{method_two}).



\subsection{$N$ vicious Brownian bridges}
\label{pbridges}

We first consider the watermelon configuration, {\it i.e.} the case of $N$ on-intersecting Brownian bridges (see definition in subsection~\ref{2bridges}). 
The periodic boundary conditions impose the starting and ending points to be at the origin. As discussed before, this needs a regularization scheme, and we take
$\gras{a}=\gras{b}=(0,\epsilon,2 \epsilon, \dots, (N-1) \epsilon)=\gras{\epsilon}$.
The joint pdf is
\begin{equation}\label{ratio_bridge}
P_{N,B}(M,\tau_M)=\lim_{\epsilon,\eta \to 0} \frac{W_{N,B}(M-\eta,\tau_M,\epsilon)}{Z_{N,B}(\eta,\epsilon)},
\end{equation}
where $W_{N,B}(M-\eta,\tau_M,\epsilon)$ is the probability weight of all paths starting and ending in $\gras{\epsilon}$
with $x_N(\tau_M)=M-\eta$, given by
\begin{multline}
W_{N,B}(M-\eta,\tau_M,\eta)=\int_{\text{ord}} \rmd \gras{y} \Big\{ \delta(y_N -(M-\eta)) \\
\times p_{<M,N}(\gras{\epsilon},1|\gras{y},\tau_M)\ p_{<M,N}(\gras{y},\tau_M|\gras{\epsilon},0) \Big\} .
\end{multline}
$p_{<M,N}$ is the propagator of $N$ indistinguishable fermions, each of them having the one-particle Hamiltonian
$H_{<M}$~(\ref{2bridges_Hamiltonian}), with eigenfunctions $\phi_k(x)$ and eigenvalues $E_k$ 
given by Eqs~(\ref{2bridges_eigenfunctions}, \ref{2bridges_energies}).
\par
The starting point of our computation is given by Eq.~(\ref{2bridges_start}) with the upper point is now $y_N$.
All the manipulations done before for 2 paths extend to $N$ paths.
A subtelty is the fact that the ordered integral over the domain $-\infty<y_1<y_2<\dots<y_{N-1}<y_N<M$
is proportional to the same unordered integral over the domain $-\infty<y_i<M$ for $1\leq i \leq N$,
because of the symmetry of the integrand. This is shown in Appendix~{\ref{ordered_unordered}}.
With the same argument of symmetry between the exchange of any $k_i \leftrightarrow k_j$, and the same for $\gras{k'}$
one could use only the diagonal terms for the two Slater determinants expressed in $\gras{y}$.
After permuting the $y$-integrals with the $k$-integrals,
this operation gives $N-1$ closure relations, \textit{i.e.} a product of $N-1$
Dirac delta functions $\prod_{i=1}^{N-1} \delta(k_i-k'_i)$, and the product $\phi_{k_N}(M-\eta) \phi_{k'_N}(M-\eta)$.
The expansion of the Slater determinants evaluated at the initial and final points involves now the function
\begin{equation}
\Theta_N(\gras{q})=\det_{1 \leq i,j \leq N} \left( q_i^{j-1} \cos \left( q_i - j\frac{\pi}{2} \right) \right) ,
\end{equation}
with the rescaled variables $q_i=k_i M$. 
After the aforementioned manipulations, one finds at lowest order
\begin{multline}\label{denominator_N}
W_{N,B}(M-\eta,\tau_M,\epsilon)=\eta^2 \epsilon^{N(N-1)} \ {\rm W}_{N,B}(M,\tau_M) \\
+ {o}\left(\eta^2 \epsilon^{N(N-1)}\right) \;,
\end{multline}
together with
\begin{equation}
\label{normalization_N}
Z_{N,B}(\eta,\epsilon)=\eta^2 \epsilon^{N(N-1)} K_{N,B} + o(\eta^2 \epsilon^{N(N-1)}) \;.
\end{equation}
Therefore one obtains from Eqs (\ref{denominator_N}) and (\ref{normalization_N}) together with Eq.~(\ref{ratio_bridge}):
\begin{multline}\label{joint_bridge_int}
P_{N,B}(M,\tau_M) = \frac{A_{N,B}}{M^{N^2+3}}   \int_0^\infty \rmd q_1 \dots \rmd q_{N-1}\ e^{-\frac{\sum_{k=1}^{N-1} q_k^2}{2M^2}} \\
\times \Upsilon_N(\{q_i\}|M,\tau_M) \Upsilon_N(\{q_i\}|M,1-\tau_M) \;,
\end{multline}
where 
\begin{equation}
\label{upsilon_N}
\Upsilon_N(\{q_i\}|M,t)= \int_0^\infty \rmd q_N\ q_N\ e^{-t \frac{q_N^2}{2 M^2}} \Theta_N(\gras{q}) \;,
\end{equation}
depends on $N-1$ variables $\{q_i\}=(q_1,\dots,q_{N-1})$. In Eq.~(\ref{joint_bridge_int}) the amplitude $A_{N,B}$, whose computation is left in Appendix~\ref{normalization}, is given by
\begin{eqnarray}\label{amplitude_AN}
A_{N,B} = \frac{2^{2N-N/2}\, N}{\pi^{N/2+1} \prod_{j=1}^{N} j!} \;.
\end{eqnarray}

Moreover the function $\Upsilon_N$ defined by Eq.~(\ref{upsilon_N}) can be written as a determinant:
$\Theta_N$ is a determinant, and the integration over $q_N$ can be absorbed in its last line.
The elements of this last line are indexed by the column index~$j$:
\begin{multline}
\int_0^\infty \rmd q_N\ {q_N}^j \cos \left( q_N-j\frac{\pi}{2} \right) e^{-t\frac{q_N^2}{2M^2}}\\
=\sqrt{\pi} M^{j+1} \mathrm{U}_j(M,t) \;,
\end{multline}
with the help of the family of functions 
\begin{equation}
\mathrm{U}_j(t) \equiv \mathrm{U}_j(M,t)= t^{-\frac{j+1}{2}} H_j\left( \frac{M}{\sqrt{2t}}\right) e^{-\frac{M^2}{2t}} \;,
\end{equation}
where $H_j$ is the Hermite polynomial of degree $j$ (see formula~\ref{Hermite_watermelons}).
Then $\Upsilon_N(\{q_i\}|M,t)$ is the determinant of a matrix, whose elements indexed by $(a,b)$ are given by
\begin{equation}
\begin{cases}
q_a^{b-1} \cos \left( q_a - b\frac{\pi}{2} \right) & \text{if $1\leq a\leq N-1$ and $1 \leq b \leq N$,} \\
\sqrt{\pi} M^{b+1} \mathrm{U}_b(M,t) & \text{if $a=N$ and $1\leq b \leq N$.}
\end{cases}
\end{equation}
%
%
The next step is to expand the determinant $\Upsilon_N$ in minors with respect to its last line:
\begin{multline}
\Upsilon_N(q_1,\dots,q_{N-1}|M,t)= \\
\sum_{j=1}^{N} (-1)^{N+j} \sqrt{\pi} M^{j+1} \mathrm{U}_j(M,t) \det \left[\grasrm{M}_{N,j}(\Theta_N) \right],
\end{multline}
where $\det\left[ \grasrm{M}_{N,j} (\Theta_N) \right]$ denotes the determinant of the
minor $(N,j)$ of the matrix $\Theta_N$, obtained by removing its $N^{\textrm{th}}$ line and its $j^{\textrm{th}}$ column.
These minors only involves the $q_1,\dots,q_{N-1}$ and $M$, not $\tau_M$. Expanding the product of the two functions $\Upsilon_N$ entering into the expression~(\ref{joint_bridge_int}) yields 
\begin{multline}
\label{double_sum}
\Upsilon_N(\{q_i\}|M,\tau_M) \times \Upsilon_N(\{q_i\}|M,1-\tau_M) =\\
\sum_{i,j=1}^{N} (-1)^{i+j} \pi M^{i+j+2} \mathrm{U}_i(\tau_M) \mathrm{U}_j(1-\tau_M)\\
\times \det \left[\grasrm{M}_{N,i}(\Theta_N) \right] \times \det \left[\grasrm{M}_{N,j}(\Theta_N) \right].
\end{multline}
The dependence on the variables $q_i$'s appears only in the last line of that expression (\ref{double_sum}), and
the Cauchy-Binet formula (\ref{nice_formula_det}) allows to compute the integration with respect to
$q_1,\dots,q_{N-1}$ of the product of determinants of minors:
\begin{eqnarray}
\label{integration_minors}
&&\int_0^\infty \rmd q_1 \dots \rmd q_{N-1} \ e^{-\frac{\sum_{k=1}^{N-1} q_k^2}{2M^2}}  \\
&&\times \det  \left[\grasrm{M}_{N,i}(\Theta_N) \right]
 \det  \left[\grasrm{M}_{N,j}(\Theta_N) \right] \nn
 \\
&&=(N-1)! \det \left[ \grasrm{M}_{i,j} \left( M^{a+b-1} \sqrt{\frac{\pi}{2^{a+b+1}}} (\mathrm{D}_{N,B})_{a,b} \right) \right] \,, \nn
 \end{eqnarray}
where $a,b$ are respectively the line and column indices of the matrix of which we take the determinant,
and $\mathrm{D}_{N,B}$ is the $N\times N$ matrix with elements
\begin{equation}
(\mathrm{D}_{N,B})_{a,b}=(-1)^{a-1} H_{a+b-2}(0)-e^{-2M^2} H_{a+b-2}(\sqrt{2}M) \,.
\end{equation}
This matrix enters into the expression of the cumulative distribution of the maximal height of $N$ watermelons~\cite{Feierl09}:
\begin{equation}
\mathrm{Proba}\left[ \max_{0 \leq \tau \leq 1} x_N(\tau) \leq M \right] = \frac{1}{\prod_{j=1}^{N-1} (j!2^j)} \det \mathrm{D}_{N,B} .
\end{equation}
Putting together~(\ref{double_sum}) and~(\ref{integration_minors}) in formula~(\ref{joint_bridge_int}), and
dividing by $K_{N,B}$ given in Eq.~(\ref{normalization_N}), one finds the joint pdf of $M$ and $\tau_M$
\begin{subequations}
\begin{multline}
P_{N,B}(M,\tau_M)= 
B_{N,B} \sum_{i,j=1}^{N}
(-1)^{i+j} \mathrm{U}_i(\tau_M) \mathrm{U}_j(1-\tau_M) \\ \times \det \left[ \grasrm{M}_{i,j} \left( \mathrm{D}_{N,B} \right) \right] \;,
\end{multline}
or equivalently
\begin{multline}
P_{N,B}(M,\tau_M) = - B_{N,B} \det
\left( \begin{array}{cc}
\mathrm{D}_{N,B} & \mathrm{U}(\tau_M) \\
^t\mathrm{U}(1-\tau_M) & 0
\end{array}
\right) 
\\
= B_{N,B} \det \left[\mathrm{D}_{N,B}\right]\
{}^t\mathrm{U}(1-\tau_M) \mathrm{D}^{-1}_{N,B} \mathrm{U}(\tau_M) \;,
\end{multline}
\end{subequations}
with the normalization constant $B_{N,B}^{-1}=\sqrt{2\pi} \prod_{j=1}^{N-1} j!2^j$.

It is also interesting to characterize the small $\tau_M$ behavior of the distribution of ${P}_{N,B}(\tau_M) = \int_0^\infty P_{N,B}(M, \tau_M) \rmd M$. To study it, it is useful to start from Eq. (\ref{joint_bridge_int}) which, after integration over $M$ yields
\begin{multline}
{P}_{N,B}(\tau_M) = \tilde A_N\prod_{i=1}^{N-1} \int_0^\infty \rmd q_i \int_0^\infty \rmd q_N q_N  \int_0^\infty \rmd q'_N q'_N  \\
\times \frac{\Theta_N(q_1, \cdots, q_N) \Theta(q_1, \cdots, q'_N)}{\left[ (1-\tau_M) {q'_N}^2 + \tau_M {q_N}^2 + \sum_{i=1}^{N-1}q_i^2 \right]^{1 + \frac{N^2}{2}} } \;,
\end{multline}
with $\tilde A_N = 2^{N^2/2} \Gamma \left(1 + \frac{N^2}{2} \right) A_{N,B}$. To study the behavior of ${P}_{N,B}(\tau_M)$ when $\tau_M \to 0$, we perform the changes of variables $q_i = \sqrt{\tau_M} y_i$, for $1 \leq i \leq N$ so that when $\tau_M \to 0$, one has to leading order 
\begin{eqnarray}\label{leading_marginal}
&&{P}_{N,B}(\tau_M) \sim \tau_M^{\frac{-N^2+N-1}{2}} \prod_{i=1}^{N} \int_0^\infty \rmd y_i y_N \int_0^\infty \rmd q'_Nq'_N   \nn \\ 
&&\times \frac{\Theta_N(\sqrt{\tau_M}y_1, \cdots, \sqrt{\tau_M}y_N) \Theta(\sqrt{\tau_M}y_1, \cdots, q'_N)}{\left[ {q'_N}^2 + \sum_{i=1}^{N}y_i^2 \right]^{1 + \frac{N^2}{2}} }\;. \nn \\
\end{eqnarray}
Besides, one has, also to leading order
\begin{eqnarray}\label{exp_1}
\Theta_N(\sqrt{\tau_M}y_1, \cdots, \sqrt{\tau_M}y_N)  &\sim& \tau_M^{\frac{N^2}{2}} \prod_{i=1}^N y_i \prod_{i < j }^N (y_i^2 - y_j^2) \nonumber \\
&+& {\cal O}(\tau_M^{\frac{N^2}{2}+1}) \;, 
\end{eqnarray}
and similarly
\begin{eqnarray}\label{exp_2}
\Theta(\sqrt{\tau_M}y_1, \cdots, q'_N) &\sim& \tau_M^{\frac{(N-1)^2}{2}} \prod_{i=1}^{N-1} y_i \prod_{i < j }^{N-1} (y_i^2 - y_j^2) f(q'_N) \nonumber \\
&+& {\cal O}(\tau_M^{\frac{(N-1)^2}{2}+1}) \;.
\end{eqnarray}
with some regular function $f(q'_N)$. However, being anti-symmetric under the permutation of the variables $y_i$'s it is easy to see that this leading order term (\ref{exp_1}, \ref{exp_2}) vanishes after the integration over $y_1, \cdots, y_N$ in Eq. (\ref{leading_marginal}). In fact if one writes the low $\tau_M$ expansion of the product as
\begin{eqnarray}
&&\Theta_N(\sqrt{\tau_M}y_1, \cdots, \sqrt{\tau_M}y_N) \Theta(\sqrt{\tau_M}y_1, \cdots, q'_N)  \nn \\
&&= \tau_M^{\frac{N^2 + (N-1)^2}{2}} \left(\nu_0 + \nu_1 \tau_M + \nu_2 \tau_M^2 + \cdots\right)
\end{eqnarray}  
where $\nu_k$'s are functions of $y_1, \cdots y_N, q'_N$ one can show that the first non-vanishing term, after integration over these variables, corresponds
to $\nu_k$ with $k=1$ for $N=2$, $k=2$ for $N=3$ and more generally $k=N-1$ for generic $N$. Therefore one has, to leading order
\begin{eqnarray}
P_{N,B}(\tau_M) \sim \tau_M^{\frac{N^2+N-2}{2}} \;,
\end{eqnarray}
although the explicit calculation of the amplitude is a very hard task. 


\subsection{$N$ vicious Brownian excursions}

We now consider watermelons with a wall, {\it i.e.} $N$ excursions under the non-crossing condition
(see the definition in paragraph~\ref{2excursions}). Again the starting and ending points are at the origin,
but with this configuration we must use $\gras{a}=\gras{b}=(\epsilon,2 \epsilon, \dots, N \epsilon)=\gras{\varepsilon}$,
which is slightly different from the cases of bridges, because now, all paths must be positive.
The joint pdf of $M$ and $\tau_M$ is
\begin{equation}
P_{N,E}(M,\tau_M)=\lim_{\eta,\epsilon \to 0} \frac{W_{N,E}(M-\eta,\tau_M,\epsilon)}{Z_{N,E}(\eta,\epsilon)},
\end{equation}
where the probability weight of all paths starting and ending in $\gras{\varepsilon}$, verifying for all $0\leq \tau \leq 1$
the non-crossing condition $0<x_1(\tau)<\dots<x_N(\tau)<M$, and for which $x_N(\tau_M)=y_N=M-\eta$, reads
\begin{multline}
W_{N,E}(M-\eta,\tau_M,\epsilon)= \int_{\text{ord}} \rmd \gras{y} \Big\{ \delta \left( y_N-(M-\eta) \right) \\
\times p_{\text{Box}}(\gras{\varepsilon},1|\gras{y},\tau_M) \ p_{\text{Box}}(\gras{y},\tau_M|\gras{\varepsilon},0) \Big\} .
\end{multline}
Here the integration is over the ordered domain $0<y_1<\dots<y_N<M$.
The boundary conditions of the excursions configuration enforce the use of the propagator $p_{\text{Box}}$ associated to $N$ fermions confined
in the segment $[0,M]$~(\ref{2excursions_propagator}): the Hamiltonian is of the form~(\ref{Hamiltonian_N}), 
with the one particle Hamiltonian~(\ref{2excursions_hamiltonian}).
\par
One can follow the first steps of the paragraph~\ref{2excursions}, replacing now $2$-vectors by $N$-vectors.
As in the bridge case, the ordered integration can be set unordered (\emph{i.e.} integrating over $0\leq y_i \leq M-\eta$
for all $1 \leq i \leq N-1$), with a numerical prefactor (see appendix~\ref{ordered_unordered}). The spectral decomposition is done with sums over $2N$ indices $\gras{n}=(n_1,\dots,n_N)$ and $\gras{n'}=(n'_1,\dots,n'_N)$, as in formula~(\ref{2excursions_weightstep1}).
Using the symmetry of the integrand under the exchanges of any couple $n_i \leftrightarrow n_j$ and  $n'_i \leftrightarrow n'_j$
(for $1 \leq i,j \leq N$) one can write the Slater determinants of the intermediate point (at $\tau=\tau_M$)
under their diagonal form, with a numerical prefactor, irrelevant in the computation.
Hence one has the formula
\begin{multline}
W_{N,E}(M-\eta,\tau_M,\epsilon)= \\
N! \, 2^{-\frac{N(N-1)}{2}} \sum_{\gras{n},\gras{n'}}
e^{-\frac{\pi^2}{2 M^2} \left[(1-\tau_M) \gras{n'}^2 + \tau_M \gras{n}^2 \right]} \\
\times  \Psi_{\gras{n'}}(\gras{\varepsilon}) \Psi_{\gras{n}}(\gras{\varepsilon}) \
\psi_{n'_N}(M-\eta) \psi_{n_N}(M-\eta) \\
\times  \prod_{i=1}^{N-1} \left\{\int\limits_0^{M-\eta} \rmd y_i \ \psi_{n'_i}(y_i) \psi_{n_i}(y_i) \right\}
\end{multline}
where $\psi_n(x)$ is given in Eq. (\ref{psi_n}). The expansion to lowest order in $\eta$ gives $N-1$ closure relations (the last line of the preceding expression)
\begin{equation}
\int_0^M \rmd y_i\ \psi_{n'_i}(y_i) \psi_{n_i}(y_i) = \delta_{n'_i,n_i} \;,
\end{equation}
and we use formula~(\ref{box_wavefunction_expansion_eta}) to get
\begin{multline}
W_{N,E}(M-\eta,\tau_M,\epsilon)=\\
 \eta^2 2^{-\frac{N(N-1)}{2}} \frac{2 \pi^2}{M^3}
\sum_{\gras{n},n'_N} (-1)^{n_N+n'_N} n_N n'_N \\
\times
e^{-\frac{\pi^2}{2 M^2} \sum\limits_{i=1}^{N-1} n_i^2} \
e^{-\frac{\pi^2}{2M^2} \left[ (1-\tau_M) {n'_N}^2 + \tau_M n_N^2 \right]} \\
\times 
\det \psi_{\gras{n}} (\gras{\epsilon}) \left. \det \psi_{\gras{\tilde n}}(\gras{\epsilon}) \right. \;,
\end{multline}
where $\gras{\tilde n} \equiv (n_1, n_2, \cdots, n'_N)$. One can then use the expansion to lowest order in $\epsilon$
\begin{multline}
\det \psi_{\gras{n}}(\gras{\epsilon}) = (-1)^{\frac{N(N-1)}{2}} \left( \prod_{j=1}^{N} \frac{j^{2j-1}}{(2j-1)!} \right)
2^{\frac{N}{2}} \pi^{N^2}\\ 
\times 
 \left(\frac{1}{M} \right)^{N(N+\frac{1}{2})}
\prod_{i=1}^N n_i \ \Delta_N(n_1^2,\dots,n_N^2) \ \epsilon^{N^2} + o(\epsilon^{N^2}) ,
\end{multline}
where $\Delta_N(\lambda_1,\dots,\lambda_N)$ is the $N\times N$ Vandermonde determinant. In this formula
the first line contains only numerical constants, irrelevant for the computation (they will be absorbed in the normalization constant),
only the second line is relevant, containing the dependence in $M$ and in the indices of the sums $n_i$, 
and the lowest order in $\epsilon^{N^2}$. Hence one has to lowest order
\begin{multline}
W_{N,E}(M-\eta,\tau_M,\epsilon)=\eta^2 \epsilon^{2N^2}\ {\rm W}_{N,E}(M,\tau_M) \\ 
+\mathcal{O}(\eta^3 \epsilon^{2N^2},\eta^2 \epsilon^{2N^2+1}) \;,
\end{multline}
as well as, for the normalization,
\begin{equation}
Z_{N,E}(\eta,\epsilon)=\eta^2 \epsilon^{2N^2}\ K_{N,E} + o(\eta^2 \epsilon^{2N^2}) \;,
\end{equation}
with $K_{N,E}$ a number. Taking the double limit $\epsilon, \eta \to 0$ one obtains the joint probability
\begin{multline}
\label{Nexcursions_joint_brut}
P_{N,E}(M,\tau_M)= \frac{{\rm W}_{N,E}(M,\tau_M)}{K_{N,E}} \\
=\frac{A_{N,E}}{M^{N(2N+1)+3}} \sum_{\gras{n},n'_N} \Bigg\{
(-1)^{n_N+n'_N} \ n_N^2 {n'_N}^{2}  \prod_{i=1}^{N-1} n_i^2 \\
\times \Delta_N(n_1^2,\dots,n_{N-1}^2, n_N^2) \ \Delta_N(n_1^2,\dots,n_{N-1}^2, {n'_N}^2)\\
\times e^{-\frac{\pi^2}{2 M^2} \sum\limits_{i=1}^{N-1} n_i^2} \
e^{-\frac{\pi^2}{2M^2} \left[ (1-\tau_M) {n'_N}^2 + \tau_M n_N^2 \right]} \Bigg\}\;,
\end{multline}
where the numerical constant is given by (see appendix~\ref{normalization_excursions})
\begin{equation}
\label{Nexcursions_normalization}
A_{N,E}=\frac{N \pi^{2N^2+N+2}}{2^{N^2-N/2} \prod_{j=0}^{N-1} \Gamma(2+j) \Gamma\left(\frac{3}{2}+j \right)} .
\end{equation}
\par
Following the notations introduced in the $N=2$ case, 
we define [writing for compactness the $N-1$-uplet $\{n_i\}=(n_1,\dots,n_{N-1})$]
\begin{align}
\Omega_p(\{n_i\} |M,t)
=&\sum_{k=1}^{\infty} (-1)^k k^2 \Delta_N(\{n_i^2\},k^2)
e^{-\frac{\pi^2}{2M^2}t k^2} \nn\\
=&\left| \begin{array}{cccc}
1&n_1^2&\dots&n_1^{2(N-1)} \\
1&n_2^2&\dots&n_2^{2(N-1)}\\
\vdots & \vdots & & \vdots \\
\omega_1(t)&\omega_2(t)&\dots&\omega_N(t)
\end{array}
\right|
\end{align}
where we used the property of the Vandermonde determinant to factorize the sum with respect to $k$ in the last line of the determinant,
which plays the role of $n_N$ associated to $t=\tau_M$, or the role of $n'_N$ with $t=1-\tau_M$,
resulting in elements like $\omega_i(t) \equiv \omega_i(M,t)$, for $1\leq i \leq N$
\begin{equation}
\omega_i(t)=\sum_{k=1}^{\infty}(-1)^k k^{2i} e^{-\frac{\pi^2}{2M^2}t k^2} \;.
\end{equation}
With this notation, we obtain the compact formula
\begin{multline}
\label{Nexcursions_first}
P_{N,E}(M,\tau_M)= \frac{A_{N,E}}{M^{N(2N+1)+3}} \prod_{i=1}^{N-1} \left\{ \sum_{n_i=1}^{\infty}
n_i^2 e^{-\frac{\pi^2}{2M^2}n_i^2} \right\} \\
\times \Omega_N(\{n_i\} |M,\tau_M) \Omega_N(\{n_i\}|M,1-\tau_M) \;,
\end{multline}
which is the ``discrete'' analogue of formula~(\ref{joint_bridge_int}) obtained in the case
of $N$ bridges.
\par
The ultimate step is to expand each determinant $\Omega_N$ with respect to their last line. 
Then, using the discrete Cauchy-Binet identity with respect to the sums over $n_i$, one has
\begin{multline}
P_{N,E}(M,\tau_M)=\frac{B_{N,E}}{M^2} \sum_{i,j=1}^{N} \left( \frac{2 \pi^2}{M^2} \right)^{i+j} \\
\times \omega_i(\tau_M) \omega_j(1-\tau_M) \det \left[ \grasrm{M}_{i,j}(\mathrm{D}_{N,E}) \right] ,
\end{multline}
with the normalization
\begin{equation}
B_{N,E}= \frac{(-1)^{N+1}\pi^{N/2}\sqrt{2}\ 2^{-2N^2}}{\prod_{j=1}^{N-1} j! \Gamma\left(\frac{3}{2}+j\right) } ,
\end{equation}
and where appear the minors of the matrix $\mathrm{D}_{N,E}\equiv \mathrm{D}_{N,E}(M)$
whose elements are, for $1 \leq i,j \leq N$
\begin{equation}
{\mathrm{D}_{N,E}}_{i,j}=\sum_{n=-\infty}^{+\infty} H_{2(i+j-1)}(\sqrt{2}M n) e^{-2 M^2 n^2}.
\end{equation}
This matrix enters into the expression of the cumulative probability of the maximum~\cite{Katori08}:
\begin{equation}
\mathrm{Proba}\left[\max_{0 \leq \tau \leq 1} x_N(\tau) \leq M\right] = 
(-1)^N\frac{\det \mathrm{D}_{N,E}}{2^{N^2} \prod_{j=1}^{N}(2j-1)!} \;,
\end{equation}
Defining the vector with elements, for $1\leq i \leq N$, 
\begin{equation}
{\mathrm{U}_E}_i(t) \equiv {\mathrm{U}_E}_i(M,t)=\frac{1}{M} \left(-\frac{2 \pi^2}{M^2} \right)^i \omega_i(t) \;,
\end{equation}
one is able to express the result as the determinant
\begin{subequations}
\begin{equation}
P_{N,E}(M,\tau_M)= -B_{N,E} \det \left(
\begin{array}{ccc}
\mathrm{D}_{N,E} & \mathrm{U}_E(\tau_M) \\
{}^t\mathrm{U}_E(1-\tau_M) & 0
\end{array}
\right)\;,
\end{equation}
or as the matrix product
\begin{multline}
P_{N,E}(M,\tau_M) \\
=B_{N,E} \det\left[ \mathrm{D}_{N,E} \right] {}^t \mathrm{U}_E(1-\tau_M) \mathrm{D}_{N,E}^{-1} \mathrm{U}_E(\tau_M) \;.
\end{multline}
\end{subequations}
\par
The behavior of the marginal distribution $P_{N,E}(\tau_M)$ when $\tau_M \to 0$ 
can be obtained as explained in paragraph~\ref{2excursions} for $N=2$ vicious excursions.
The leading behavior, when  $\tau_M \to 0$, is given by
\begin{equation}
P_{N,E}(\tau_M) \sim e^{-\frac{\pi}{\sqrt{\tau_M} }\sqrt{\frac{N(N+1)(2N+1)}{6}}} \;.
\end{equation}
However the algebraic correction to this leading behavior is hard to evaluate.



\section{Numerical simulations}

In this section we present the results of our numerical simulations of the distribution $P_{N, B/E}(\tau_M)$  both for the
cases of bridges and excursions. To generate numerically such watermelons configurations we exploit
the connection between these vicious walkers problems and Dyson's Brownian motion, which we first recall. 

\subsection{Relationship with Dyson's Brownian motion}

To make the connection between the vicious walkers problem and Dyson's Brownian motion, we consider the propagator
of the $N$ vicious walkers
\begin{equation}\label{start_propag}
{\cal P}_N({\mathbf v},t_2|{\mathbf u},t_1) \equiv {\cal
P}_N(v_1,\cdots,v_N,t_2|u_1,\cdots,u_N,t_1) \;,
\end{equation}
which is the probability that the process reaches the configuration 
$x_1(t_2) = v_1, \cdots, x_N(t_2) = v_N$ at time $t_2$ given that $x_1(t_1) = u_1, \cdots,
x_N(t_1) = u_N$ at time $t_1$: this is thus a conditional probability. This propagator satisfies the Fokker-Planck equation:
\begin{equation}\label{fp_vicious}
\frac{\partial}{\partial t_2} {\cal P}_N({\mathbf v},t_2|{\mathbf u},t_1) = \frac{1}{2} \sum_{i=1}^N \frac{\partial^2}{\partial v_i^2} {\cal P}_N({\mathbf v},t_2|{\mathbf u},t_1) \;,
\end{equation}
together with the initial condition
\begin{eqnarray}\label{fp_ci_vicious}
{\cal P}_N({\mathbf v},t_2=t_1|{\mathbf u},t_1) = \delta^{(N)}({\mathbf v}-{\mathbf u}) \;,
\end{eqnarray}
and the non-crossing condition, which is specific to this problem, 
\begin{eqnarray}\label{no_crossing}
{\cal P}_N({\mathbf v},t_2|{\mathbf u},t_1) = 0 \;, \; {\rm si} \; v_i = v_j \;, \forall \; (t_1, t_2) \;.
\end{eqnarray}

On the other hand, let us focus on Dyson's Brownian motion. For this purpose we consider random matrices whose elements are time dependent and are themselves Brownian motions. For instance, for random matrices from GUE, {\it i.e.} with $\beta = 2$, we consider random Hermitian matrices $H\equiv H(t)$, of size $N \times N$, which elements $H_{mn}(t)$ are given by  
\begin{eqnarray}\label{hermite_time}
H_{mn}(t) = 
\begin{cases}
\frac{1}{\sqrt{2}} \left(b_{mn}(t) + i  \, \tilde b_{mn}(t)\right) \;,& m < n \;, \\
b_{mm}(t) \;, & m = n \\
\frac{1}{\sqrt{2}} \left(b_{nm}(t) - i  \, \tilde b_{nm}(t)\right) \;, & m > n \;, \\
\end{cases} 
\end{eqnarray}
where $b_{mn}(t)$ and $\tilde b_{mn}(t)$ are independent Brownian motions (with a diffusion constant $D=1/2$). We denote 
$\lambda_1(t) < \lambda_2(t) < \cdots < \lambda_N(t)$ the $N$ eigenvalues of $H(t)$ (\ref{hermite_time}) [or more generally of a matrix belonging to a $\beta$ ensemble with $\beta = 1, 2, 4$ which is constructed as in Eq.~(\ref{hermite_time}) with the appropriate symmetry]. One can then show that the $\lambda_i$'s obey the following equations of motion (which define Dyson's Brownian motion)~\cite{Mehta91, Dyson62}
\begin{eqnarray}\label{Dyson}
\frac{d \lambda_i(t)}{dt} = \frac{\beta}{2} \sum_{1 \leq j\neq i \leq N} \frac{1}{\lambda_i(t) - \lambda_j(t)} + \eta_i(t) \;,   
\end{eqnarray}
where $\eta_i$'s are independent Gaussian white noises, $\langle \eta_i(t) \eta_j(t') \rangle = \delta_{ij} \delta(t-t')$. Let ${\cal P}_{\rm Dyson}(\boldlbd,t|\boldmu,0) \equiv {\cal P}_{\rm Dyson}(\lambda_1, \cdots, \lambda_N,t | \mu_1, \cdots, \mu_N, t=0)$ be the propagateur of this Dyson's Brownian motion~(\ref{Dyson}). It satisfies the Fokker-Planck equation
\begin{eqnarray}\label{fp_dyson}
&&\frac{\partial}{\partial t} {\cal P}_{\rm Dyson} = \frac{1}{2} \sum_{i=1}^N \frac{\partial^2}{\partial \lambda_i^2} {\cal P}_{\rm Dyson} \nonumber \\
&&- \frac{\beta}{2} \sum_{i=1}^N \frac{\partial}{\partial \lambda_i} \left[\sum_{1\leq j \neq i \leq N} \frac{1}{\lambda_i-\lambda_j} {\cal P}_{\rm Dyson}\right] \;.
\end{eqnarray}
Of course, one has also here ${\cal P}_{\rm Dyson}(\boldlbd,t|\boldmu,0) = 0$ if $\lambda_i = \lambda_j$, for any time $t$ and from Eq. (\ref{fp_dyson}) one actually obtains
\begin{eqnarray}\label{repulsion_beta}
{\cal P}_{\rm Dyson}(\boldlbd,t|\boldmu,0) \sim (\lambda_j - \lambda_i)^{\beta} \;, \; \lambda_i \to \lambda_j \;.
\end{eqnarray} 
This Fokker-Planck equation can be transformed into a Schr\"odinger equation by applying the standard transformation~\cite{risken_book}:
\begin{eqnarray}\label{transform_fp}
{\cal P}_{\rm Dyson}(\boldlbd,t|\boldmu,0)  &=& \frac{\exp{\left[ \frac{\beta}{2} \sum_{1\leq i<j \leq N} \log{(\lambda_j-\lambda_i)}\right]}}{\exp{\left[ \frac{\beta}{2} \sum_{1\leq i<j \leq N} \log{(\mu_j-\mu_i)}\right]}} \nonumber \\
&\times& {\cal W}_{\rm Dyson}(\boldlbd,t|\boldmu,0) \;, 
\end{eqnarray} 
where ${\cal W}_{\rm Dyson}(\boldlbd,t|\boldmu,0)$ is such that
\begin{eqnarray}\label{fp_ci_W}
{\cal W}_{\rm Dyson}(\boldlbd,t=0|\boldmu,0) = \delta^{(N)} ({\boldlbd}-{\boldmu}) \;.
\end{eqnarray}

On the other hand, given Eqs~(\ref{repulsion_beta}) and~(\ref{transform_fp}) one has 
\begin{eqnarray}
{\cal W}_{\rm Dyson}(\boldlbd,t|\boldmu,0) \sim (\lambda_j - \lambda_i)^{\beta/2} \;, \; \lambda_i \to \lambda_j \;.
\end{eqnarray}
From Eqs~(\ref{fp_dyson}) and (\ref{transform_fp}), one obtains that ${\cal W}_{\rm Dyson}$ satisfies the following Schr\"odinger equation
\begin{eqnarray}
&&\frac{\partial}{\partial t} {\cal W}_{\rm Dyson} = \frac{1}{2} \sum_{i=1}^N \frac{\partial^2}{\partial \lambda_i^2} {\cal W}_{\rm Dyson} \nn \\
&&- \frac{\beta}{8}(\beta-2) \sum_{i=1}^N\sum_{1 \leq j \neq i \leq N} \frac{1}{(\lambda_j - \lambda_i)^2} {\cal W}_{\rm Dyson} \;.\label{schrod_dyson}
\end{eqnarray}
Hence for a generic value of $\beta$, (\ref{schrod_dyson}) is the Schr\"odinger equation associated to a Calogero-Sutherland Hamiltonian
(on an infinite line) \cite{calogero, sutherland}. However for the special value $\beta = 2$, the strength of the interaction vanishes exactly and in that case
the equation satisfied by ${\cal W}_{\rm Dyson}$ (\ref{repulsion_beta}, \ref{fp_ci_W}, \ref{schrod_dyson}) is identical to the one satisfied by the propagator in the vicious walkers problem~(\ref{fp_vicious}, \ref{fp_ci_vicious}, \ref{no_crossing})~: in that case Eq. (\ref{schrod_dyson}) corresponds to the Schr\"odinger  equation for free fermions. In other words, for the special value $\beta = 2$ there exists a simple relation between vicious walkers and Dyson Brownian motion which can be simply written as 
\begin{equation}\label{dyson_vicious}
{\cal P}_{\rm Dyson}(\boldlbd,t|\boldmu,0)  = \frac{\prod_{i<j} (\lambda_j - \lambda_i)}{\prod_{i<j}(\mu_j - \mu_i)} {\cal P}_N(\boldlbd,t|\boldmu,0) \;.
\end{equation}
In particular, for watermelons without wall on the unit time interval, for which the initial and final positions coincide, this relation (\ref{dyson_vicious}) tells us that this process is identical to Dyson's Brownian motion where the elements of the matrices are themselves Brownian bridges {\it i.e.} $b_{ij}\to B_{ij}$, $\tilde b_{ij} \to \tilde B_{ij}$ in Eq.~(\ref{hermite_time}) such that $B_{ij}(0) = B_{ij}(1) = 0$ and $\tilde B_{ij}(0) = \tilde B_{ij}(1) = 0$. This then allows to generate easily watermelons because one can generate a Brownian bridge $B_{ij}(t)$ on the interval $[0, 1]$ from a standard Brownian motion 
$b_{ij}(t)$ via the relation $B_{ij}(t) = b_{ij}(t) - t \, b_{ij}(1)$.

For non-intersecting excursions, one can show that the dynamics corresponds to Dyson's Brownian motion associated to random symplectic and Hermitian matrices, as explained in Ref.~\cite{Katori08}.

\subsection{Numerical results}

To sample non-intersecting Brownian bridges we use the aforementioned equivalence to Dyson's Brownian motion. We consider the discrete-time version of the matrix in Eq.~(\ref{hermite_time}), at step $0\leq kÊ\leq T$:
\begin{equation}
H_{i,j}(k) =
\begin{cases}
\frac{1}{\sqrt{2}} \left(B_{ij}(k) + i B_{ij}(k)\right) & \text{$1\leq i < j \leq N$,} \\
B_{ii}(k) & \text{$i=j$, $1\leq i \leq N$,} \\
\frac{1}{\sqrt{2}} \left(B_{ij}(k) - i B_{ij}(k)\right) & \text{$1\leq i < j \leq N$,}
\end{cases}
\end{equation}
where $B_{ij}(k)$ are independent discrete-time Brownian bridges. They are constructed from ordinary discrete-time and continuous space
random walks $b_{ij}(k)$ where each jump is drawn from a Gaussian distribution. 
The aforementioned relation, to construct Brownian bridges, reads in discrete time $B_{ij}(k)=b_{ij}(k)-(k/T)b_{ij}(T)$.
\par
Diagonalizing the matrix $H_{i,j}(k)$ at each step $k$ gives the ordered ensemble of $N$ distinct eigenvalues
$\lambda_1(k)<\lambda_2(k)<\dots<\lambda_N(k)$, which is equivalent to a sample of non-intersecting Brownian bridges, see Eq.~\ref{dyson_vicious}.
For each sample,
we keep in memory the maximal value of $\lambda_N(k)$, 
and the step time $k_M$ at which $\lambda_N(k_M)=\max_{0\leq k \leq T} \lambda_N(k)$.
The plots shown here are obtained by making the histogram of
the rescaled time $k_M/T$ (with a bin of size $8$ to smoothen the curve), with $T=256$ steps averaged over $10^{6}$ samples, for $N=2,3,\dots,16$. For Brownian excursions, one has to use another symmetry for the matrix, which is explained in Ref.~\cite{Katori08}.

\begin{figure}[t]
\includegraphics[width=\linewidth]{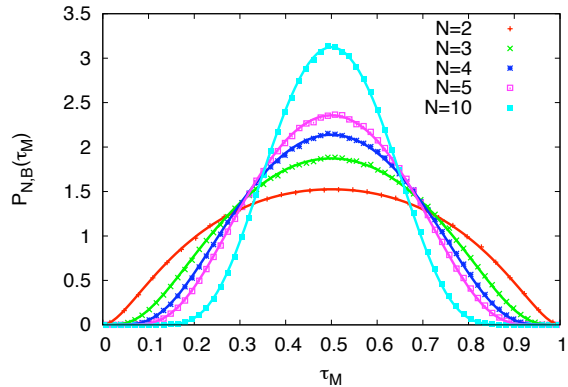}
\caption{Plots of the distribution $P_{N,B}(\tau_M)$ of the time $\tau_M$ to reach the maximum for $N=2,3,4,5,10$
non-intersecting Brownian bridges. The dots correspond to our numerical data while the solid lines correspond to our analytical predictions. 
There is no fitting parameter.}
\label{numerical_Nbridges}
\end{figure}

\begin{figure}[h]
\includegraphics[width=\linewidth]{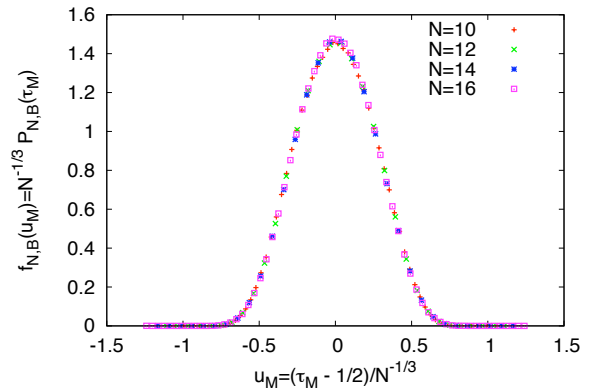}
\caption{Scaling form $f_{N,B}(u_M)$ of the distribution $P_{N,B}(\tau_M)$ with $N=10,12, 14, 16$. 
The $x$-axis is the rescaled time $u_M=(\tau_M-\half) \times N^{1/3}$, and the $y$-axis is the rescaled distribution
$f_{N,B}(u_M)= N^{-1/3} \times P_{N,B}(\half+u_M N^{-1/3})$.}
\label{numerical_scaling}
\end{figure}

First we present the comparison between our exact result for Brownian bridges and numerical simulations, for $N=2,3,4,5,10$.
The agreement between our numerical results and our exact analytical formula is perfect, see Fig.~\ref{numerical_Nbridges}. Then, anticipating on the applocation of our results to growing interfaces,
we show a plot of the rescaled distribution as a function of the rescaled time $u_M = (\tau_M-1/2)/N^{-1/3}$. Despite the fact that this scaling requires in principle that $N\gg 1$,
one sees that the rescaled distributions for moderate values of $N = 10, 12, 14, 16$ fall on a single master curve, see Fig.~\ref{numerical_scaling}. However, given the relatively narrow range of $N$ explored in these simulations, one can not be sure that the asymptotic, large $N$, behavior is reached. 


\par
\begin{figure}[h]
\includegraphics[width=\linewidth]{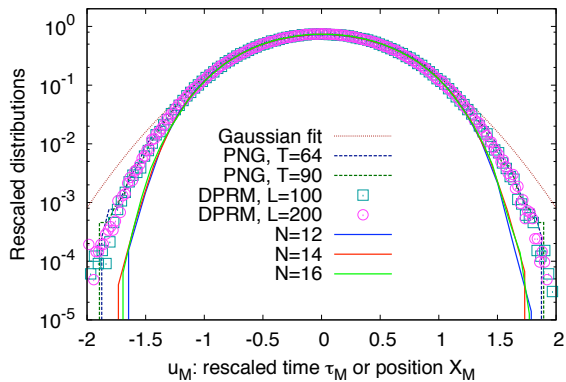}
\caption{Distribution of the rescaled time (respectively position) $u_M$ for $N$ vicious Bridges (respectively for PNG and DPRM models, see Fig. \ref{fig_intro}) on a log-lin plot. The dashed lines correspond to the PNG model, the symbols to the DPRM, the solid lines to the watermelons and the dotted line is a Gaussian fit. See the text for more details.}
\label{png}
\end{figure}

To compare our analytical results for $N$-watermelons to growing interfaces models and the DPRM, we have simulated both the PNG model and a discrete model
of DPRM on a lattice. To sample the PNG model, we have applied the rules that were discussed before (see Fig. \ref{cartoon_png}) where the time interval is discretized. In particular we use a discretization of the nucleation process: at each time step, a nucleation occurs at site $x$ such that $|x| < t$ with a 
small probability $p$, whose value is chosen such that one recovers a uniform density of nucleation events $\rho=2$ in the continuum limit. The interface evolves during a time $T$. We compute the distribution of the position $X_M$ of the maximal height of the droplet (see Fig. \ref{fig_intro} left), averaged over $10^5$ samples. In Fig. \ref{png} we have plotted the distribution of the rescaled position $u_M = X_M/T^{2/3}$ for $T=64$ and $T=90$ (both of them being statistically independent): the collapse of these data on a single master curve is consistent with the expected KPZ scaling in $T^{2/3}$.  

On the other hand, we have simulated a directed polymer model in $1+1$ dimensions as depicted in Fig. \ref{fig_mapping_dprm} (left), where on each site,
there is a random energy variable, with Gaussian fluctuations. We find the optimal path of length $L$ , which is the polymer with minimal energy, with one free end-point. And we compute the distribution of the position $X_M$ of this end-point, averaged over $10^5$ samples. In Fig. \ref{png} we have plotted the distribution of the rescaled position $u_M = X_M/(a L^{2/3})$ for $L=100$ and $L=200$ (both of them being statistically independent): the collapse of these data, for different $L$, on a single master curve is consistent with the expected KPZ scaling in $L^{2/3}$, while the value of $a \simeq \sqrt{2}$ was adjusted such that these data coincide with the ones for the PNG. 

For comparison we have plotted on this same graph, on Fig. \ref{png}, the data for the rescaled time $u_M$ of the maximum for $N$ non-intersecting Brownian bridges, {\it i.e.} the same data as in Fig. \ref{numerical_scaling} (except $N=10$), without any additional fitting parameter. 
One sees that the finite $N$ results describe quite accurately the center of the distribution, whereas there is 
a systematic deviation in the tails of the distribution (as expected since $N$ is relatively small). Both data, for watermelons on the one hand and for
the PNG on the other hand, suffer respectively from finite $N$ and finite $T$ effects. Our data for the PNG seem to indicate that when $T$ increases, the probability
weight is slightly displaced from the tails to the center of the (rescaled) distribution: this suggests that the limiting distribution is not Gaussian (the best Gaussian fit is plotted with a dotted line in Fig. \ref{png}). This convergence to the asymptotic distribution is however quite slow.


%
%

%

\section{Conclusion}

In conclusion, we have presented a detailed derivation of the analytic computation of
the distribution of the time at which the maximal height of $N$ vicious walkers is reached. Our analytic
approach is based on a path integral approach for free fermions, which incorporates in a rather physical way
the non-colliding condition. We have considered three different types of vicious walkers configurations: non-intersecting free 
Brownian motions, "stars" configuration, for $N=2$, non-intersecting Brownian bridges, "watermelons" configurations, for any number $N$ of vicious walkers and finally non-intersecting Brownian excursions, "watermelons with a wall" also for any $N$. We have also checked our results using numerical
simulations of Dyson's Brownian, which uses the deep connection between vicious walkers and Random Matrix Theory, which we have reminded in detail.

Thanks to the connection between vicious walkers and the Airy process (\ref{rel_wt_airy}), our results yields, in the large $N$ limit, results for the distribution 
of $X_M$, the position of the maximal height of a curved growing interface in the KPZ universality class (Fig.~\ref{fig_intro} left) or the transverse coordinate of the end-point of the optimal directed polymer (Fig.~{\ref{fig_intro} right). We have shown that our analytical results for finite $N \sim 15$, correctly rescaled, are in good agreement with our numerical data for the PNG model and DPRM on the square lattice. Performing the large $N$ asymptotic
analysis of our formula (\ref{joint_bridge_int}, \ref{Nexcursions_first}) remains a formidable challenge, which will hopefully motivate further study of this problem.

\acknowledgments

We thank the Centro Atomico in Bariloche for hospitality where part of this article was written.

\appendix

\section{Integration over ordered variables}
\label{ordered_unordered}

In the main text, we have to sum over realizations with the constraint $x_1<x_2<\dots<x_{N-1}<x_N$.
Here we study how integrals may be simplified by the symmetry of the integrand.

\subsection{$N=2$, two variables}

Consider a box $[a,b]$ in one dimension, and a function of two variables $f(x_1,x_2)$.
In the main text, we want to compute ordered integrals like
\begin{equation}
\label{integration_2_ordered}
\int_a^b \rmd x_2 \int_a^{x_2} \rmd x_1 f(x_1,x_2) .
\end{equation}
This is usually easier to compute the integral over the whole domain, that is to say
\begin{equation}
\label{integration_2_domain}
\int_a^b \rmd x_2 \int_a^{b} \rmd x_1 f(x_1,x_2) \;.
\end{equation}
Here we make a link between those two quantities, 
when $f$ is symmetric in the exchange of $x_1$ and $x_2$ $f(x_2,x_1)=f(x_1,x_2)$.
Let us start from equation~(\ref{integration_2_domain}) with a symmetric function $f$
\begin{align}
&\int_a^b \rmd x_2 \int_a^{b} \rmd x_1 f(x_1,x_2) \nn \\
=&\int_a^b \rmd x_2 \int_a^{x_2} \rmd x_1 f(x_1,x_2) + \int_a^b \rmd x_2 \int_{x_2}^b \rmd x_1 f(x_1,x_2) \nn \\
=&  \int_a^b \rmd x_2 \int_a^{x_2} \rmd x_1 f(x_1,x_2) + \int_a^b \rmd x_1 \int_a^{x_1} \rmd x_2 f(x_1,x_2) \nn \\
=& \int_a^b \rmd x_2 \int_a^{x_2} \rmd x_1 f(x_1,x_2) +  \int_a^b \rmd x_1 \int_a^{x_1} \rmd x_2 f(x_2,x_1)  \nn \\
=& 2 \int_a^b \rmd x_2 \int_a^{x_2} \rmd x_1 f(x_1,x_2)
\end{align}
From the second to the third line we use triangular integration $\{a<x_2<b,\ x_2<x_1<b\} = \{a<x_1<b,\ a<x_2<x_1\}$
in the second term. 
From the third line to the forth, we relabelled the variables in the second term. 
In this operation if $f$ was antisymmetric, we would have a minus sign, and hence a cancellation of the two terms.
The conclusion for two variables is that if $f$ is antisymmetric, the integration over the whole domain is zero, 
but if $f$ is symmetric in the exchange of the two variables, we have the useful formula
\begin{equation}
\label{integration_2_formula}
\int_a^b \rmd x_2 \int_a^{x_2} \rmd x_1 f(x_1,x_2) = 
\frac{1}{2} \int_a^b \rmd x_2 \int_a^{b} \rmd x_1 f(x_1,x_2) .
\end{equation}

\subsection{$N$ variables}

The previous result can be extended for $N$ variables, and the result is, 
for $f(x_1,\dots,x_N)$ symmetric in the exchange of any of two of its arguments
\begin{equation}
\label{integration_n_formula}
\int_{\text{ord}} \rmd \gras{x} f(\gras{x})= 2^{-\frac{N(N-1)}{2}} \int \rmd \gras{x} f(\gras{x}) \;,
\end{equation}
with $\gras{x}=(x_1,\dots,x_N)$ and an integration on the domain $a<x_1<x_2<\dots<x_N<b$ in the left-hand side (lhs),
and an integration over the whole domain $a<x_i<b$ for $1\leq i \leq N$ in the right-hand side (rhs).
\par
The proof is done by applying iteratively the formula~(\ref{integration_2_formula}) first to the $N-1$ couples
$(x_N,x_{N-1}),\ (x_{N},x_{N-2}), \ \dots, (x_N,x_1)$ changing the upper limit of integrations from $b$ to $x_N$.
This operation results gives a factor $2$ for each couple so that:
\begin{align}
&\int_a^b \rmd x_N \int_a^{b} \rmd x_{N-1} \dots \int_a^{b} \rmd x_2 \int_a^{b} \rmd x_1 \ f(\gras{x}) \nn \\
=& 2^{N-1} \int_a^b \rmd x_N \int_a^{x_N} \rmd x_{N-1} \dots \int_a^{x_N} \rmd x_2 \int_a^{x_N} \rmd x_1 \ f(\gras{x}) \;.
\end{align}
Then one repeats this procedure for the $N-2$ couples $(x_{N-1},x_{N-2})$ to $(x_{N-1},x_1)$,
which gives a factor $2^{N-2}$, and continues until step $k$ with the $N-k$ couples $(x_{N-k+1},x_{N-k})$ to $(x_{N-k+1},x_1)$
to have
\begin{align}
&\int_a^b \rmd x_N \int_a^{b} \rmd x_{N-1} \dots \int_a^{b} \rmd x_2 \int_a^{b} \rmd x_1 \ f(\gras{x}) \nn \\
=& 2^{N-1} 2^{N-2} \dots 2^{N-k} \nn \\
&\times 
\underbrace{\int_a^b \rmd x_N
\int_a^{x_N} \rmd x_{N-1} \dots
\int_a^{x_{N-k+1}} \rmd x_{N-k} }_{\textrm{ordered}} \nn \\
& \times 
\underbrace{
\int_a^{x_{N-k+1}} \rmd x_{N-k-1} \dots
\int_a^{x_{N-k+1}} \rmd x_1}_{\textrm{unordered}} f(\gras{x})
\end{align}
Doing so until $k=N-1$ one obtains the result~(\ref{integration_n_formula}).

\section{Cauchy-Binet identity}
\label{Cauchy-Binet}

\subsection{Continuous version}

In the main text, we are confronted with integrals of two determinants that takes the form:
\begin{equation}
I^{(N)}=\int \rmd \gras{x} \ \det_{1\leq i,j \leq N} \left( f_i (x_j) \right) \ \det_{1\leq k,l \leq N} \left( g_k (x_l) \right) ,
\end{equation}
where $\rmd \gras{x}=\prod_{i=1}^N \rmd x_i$, $f_i, g_i$, $1\leq i \leq N$ are functions of one variable, 
and each variable $x_i$ has the same domain of integration
(in the main text we usually integrate over the momenta $k_i$ from $0$ to $+\infty$).
We have
\begin{align}
I^{(N)} 	= \int \prod_{i=1}^{N} \rmd x_i \ &\left[ \sum_{\sigma \in S_N} \varepsilon (\sigma) \prod_{i=1}^{N} f_i(x_{\sigma(i)}) \right] \nn \\
		& \times \left[ \sum_{\sigma' \in S_N} \varepsilon (\sigma') \prod_{k=1}^{N} g_k(x_{\sigma'(k)}) \right] \nn \\
		= \sum_{\sigma \in S_N} \varepsilon (\sigma) &\int \prod_{i=1}^{N} \rmd x_i \prod_{i=1}^{N} f_i(x_{\sigma(i)}) \nn \\
		& \times \left[ \sum_{\sigma' \in S_N} \varepsilon (\sigma') \prod_{k=1}^{N} g_k(x_{\sigma'(k)}) \right]
\end{align}
$S_N$ is the symmetric group, $\epsilon (\sigma)$ is the signature of the permutation $\sigma \in S_N$. 
Now we make the change of variables $x'_i=x_{\sigma(i)}$. That is $x_j=x'_{\sigma^{-1}(j)}$ and the Jacobian is unity.
So we have 
\begin{multline}
I^{(N)} 	= \sum_{\sigma \in S_N} \varepsilon (\sigma) \int \prod_{i=1}^{N} \rmd x'_i \prod_{i=1}^{N} f_i(x'_i) \\
			\times \left[ \sum_{\sigma' \in S_N} \varepsilon (\sigma') \prod_{k=1}^{N} g_i(x'_{\sigma^{-1}(\sigma'(k))}) \right] \;.
\end{multline}
$S_N$ is a group, so it exists $\sigma'' \in S_N$ so that $\sigma''=\sigma^{(-1)} \circ \sigma'$, 
and $\varepsilon(\sigma'')=\varepsilon(\sigma^{-1}) \varepsilon(\sigma')=\varepsilon(\sigma) \varepsilon(\sigma')$.
Moreover $\sum_{\sigma \in S_N}=\sum_{\sigma^{-1} \circ \sigma \in S_N}= \sum_{\sigma'' \in S_N}$.
Getting back to the $x_i$ variables we have
\begin{align}
I^{(N)}	= \sum_{\sigma \in S_N} & \varepsilon(\sigma) \int \prod_{i=1}^{N} \rmd x_i \prod_{i=1}^{N} f_i(x_i) \nn \\
			& \times \left[ \sum_{\sigma' \in S_N} \varepsilon (\sigma') \prod_{i=k}^{N} g_k(x_{\sigma^{-1}(\sigma'(k))}) \right] \nn \\
		= \sum_{\sigma \in S_N}  & \int \prod_{i=1}^{N} \rmd x_i \prod_{i=1}^{N} f_i(x_i) \nn\\
		 & \times
			\left[ \sum_{\sigma' \in S_N} \varepsilon(\sigma) \varepsilon (\sigma')
				\prod_{k=1}^{N} g_k(x_{\sigma^{-1}(\sigma'(k))}) \right] \nn \\
		= \sum_{\sigma \in S_N} &  \int \prod_{i=1}^{N} \rmd x_i \prod_{i=1}^{N} f_i(x_i) \nn \\
		& \times 
			\left[ \sum_{\sigma'' \in S_N} \varepsilon(\sigma'') \prod_{k=1}^{N} g_k(x_{\sigma''(k)}) \right] \;.
\end{align}
In the last line, the term in bracket is simply the determinant $\det_{1\leq k , l \leq N}(g_k(x_l))$. 
The permutation $\sigma$ does not appear anymore, so we have $\sum_{\sigma \in S_N} 1 = {\rm{Card}}(S_N) =N!$.
Hence
\begin{equation}
I^{(N)} = N! \int \prod_{i=1}^{N} \rmd x_i \prod_{i=1}^{N} f_i(x_i) \ \det_{1\leq k,l \leq N} (g_k(x_l)) .
\end{equation}
At this stage, the basic result is that we have expressed the first determinant only as a product of the diagonal terms.
This is used in the main text.
\par
To go further we use the fact that the determinant of one matrice is the same as the determinant of its transpose:
\begin{align}
I^{(N)} 	&= N! \int \prod_{i=1}^{N} \rmd x_i \prod_{k=1}^{N} f_k(x_k) 
			\left[ \sum_{\sigma \in S_N} \epsilon(\sigma) \prod_{k=1}^{N} g_{\sigma(k)}(x_k) \right] \nn \\
		&= N! \sum_{\sigma \in S_N} \epsilon(\sigma) \int \prod_{i=1}^{N} \rmd x_i \prod_{k=1}^{N}
		\left(f_k(x_k) g_{\sigma(k)}(x_k) \right) \nn \\
		&= N! \sum_{\sigma \in S_N} \epsilon(\sigma) \prod_{k=1}^{N} \left\{ \int \rmd x f_k(x) g_{\sigma(k)}(x) \right\} .
\end{align}
From the second to the third line, we factorized the integrals, and we wrote $x=x_k$ for each $k$.
We recognize the determinant
of the matrix whose elements indexed by $(i,j)$ are given by an integral of the product $f_i \times g_j$, yielding the so-called Cauchy-Binet identity:
\begin{multline}
\label{nice_formula_det}
\int \rmd \gras{x} \ \det_{1\leq i,j \leq N} \left( f_i (x_j) \right) \ \det_{1\leq k,l \leq N} \left( g_k (x_l) \right) \\
= N! \det_{1\leq i,j \leq N} \left\{ \int \rmd x f_i(x) g_j(x) \right\} .
\end{multline}
\par
Moreover, one can easily deduce from Eq. (\ref{nice_formula_det}) the following relation, valid for any integrable function $h(x)$:
\begin{multline}
\int  \rmd \gras{x} \ \prod_{i=1}^N h(x_i) \
\left( \det_{1\leq i,j \leq p} f_i (x_j) \right) \left( \det_{1\leq i,j \leq p} g_i (x_j) \right) \\
= N! \det_{1 \leq i,j \leq N} \left\{ \int \rmd {x} \ h(x) f_i(x)  g_j(x) \right\} .
\end{multline}
This relation is used in the main text.

\subsection{Discrete version}

Here we consider the following multiple sum:
\begin{align}
I_N=&\sum_{n_1} \sum_{n_2} \dots \sum_{n_N} \det_{1\leq i,j,\leq N} \left( f_{i}(n_j) \right)
\det_{1\leq i,j,\leq N} \left( g_{i}(n_j) \right)  \nn \\
= & \sum_{\sigma \in S_N} \epsilon(\sigma) \sum_{n_1} \sum_{n_2} \dots \sum_{n_N}
\prod_{i=1}^{N}f_{i}(n_{\sigma(i)}) \nn \\
& \times \det_{1\leq i,j,\leq N} \left( g_{n_i}(x_j) \right)  \;.
\end{align}
Changing the indices of the sums $m_i=n_{\sigma(i)}$, or $n_k=m_{\sigma^{-1}(k)}$, one gets
\begin{align}
I_N=&\sum_{\sigma \in S_N} \epsilon(\sigma) \sum_{m_{\sigma^{-1}(1)}} \sum_{m_{\sigma^{-1}(2)}} \dots \sum_{m_{\sigma^{-1}(p)}}
\prod_{i=1}^{N}f_{i}(m_i) \nn \\
& \times   \det_{1\leq i,j,\leq N} \left( g_{i}(m_{\sigma^{-1}(i)}) \right) \nn \\
=& \sum_{\sigma \in S_N} \epsilon(\sigma) \sum_{m_1} \sum_{m_2} \dots \sum_{m_N} 
\prod_{i=1}^{N}f_{m_i}(x_i)  \nn \\
& \times \left( \epsilon(\sigma)  \det_{1\leq i,j,\leq N} \left( g_{i}(m_j) \right) \right) \;,
\end{align}
where we re-arranged the columns in the remaining determinant in the `natural' order, 
thus factorizing the signature of the permutation $\sigma$.
Hence there is a sum over the permutations $\sigma \in S_N$ of $\epsilon^2(\sigma)$, giving the cardinal $N!$.
We obtain as in the continuous case that the first determinant is written as the product of its diagonal terms  with a factor $N!$:
\begin{align}
&I_N=N!  \sum_{m_1} \sum_{m_2} \dots \sum_{m_N} \prod_{i=1}^{N}f_{i}(m_i) \det_{1\leq i,j,\leq N} \left( g_{i}(m_j) \right) \nn \\
&= N!  \sum_{m_1} \sum_{m_2} \dots \sum_{m_N} \prod_{i=1}^{N}f_{m_i}(x_i) 
\sum_{\sigma \in S_N} \epsilon(\sigma) \prod_{i=1}^{N} g_{\sigma(i)}(m_i) \nn \\
&=  N! \sum_{\sigma \in S_N} \epsilon(\sigma) \sum_{m_1} \sum_{m_2} \dots \sum_{m_N}  
\prod_{i=1}^{N} \left( f_{i}(m_i) g_{\sigma(i)}(m_i) \right) \;.
\end{align}
Factorizing the sums, one gets
\begin{align}
I_N &=  N! \sum_{\sigma \in S_N} \epsilon(\sigma) \prod_{i=1}^{N} \left( \sum_{m} f_i(m) g_{\sigma(i)}(m) \right) \nn \\
&= N! \det_{1\leq i, j \leq N} \left( \sum_{m} f_i(m) g_j(m) \right) \;.
\end{align}
In the same way as in the continuous case, if one considers
\begin{multline}
J_N=\sum_{n_1} \sum_{n_2} \dots \sum_{n_N} \prod_{i=1}^N h(n_i) \\
\times
\det_{1\leq i,j,\leq N} \left( f_{i}(n_j) \right)
\det_{1\leq i,j,\leq N} \left( g_{i}(n_j) \right) \;,
\end{multline}
one finally obtains
\begin{equation}
J_N= N! \det_{1 \leq i , j \leq N} \left( \sum_{m} h(m) f_i(m) g_j(m) \right) .
\end{equation}

\section{Normalization constant}
\label{normalization}

In this section, we compute the normalization constant of the joint distributions $P_{N,B}(M, \tau_M)$ 
and $P_{N,E}(M, \tau_M)$ for non-interesecting bridges and excursions respectively.

\subsection{Non-intersecting Brownian bridges}\label{normalization_bridge}

To compute the amplitude $A_{N,B}$ in Eq. (\ref{amplitude_AN}) we use the fact that the integration of $P_{N,B}(M, \tau_M)$
in Eq. (\ref{joint_bridge_int}), {\it i.e.} $\int_0^1 P_{N,B}(M, \tau_M) \rmd \tau_M$ must yield back the pdf of $M$ which was
computed in Ref. \cite{Schehr08}. Indeed in Ref. \cite{Schehr08} it was shown that
\begin{eqnarray}
F_{N,B}(M)&=& {\rm Proba}\left[\max_{0\leq \tau \leq 1} x_N(\tau) \leq M \right] \\
&=& \frac{ a_{N,B}}{M^{N^2}} \int_0^\infty \rmd {\bf q} \; e^{-\frac{{\bf q}^2}{2M^2}} \left[\Theta_N({\bf q}) \right]^2 \;,
\end{eqnarray}
where the amplitude $a_{N,B}$ can be computed using a Selberg integral \cite{Schehr08}
\begin{eqnarray}
a_{N,B} = \frac{2^{2N}}{(2 \pi)^{N/2} \prod_{j=1}^N j!} \;.
\end{eqnarray}
The pdf of the maximum, $F_{N,B}'(M) = \partial_M F_{N,B}(M)$ is thus given by
\begin{eqnarray}\label{Fn_prime_1}
&&F_{N,B}'(M) = -N^2\frac{a_{N,B}}{M^{N^2+1}} \int_0^\infty \rmd {\bf q} \; e^{-\frac{{\bf q}^2}{2M^2}} \Theta_N({\bf q})^ 2 \\
&& + \frac{N\,a_{N,B} }{M^{N^2+1}} \int_0^\infty \rmd {\bf q} \; e^{-\frac{{\bf q}^2}{2M^2}}\sum_{j,l = 1}^N (-1)^{j+l} q_N^{j+l}\cos{\left(q_N - j\frac{\pi}{2}\right)}  \nonumber \\
&&\times  \cos{\left(q_p-l\frac{\pi}{2}\right)}  \det[{\bf M}_{j,N}(\Theta_N)] \det[{\bf M}_{l,N}(\Theta_N)] \;,
\end{eqnarray}
where, in the last two lines, we have expanded both determinants in minors with respect to the last column (we recall that ${\bf M}_{j,N}(\Theta_N)$ denotes the minor $(j,N)$ of the matrix $\Theta_N$, obtained by removing its $j^{\rm th}$ line and its $N^{\rm th}$ column). Note that these minors do not depend on $q_N$. This suggests to consider, in the last two lines of Eq. (\ref{Fn_prime_1}), only the integral over the variable $q_N$ where we perform an integration by part. This yields:
\begin{eqnarray}\label{result_ipp}
&&\int_0^\infty \rmd q_N \frac{q_N}{M^2} e^{-\frac{q_N^2}{2M^2}} q_N^{j+l-1} \cos{\left(q_N-j\frac{\pi}{2}\right)}\cos{\left(q_N-l\frac{\pi}{2}\right)} \nonumber \\
&&= (j+l-1) \int_0^\infty \rmd q_N \, q_N^{j+l-2} \cos{\left(q_N-j\frac{\pi}{2}\right)}\nonumber \\
&&\times \cos{\left(q_N-l\frac{\pi}{2}\right)}  e^{-\frac{q_N^2}{2M^2}} + \int_0^\infty \rmd q_N q_N^{l+j-1} e^{-\frac{q_N^2}{2M^2}} \nn \\
&& \times \partial_{q_N}\left[ \cos{\left(q_N-j\frac{\pi}{2}\right)}\cos{\left(q_N-l\frac{\pi}{2}\right)}\right] \;.
\end{eqnarray}
Inserting this result of the integration by part (\ref{result_ipp}) in Eq.~(\ref{Fn_prime_1}) and using the following identity (which we have explicitly checked for $N=2,3,4,5$ but can only conjecture for any integer $N > 5$):
\begin{widetext}
\begin{multline}
(N^2+N) \int_0^\infty \rmd {\bf q} \; e^{-\frac{{\bf q}^2}{2M^2}} \left[\Theta_N({\bf q}) \right]^2 
=
2 N \int_0^\infty \rmd {\bf q} \; e^{-\frac{{\bf q}^2}{2M^2}} \left|
\begin{array}{ccc}
\sin{\left(q_1\right)} &  
\cdots &\sin{\left(q_N\right)} \\
q_1 \cos{\left(q_1\right)}  & \cdots
& q_N \cos{\left(q_N\right)}\\
q_1^2 \sin{\left(q_1\right)}  & \cdots & q_N^2 \sin{\left(q_N\right)}\\
\cdot & \cdot   & \cdot\\
\cdot & \cdot   & \cdot\\
q_1^{N-1} \cos{\left(q_1 - N\frac{\pi}{2}\right)} & \cdots & q_N^{N-1} \cos{\left(q_N -
N\frac{\pi}{2}\right)} \nonumber
\end{array}
\right| \\
\times \left|
\begin{array}{ccc}
\sin{\left(q_1\right)} &  
\cdots &\sin{\left(q_N\right)} \\
q_1 \cos{\left(q_1\right)}  & \cdots
& 2 q_N \cos{\left(q_N\right)}\\
q_1^2 \sin{\left(q_1\right)}  & \cdots & 3 q_N^2 \cos{\left(q_N\right)} \\
\cdot & \cdot   & \cdot \\
\cdot & \cdot   & \cdot\\
q_1^{N-1} \cos{\left(q_1 - N\frac{\pi}{2}\right)} & \cdots & N\,q_N^{N-1} \cos{\left(q_N - 
N\frac{\pi}{2}\right)} \nonumber
\end{array}
\right| \;,
\end{multline}
\end{widetext}
one obtains finally
\begin{eqnarray}\label{Fn_prime_2}
&&F'_{N,B}(M) = \frac{-N a_{N,B}}{M^{N^2+1}} \int_0^\infty \rmd {\bf q} \; e^{-\frac{{\bf q}^2}{2M^2}} \sum_{j,l=1}^N (-1)^{j+l} q_N^{j+l-1} \nn \\ 
&&\times \det[{\bf M}_{j,N}(\theta_N)] \det[{\bf M}_{l,N}(\theta_N)] \sin{\left(2 q_N - (l+j) \frac{\pi}{2}\right)} \;. \nn \\
\end{eqnarray}

On the other hand, from the definition of $P_{N,B}(M, \tau_M)$ given in Eq.~(\ref{joint_bridge_int}) one has
\begin{eqnarray}\label{int_taum}
\int_0^1 P_{N,B}(M, \tau_M) \rmd \tau_M = F'_{N,B}(M) \;,
\end{eqnarray}
with
\begin{multline}\label{joint_bridge_int_app}
P_{N,B}(M,\tau_M) = \frac{A_{N,B}}{M^{N^2+3}}  \int_0^\infty \rmd q_1 \dots \rmd q_{N-1}\ e^{-\frac{\sum_{k=1}^{N-1} q_k^2}{2M^2}} \\
\times \Upsilon_N(\{q_i\}|M,\tau_M) \Upsilon_N(\{q_i\}|M,1-\tau_M)  \;.
\end{multline}
To compute this integral over $\tau_M$ (\ref{int_taum}) we compute instead
\begin{eqnarray}\label{def_pmL}
&&{\rm P}(M,T) = \frac{A_{N,B}}{M^{N^2+3}}  \int_0^T  \rmd \tau_M \int_0^\infty \rmd q_1 \dots \rmd q_{N-1}\ e^{-\frac{\sum_{k=1}^{N-1} q_k^2}{2M^2}} \nonumber
\\
&&\times \Upsilon_N(\{q_i\}|M,\tau_M) \Upsilon_N(\{q_i\}|M,T-\tau_M) \;.
\end{eqnarray} 
and clearly
\begin{eqnarray}
\int_0^1 P_{N,B}(M, \tau_M) \rmd \tau_M = {\rm P}(M,T=1) \;. 
\end{eqnarray}
The structure of ${\rm P}(M,T)$ in Eq. (\ref{def_pmL}) suggests to compute its Laplace transform with respect to $T$
\begin{eqnarray}\label{p_hat_1}
&&\hat {\rm P}(M,s) = \int_0^\infty {\rm P}(M,T) e^{-s T} \rmd T \nn \\
&& = \lim_{\epsilon_1 \to 0, \epsilon_2\to 0} \frac{A_{N,B}}{M^{N^2+3}} \prod_{i=1}^{N-1} \int_0^\infty \rmd q_i e^{-\frac{q_i^2}{2M^2}} e^{-\epsilon_1 |q_N| - \epsilon_2 |q'_N|} \nonumber \\
&&\times \int_0^\infty \rmd q_N  \int_0^\infty \rmd {q_N}' q_N {q_N}' \frac{\Theta_N(q_1,\cdots, q_N)}{\frac{q_N^2}{2M^2}+s} \nn \\
&& \times \frac{\Theta_N(q_1,\cdots, {q_N}')}{\frac{{q_N}^{'2}}{2M^2}+s} \;,
\end{eqnarray}
where we have regularized the integrals over $q_N$ and $q_N'$ with the help of this exponential term $e^{-\epsilon_1 |q_N| - \epsilon_2 |q'_N|}$. The next step is to expand the determinants in Eq. (\ref{p_hat_1}) with respect to their last column. This yields
\begin{eqnarray}\label{p_hat_2}
&&\hat {\rm P}(M,s) = \frac{4 A_{N,B}}{M^{N^2-1}}\prod_{i=1}^{N-1} \int_0^\infty \rmd q_i e^{-\frac{q_i^2}{2M^2}} \sum_{j,l=1}^N (-1)^{j+l} \nonumber \\
&&\times \det[{\bf M}_{j,N}(\Theta_N)] \det[{\bf M}_{l,N}(\Theta_N)] g_j(s) g_l(s) \;,
\end{eqnarray}
where 
\begin{eqnarray}\label{integral_g}
g_j(s) = \lim_{\epsilon \to 0} \int_0^\infty \rmd q  \frac{q^j \cos{\left(q - \frac{j \pi}{2}\right)}}{q^2 + 2 M^2 s} e^{-\epsilon |q|} \;.
\end{eqnarray}
This integral (\ref{integral_g}) can be easily evaluated using residues, this yields
\begin{eqnarray}\label{integral_g_explicit}
g_j(s) = \frac{\pi}{2} (M \sqrt{2s})^{j-1} e^{-M \sqrt{2s}} \;.
\end{eqnarray}
Using this result (\ref{integral_g_explicit}) in the expression above (\ref{p_hat_2}), one obtains
\begin{multline}
\hat {\rm P}(M,s) = \frac{\pi^2 A_{N,B}}{M^{N^2-1}}\prod_{i=1}^{N-1} \int_0^\infty \rmd q_i e^{-\frac{q_i^2}{2M^2}} \sum_{j,l=1}^N (-1)^{j+l}\\
\det[{\bf M}_{j,N}(\theta_N)] \det[{\bf M}_{l,N}(\theta_N)] (M\sqrt{2s})^{j+l-2} e^{-2M\sqrt{s}} \;.
\end{multline}
Using the results identity shown above (\ref{integral_g}, \ref{integral_g_explicit}) we can write $\hat {\rm P}(M,s)$ as
\begin{eqnarray}\label{p_hat_3}
&&\hat {\rm P}(M,s) = -\frac{\pi A_{N,B}}{M^{N^2+1}}\prod_{i=1}^{N-1} \int_0^\infty \rmd q_i e^{-\frac{q_i^2}{2M^2}} \nn \\
&&\times \sum_{j,l=1}^N (-1)^{j+l}\det[{\bf M}_{j,N}(\theta_N)] \det[{\bf M}_{l,N}(\theta_N)] \nn \\
&&\times \int_0^\infty \rmd q q^{j+l-1} \frac{\sin{\left(2q - (j+l) \frac{\pi}{2} \right)}}{s + \frac{q^2}{2M^2}} \;.
\end{eqnarray}
Under this form (\ref{p_hat_3}) it is easy to invert the Laplace transform to obtain ${\rm P}(M,T)$ and finally one obtains
\begin{multline}\label{final_p_hat}
\int_0^1 P_{N,B}(M, \tau_M) \rmd \tau_M \\
= - \frac{\pi A_{N,B}}{M^{N^2+1}} \int_0^\infty \rmd {\bf q} \; e^{-\frac{{\bf q}^2}{2M^2}} \sum_{j,l=1}^N (-1)^{j+l} q_N^{j+l-1} \\ 
\times \det[{\bf M}_{j,N}(\theta_N)] \det[{\bf M}_{l,N}(\theta_N)] \sin{\left(2 q_N - (l+j) \frac{\pi}{2}\right)} \;.
\end{multline}
Finally, using the identity (\ref{int_taum}) together with the explicit expressions (\ref{Fn_prime_2}, \ref{final_p_hat}) one obtains
\begin{eqnarray}
A_{N,B} = \frac{1}{\pi} N a_{N,B}= \frac{2^{2N-N/2}\, N}{\pi^{N/2+1} \prod_{j=1}^{N} j!} \;,
\end{eqnarray}
as given in the text in Eq.~(\ref{amplitude_AN}).

\subsection{Non-intersecting Brownian excursions}
\label{normalization_excursions}

In this subsection we compute the constant $A_{N,E}$ of the formula~(\ref{Nexcursions_first}) with the same method.
Let us recall the joint distribution of $M$ and $\tau_M$ for the $N$ vicious excursions
\begin{multline}
\label{normalization_excursions_start}
P_{N,E}(M,\tau_M)= \frac{A_{N,E}}{M^{N(2N+1)+3}} \prod_{i=1}^{N-1} \left\{ \sum_{n_i=1}^{\infty}
n_i^2 e^{-\frac{\pi^2}{2M^2}n_i^2} \right\} \\
\times \Omega_N(\{n_i\} |M,\tau_M) \Omega_N(\{n_i\}|M,1-\tau_M) ,
\end{multline}
and where
\begin{equation}
\label{normalization_excursions_omega}
\Omega_N(\{n_i\} |M,t) = \sum_{k=1}^{\infty} (-1)^k k^2 \Delta_N(\{n_i^2\},k^2) e^{-\frac{\pi^2}{2M^2}t k^2} ,
\end{equation}
with the same notation as in the main text $\{n_i\}=(n_1,\dots,n_{N-1})$ and $\{n_i^2\}=(n_1^2,\dots,n_{N-1}^2)$.
The normalization of this joint pdf is
\begin{equation}
\int_0^\infty \rmd M \int_0^1 \rmd \tau_M\ P_{N,E}(M,\tau_M) = 1.
\end{equation}
As in the case of bridges we can use the result of Ref.~\cite{Schehr08} of the cumulative distribution of the maximum of $N$ vicious excursions
\begin{multline}
F_{N,E}(M)=\frac{a_{N,E}}{M^{2N^2+N}}\\
\times \sum_{n_1=1}^{\infty} \dots \sum_{n_N=1}^{\infty} \left\{ \prod_{i=1}^N {n_i}^2\ \Delta_N^2(n_1^2,\dots,n_N^2)
e^{-\frac{\pi^2}{2M^2} \gras{n}^2} \right\},
\end{multline}
where $\Delta_N^2$ is the square of the Vandermonde determinant
\begin{equation}
\Delta_N(n_1^2,\dots,n_N^2)= \det_{1\leq i,j \leq N} n_i^{2(j-1)} ,
\end{equation}
and $a_{N,E}$ is a normalization constant obtained by imposing $\lim_{M\to \infty} F_{N,E}(M)=1$:
\begin{equation}
a_{N,E}=\frac{\pi^{2N^2+N}}{2^{N^2-N/2} \prod_{j=0}^{N-1} \Gamma(2+j) \Gamma\left(\frac{3}{2} +j \right)} .
\end{equation}
The derivative of the cumulative gives the pdf of the maximum
\begin{multline}
\label{normalization_excursions_tocompare}
P_{N,E}(M)=\frac{\rmd F_{N,E}(M)}{\rmd M} \\
=\frac{a_{N,E}}{M^{2N^2+N+1}} \sum_{n_1} \dots \sum_{n_N} \Bigg\{ \left( -(2N^2+N)+\frac{N \pi^2}{M^2} {n_N}^2\right) \\
\times  \prod_{i=1}^N {n_i}^2 \Delta_N^2({n_1}^2,\dots,{n_N}^2)
e^{-\frac{\pi^2}{2M^2} \gras{n}^2} \Bigg\}.
\end{multline}
The term $\frac{N \pi^2}{M^2} {n_N}^2$ in the parenthesis of the second line comes from the derivative of the exponential,
giving a factor $\frac{\pi^2}{M^2} \sum_i {n_i}^2$ that can be simplified by the symmetry in the exchange of $n_i$ and $n_N$.
\par
Now we have to compute the marginal $P_{N,E}(M)$ by integrating the joint pdf given by Eq~(\ref{normalization_excursions_start}) 
over $\tau_M$
\begin{equation}
P_{N,E}(M)=\int_0^1 \rmd \tau_M \ P_{N,E}(M,\tau_M)
\end{equation}
to identify our constant $A_{N,E}$.
As in the Bridges case, we compute instead the function of $T$
\begin{multline}
\label{normalization_excursions_P}
{\rm P}(M,T)=
\frac{A_{N,E}}{M^{N(2N+1)+3}}
\prod_{i=1}^{N-1} \left\{ \sum_{n_i=1}^{\infty}
n_i^2 e^{-\frac{\pi^2}{2M^2}n_i^2} \right\} \\
\times \int_0^T \rmd \tau_M \ \Omega_N\left(\{n_i\}|M,\tau_M\right) \Omega_N\left(\{n_i\}|M,T-\tau_M \right),
\end{multline}
and at the end we will recover
\begin{equation}
P_{N,E}(M)={\rm P}(M,T=1) .
\end{equation}
The formula for ${\rm P}(M,T)$ is best treated with a Laplace transform, 
because the convolution in the second line of Eq.~(\ref{normalization_excursions_P}) becomes a simple product.
Taking the Laplace transform with respect to $T$, one has
\begin{eqnarray}
&&\hat{{\rm P}}(M,s)=\int_0^\infty \rmd T \ e^{-sT}\ {\rm P}(M,T)\nn\\
&&=\frac{A_{N,E}}{M^{N(2N+1)+3}}
\prod_{i=1}^{N-1} \left\{ \sum_{n_i=1}^{\infty}
n_i^2 e^{-\frac{\pi^2}{2M^2}n_i^2} \right\} \nn \\
&& \times \left(\hat{\Omega}_N(\{n_i\}|M,s) \right)^2 \;,
\end{eqnarray}
the Laplace transform of $\Omega_N$ being
\begin{align}
\hat{\Omega}_N&(\{n_i\}|M,s) = \int_0^\infty \rmd T\ e^{-sT}\ \Omega_N(\{n_i\}|M,T) \nn \\
&=\lim_{\epsilon \xrightarrow{<} 1}
 \sum_{k=1}^\infty (-\epsilon)^k \frac{k^2 \Delta_N({n_1}^2,\dots,{n_{N-1}}^2,k^2)}{s+\frac{\pi^2}{2M^2} k^2} \;.
\end{align}
This $\epsilon<1$ has been introduced to regularize the integrand so that one can permute the integral from the Laplace transform
and the infinite sum over $k$. Writting the Vandermonde determinant as
\begin{multline}
\Delta_N({n_1}^2,\dots,{n_{N-1}}^2,k^2)=\prod_{i=1}^{N-1} (k^2-{n_i}^2) \\
\times \prod_{1\leq i < j \leq N-1} ({n_j}^2-{n_i}^2),
\end{multline}
one can use the identity, shown below in Eq.~(\ref{identity}), with $p^2=2M^2 s/\pi^2$ with $s>0$ to get
\begin{multline}
\tilde{\Omega}_N(\{n_i\}|M,s) = (-1)^N \sqrt{2s} \frac{M^3}{\pi^2} 
\frac{\prod_{i=1}^{N-1} \left(n_i^2+\frac{2M^2}{\pi^2}s\right)}{\sinh\left(M \sqrt{2s}\right)} \\
\times  \prod_{1\leq i < j \leq N-1} ({n_j}^2-{n_i}^2)
\end{multline}
Under this form, we are ready to take the inverse Laplace transform of $\hat{{\rm P}}(M,s)$
using the Bromwich integral:
\begin{equation}
{\rm P}(M,T)=\frac{1}{2 i \pi} \int_{\gamma-i\infty}^{\gamma+i\infty} \rmd s \ e^{s T} \hat{{\rm P}}(M,s) ,
\end{equation}
where $\gamma$ is such that the vertical line keeps all the poles of the integrand to its left.
The poles in the complex plane are located in $s=-(k^2 \pi^2)/(2 M^2)$, for $k=1,2, \dots$
on the negative real axis, so that $\gamma=0$ is well-suited. We close the contour with a semi-circle to the left.
By the Jordan lemma, the integral on the semi-circle when $R \to \infty$ is zero.
The residue theorem then gives
\begin{align}
{\rm P}(M,T)&=
\frac{A_{N,E}}{M^{N(2N+1)+3}}
\prod_{i=1}^{N-1} \left\{ \sum_{n_i=1}^{\infty}
n_i^2 e^{-\frac{\pi^2}{2M^2}n_i^2} \right\} \nn \\
\times \Bigg\{ &\sum_{k=1}^{\infty} (-1)\frac{2 k^2 M^2}{\pi^2} e^{-\frac{\pi^2}{2 M^2} k^2 T} \nn \\
& \times  \prod_{1\leq i < j \leq N-1} ({n_j}^2-{n_i}^2)^2 \prod_{i=1}^{N-1} (k^2-n_i^2)^2 \nn \\
& \times \left[ \left(\frac{3}{2} -\frac{\pi^2}{2M^2}k^2 T\right) + \sum_{i=1}^{N-1} \frac{2k^2}{k^2-n_i^2} \right] \Bigg\} \;.
\end{align}
Taking $T=1$ and re-labelling $k=n_N$, one obtains
\begin{multline}
\label{normalization_excursions_nearfinal}
P_{N,E}(M)={\rm P}(M,T=1)=\frac{A_{N,E}}{M^{N(2N+1)+1}} \left(-\frac{2}{\pi^2} \right) \\
\times \sum_{n_1=1}^{\infty} \dots \sum_{n_N=1}^{\infty}
\prod_{i=1}^{N} n_i^2 \Delta_N^2(n_1^2,\dots,n_N^2) e^{-\frac{\pi^2}{2M^2} \gras{n}^2} \\
\times \left[\left( \frac{3}{2} -\frac{\pi^2}{2M^2}n_N^2\right) + \sum_{i=1}^{N-1} \frac{2n_N^2}{n_N^2-n_i^2} \right] \;.
\end{multline}
In order to compare this formula with Eq.~(\ref{normalization_excursions_tocompare}), we need to simplify
the last term in the brackets. To this end, we introduce the function $g(\gras{n}) \equiv g(n_1, \cdots, n_N)$:
\begin{eqnarray}
 g(\gras{n}) = \prod_{i=1}^{N} n_i^2 \Delta_N^2(n_1^2,\dots,n_N^2) \;,
\end{eqnarray}
which is totally symmetric under the exchange
of any couple of its arguments. One has
\begin{align}
&\sum_{n_1} \dots \sum_{n_N} \left( \sum_{i=1}^{N-1} \frac{2n_N^2}{n_N^2-n_i^2} \right)g(\gras{n}) \nn \\
=& \sum_{n_1} \dots \sum_{n_N} \left(  \sum_{i=1}^{N-1} \frac{2(n_N^2-n_i^2)+2n_i^2}{n_N^2-n_i^2} \right)g(\gras{n})  \nn \\
=& \sum_{n_1} \dots \sum_{n_N} \left( 2(N-1) + \sum_{i=1}^{N-1} \frac{2n_i^2}{n_N^2-n_i^2} \right) g(\gras{n}) \nn \\
=& \sum_{n_1} \dots \sum_{n_N} \left( 2(N-1) + \sum_{i=1}^{N-1} \frac{2n_N^2}{n_i^2-n_N^2} \right) g(\gras{n}) \;,
\end{align}
where we change the role of $n_i$ and $n_N$ in the last line. Equating the last line with the first, one obtains
\begin{equation}
\sum_{n_1} \dots \sum_{n_N} \left( \sum_{i=1}^{N-1} \frac{2n_N^2}{n_N^2-n_i^2} \right) g(\gras{n}) =
\sum_{n_1} \dots \sum_{n_N} (N-1) g(\gras{n}) .
\end{equation}
Inserting this identity in formula~(\ref{normalization_excursions_nearfinal}), 
the comparison with Eq.~(\ref{normalization_excursions_tocompare})
is straightforward, and one obtains
\begin{align}
A_{N,E}&=N \pi^2 a_{N,E} \nn \\
&=\frac{N \pi^{2N^2+N+2}}{2^{N^2-N/2} \prod_{j=0}^{N-1} \Gamma(2+j) \Gamma \left( \frac{3}{2} +j \right)} ,
\end{align}
as given in the text in Eq.~(\ref{Nexcursions_normalization})

\section{Identity}

We want to show the identity
\begin{multline}
\label{identity}
\lim_{\epsilon \xrightarrow{<} 1}
 \sum_{k=1}^\infty (-\epsilon)^k \frac{k^2 \prod_{i=1}^{N-1} (k^2-{n_i}^2)}{p^2+k^2} \\
 =(-1)^N \frac{\pi}{2} \ p\ \frac{\prod_{i=1}^{N-1} (p^2+ {n_i}^2)}{\sinh(\pi p)}
\end{multline}
This will be shown by induction on $N$: we thus first analyse the case $N=1$ of this identity
\begin{equation}
\label{identity_1}
\lim_{\epsilon \xrightarrow{<} 1}
 \sum_{k=1}^\infty (-\epsilon)^k \frac{k^2}{p^2+k^2}
 = - \frac{\pi}{2} \frac{p}{\sinh(\pi p)} \;,
 \end{equation}
for any real $p \neq 0$ (indeed this has a limit for $p=0$ and one can extend the result for $p=0$).
The first step is to separate the series in the lhs into two parts
\begin{align}
\lim_{\epsilon \xrightarrow{<} 1}
 \sum_{k=1}^\infty &(-\epsilon)^k \frac{k^2}{p^2+k^2} =
\lim_{\epsilon \xrightarrow{<} 1} 
 \sum_{k=1}^\infty (-\epsilon)^k \frac{k^2 +p^2-p^2}{p^2+k^2} \nn \\
 \label{identity_step1}
&= \lim_{\epsilon \xrightarrow{<} 1}
 \sum_{k=1}^\infty (-\epsilon)^k - 
 \lim_{\epsilon \xrightarrow{<} 1}
 \sum_{k=1}^\infty (-\epsilon)^k \frac{p^2}{p^2+k^2} \;.
 \end{align}
The first series is geometric and one has
\begin{equation}
\label{identity_step2}
\lim_{\epsilon \xrightarrow{<} 1} \sum_{k=1}^\infty (-\epsilon)^k
= \lim_{\epsilon \xrightarrow{<} 1} \left(\frac{1}{1+\epsilon} - 1\right) = -\frac{1}{2} \;.
\end{equation}
In the second, one can permute the limit with the sum because of the normal convergence
\begin{align}
 \lim_{\epsilon \xrightarrow{<} 1}
 \sum_{k=1}^{\infty} (-\epsilon)^k \frac{p^2}{p^2+k^2}&=
p^2 \sum_{k=1}^{\infty} (-1)^k \frac{1}{k^2+p^2} \nn \\
&=p^2 \frac{1}{2} \left( \sum_{k=-\infty}^{+\infty} (-1)^k \frac{1}{k^2+p^2} - \frac{1}{p^2} \right) \nn \\
\label{identity_step3}
&=  \frac{1}{2} \sum_{k=-\infty}^{+\infty} (-1)^k \frac{p^2}{k^2+p^2} -\frac{1}{2}
\end{align}
To compute the rhs, we consider the function on the complex plane
\begin{equation}
f(z)=\frac{p^2\ e^{i\pi z}}{z^2+p^2} \times \frac{1}{e^{i2 \pi z}-1} .
\end{equation}
This function has simple poles in $z=k$ with $k \in \mathbb{Z}$ and in $\pm i p$. 
One can verify that with $z=Re^{i\theta}$
\begin{equation}
|z f(z)| \leq \frac{Rp^2}{2|R^2-p^2|}\frac{1}{|\sinh(\pi R\sin\theta)|} \to_{R \to \infty} 0 . 
\end{equation}
Then using the Jordan lemma, one has that the contour integral of $f(z)$ along the infinite circle centered in $z=0$ is zero.
Using the residue theorem, one finds
\begin{align}
0=&\sum_{k\in\mathbb{Z}}(2i\pi)\left(-\frac{i}{2\pi}\right) \frac{p^2\ e^{ik\pi}}{k^2+p^2} \nn \\
&+(2i \pi) \frac{p^2 \ e^{-\pi p}}{2i p} \frac{1}{e^{-2\pi p}-1} \nn \\
&+(2i \pi) \frac{p^2 \ e^{\pi p}}{-2i p} \frac{1}{e^{2\pi p}-1} ,
\end{align}
or
\begin{equation}
\label{identity_step4}
\sum_{k\in\mathbb{Z}} (-1)^k \frac{p^2}{k^2+p^2} = \frac{\pi p}{\sinh(\pi p)} .
\end{equation}
Inserting Eq.~(\ref{identity_step4}) in Eq.~(\ref{identity_step3}) and Eq.~(\ref{identity_step2}) in Eq.~(\ref{identity_step1}), one has
\begin{align}
\lim_{\epsilon \xrightarrow{<} 1}
 \sum_{k=1}^\infty (-\epsilon)^k \frac{k^2}{p^2+k^2}
&= -\frac{1}{2}-\left( \frac{1}{2}  \frac{\pi p}{\sinh(\pi p)}  - \frac{1}{2} \right) \nn \\
&= -\frac{\pi}{2} \frac{p}{\sinh(\pi p)} \;, 
\end{align}
which is the announced result for $N=1$.
\par
Let us prove the general result by induction. Consider the quantity
\begin{multline}
\label{identity_induction_step}
\lim_{\epsilon \xrightarrow{<} 1}
 \sum_{k=1}^\infty (-\epsilon)^k \frac{k^2 \prod_{i=1}^{N} (k^2-{n_i}^2)}{p^2+k^2}
 \\
=\lim_{\epsilon \xrightarrow{<} 1}
 \sum_{k=1}^\infty (-\epsilon)^k \frac{k^2 (k^2+p^2-(p^2+{n_N}^2))\prod_{i=1}^{N-1} (k^2-{n_i}^2)}{p^2+k^2}
 \\
=\lim_{\epsilon \xrightarrow{<} 1}
 \sum_{k=1}^\infty (-\epsilon)^k \ k^2 \prod_{i=1}^{N-1} (k^2-{n_i}^2) \\
 -
\lim_{\epsilon \xrightarrow{<} 1}
 (p^2+{n_N}^2) \sum_{k=1}^\infty (-\epsilon)^k \frac{k^2 \prod_{i=1}^{N-1} (k^2-{n_i}^2)}{p^2+k^2} \;.
 \end{multline}
With the lemma (which we show below)
\begin{equation}
\label{identity_lemma}
\lim_{\epsilon \xrightarrow{<} 1}
 \sum_{k=1}^\infty (-\epsilon)^k \ k^{2q} = 0, \quad \forall q \in \mathbb{N}^* ,
 \end{equation}
and the induction hypothesis at rank $N$~(\ref{identity}) that we use in the second term of Eq.~(\ref{identity_induction_step}),
we have proved the formula at rank $N+1$ which ends the induction proof of the identity~(\ref{identity}).
\par
In order to prove the lemma~(\ref{identity_lemma}), we put $\epsilon=e^{-x}$ with the real $x>0$, 
and we shall take the limit $x \xrightarrow{>} 0$. Denoting for $q=0,1,2, \dots$
\begin{equation}
\label{identity_g}
g_q(x)=\sum_{k=1}^{\infty} (-1)^k e^{-kx} k^{2q},
\end{equation}
one sees that
\begin{equation}
\label{identity_derivative}
\frac{\rmd^2}{\rmd x^2} g_q(x)= g_{q+1}(x).
\end{equation}
Using the fact that $g_0(x)$ is a simple geometric series, we apply this relation to compute $g_1(x)$:
\begin{align}
g_1(x)&=\frac{\rmd^2}{\rmd x^2} g_0(x) \nnÊ\\
&=\frac{\rmd^2}{\rmd x^2} \left( \frac{1}{1+e^{-x}}-1 \right) \nn \\
&=\frac{e^{-x}\left(e^{-x}-1\right)}{\left(1+e^{-x}\right)^3}
\label{identity_function}
\end{align}
The domain of definition of $g_1(x)$ is the positive real axis, and in this domain, $g_1(x)$ coincides with
the function defined over all the real line
\begin{equation}
x \mapsto \hat{g}_1(x)=\frac{e^{-x}\left(e^{-x}-1\right)}{\left(1+e^{-x}\right)^3} .
\end{equation}
One sees easily that $\hat{g}_1(x)$ is an odd function $\hat{g}_1(-x)=-\hat{g}_1(x)$,
which implies that $\lim_{x \to 0}g_1(x)=\hat{g}_1(0)=0$, which shows the lemma for $q=1$.
To go further, the Taylor series of $\hat{g}_1(x)$ must have only odd powers of $x$. 
Because they coincide on $x>0$, this is true for $g_1(x)$:
\begin{equation}
\forall x>0,\quad g_1(x)=\sum_{p=0}^{\infty} a_{1,p} x^{2p+1} .
\end{equation}
Indeed this odd power series expansion extends to all $g_q(x)$ (with other coefficients)
thanks to the relation~(\ref{identity_derivative}) because we differentiate twice to obtain $g_{q+1}$ from $g_q$ 
\begin{equation}
\forall x>0, \quad g_q(x)=\sum_{p=0}^{\infty} a_{q,p} x^{2p+1},
\end{equation}
and this prove that
\begin{equation}
\lim_{x \xrightarrow{>} 0} g_q(x)=0,
\end{equation}
which, regarding the definition~(\ref{identity_g}), ends the proof of the lemma~(\ref{identity_lemma}).

\section{Some useful formulae for Hermite polynomials}
\label{Hermite}

For our purpose, the following integral representation of the Hermite polynomial $H_n(z)$ is useful
\begin{eqnarray}
 H_n(z) = \frac{2^{n+1}}{\sqrt{\pi}} e^{z^2} \int_0^\infty dt e^{-t^2} t^n \cos{(2zt - n \pi/2)}
\end{eqnarray}
The formula used in the text is explicitely
\begin{multline}
\label{Hermite_watermelons}
\int_0^\infty \rmd q\ q^n \cos\left(q-n \frac{\pi}{2}\right) e^{-\frac{t}{2M^2} q^2} \\
=
\sqrt{\pi} \left( \frac{M}{\sqrt{2t}} \right)^{n+1} e^{-\frac{M^2}{2t}} H_n\left(\frac{M}{\sqrt{2t}}\right)
\end{multline}
For the stars configurations we use the identity
\begin{multline}
\label{hermite_zeta}
\int_0^\infty \rmd q \ q^n \cos\left(q \zeta - n \frac{\pi}{2}\right) e^{-t \frac{q^2}{2M^2}} = \\
\sqrt{\pi} \left( \frac{M}{\sqrt{2t}}\right)^{n+1} e^{-\frac{M^2}{2t} \zeta^2} H_n\left(\frac{M}{\sqrt{2t}} \zeta \right)
\end{multline}
(in particular with $t=1-\tau_M$ and $n=1$).

\end{document}